\documentclass[11pt]{article}

\usepackage{slashed}
\usepackage{color}
\usepackage{graphicx}  
	\graphicspath{{graphs/}}
\usepackage[all]{xypic}
\usepackage{psfrag}
\usepackage{tocbibind} 
\usepackage{cases}
\usepackage{epsf}
\usepackage{epic}
\usepackage{eepic}
\usepackage{epsfig}
\usepackage{wrapfig}
\usepackage{a4}
\usepackage{amssymb,latexsym}
\usepackage[mathscr]{eucal}
\usepackage{cite}
\usepackage{citesort}
\usepackage[active]{srcltx}
\usepackage{url}
\usepackage{fancybox}
\usepackage{ntheorem}
\usepackage{mhequ}
\usepackage[g]{esvect}
\usepackage{epstopdf}

\newcommand{\ptcheck}[1]{}

\newcommand{\ptclevoca}[1]{}
\newcommand{\ptctodo}[1]{}

\newcommand{\viennan}[1]{}
\newcommand{\ptcrn}[1]{}
\newcommand{\nobeijingxn}[1]{}
\newcommand{\krakown}[1]{}
\newcommand{\nobeinjingxn}[1]{}

\newcommand{\syncx}[1]{\ptcxx{here starts a sync, or possibly a restrict environment}}
\renewcommand{\syncx}[1]{\ptc{syncx command here}{\color{red}#1}}

\newcommand{\ptcxx}[1]{\mnote{{\bf ptc:} {\color{red} #1}}}

\newcommand{\mnotex}[1]
{\protect{\stepcounter{mnotecount}}$^{\mbox{\footnotesize
$
\bullet$\themnotecount}}$ \marginpar{
\raggedright\tiny\em
$\!\!\!\!\!\!\,\bullet$\themnotecount: #1} }

\newcommand{\jamesx}[1]{}
\renewcommand{\jamesx}[1]{{\mnote{{\color{blue}{\bf jg:}
#1} }}}

\newcommand{\cref}[1]{\mbox{{\color{red}FIXME; what is the 4 for?}}4\emph{\ref{#1})}}

\newcommand{\h}[2]{#1\dotfill\ #2\\\ptc{fixme}}

\newcommand{\wascalO}{{O }}
\newcommand{\wascalOof}{{\wasCalOof}}
\newcommand{\wasCalOof}{\redOof}
\newcommand{\redOofO}{ {O(}}
\newcommand{\redOofo}{{ {o(}}}
\newcommand{\redOof}{{ {O(}}}

\newcommand{\ellhere}{{  L}}%

 \newcommand{\chitwo}{{  \iota}}
\newcommand{\hhere}{{  \tilde g}}
\newcommand{\ringlambda}{{\mathring\hhere}}%

\newcommand{\HLhere}{{  H}}%
\newcommand{\sectionofScri}%
{{ \,\,\,\,\mathring{\!\!\!\!\mcN}}}
\newcommand{\clmcN}{{  \,\,\,\,\overline{\!\!\!\!\mcN}}}
\newcommand{\sectionofscri}{\sectionofScri}

\newcommand{\Mcheckhere}{{ \check N}}
\newcommand{\HMsection}{S^1\times \Mcheckhere}

\newcommand{\Rlocal}{{\rho}}

\newcommand{\betatwo}{{ \psi}}

\newcommand{\bbh}{{ \tilde b}}

\newcommand{\chicoordinate}{{
    \alpha}}
\newcommand{\chitensor}{{\chi}}

\newcommand{\bvarepsilon}{{\varepsilon}}

\newcommand{\rnoBo}{\rbo}

\newcommand{\ptmcM}{{\partial \tmcM}}%

\newcommand{\nuoa}{\nuoasymptoticvalue}
\newcommand{\nuAa}{\nuAasymptoticvalue}
\newcommand{\nuBa}{\nuBasymptoticvalue}
\newcommand{\nuoasymptoticvalue}{\left(\nu^0_\Bo\right)_0}
\newcommand{\nuAasymptoticvalue}{\left(\nu^A_\Bo\right)_0}
\newcommand{\nuBasymptoticvalue}{\left(\nu^B_\Bo\right)_0}

\newcommand{\hring}{\mathring h}
\newcommand{\hringmeasure}{d \mu_{\mathring h}}
\newcommand{\hringvolume}{\mu_{\mathring h}(\sectionofScri)}

\newcommand{\limatrzero}{{\lim_{ r\to r_0}}}%
\newcommand{\intfromrzero}{\int_{r=r_0}}%
\newcommand{\sintfromrzero}{\int_{s= 0}}%
\newcommand{\tintfromrzero}{\int_{\tilde r=r_0}}%

\newcommand{\coneg}{\check g}
\newcommand{\coneGamma}{\check \Gamma}
\newcommand{\conenabla}{\check \nabla}
\newcommand{\coneR}{\check R}

\newcommand{\mTB}{m_{\mbox{\rm\scriptsize TB}} }

\newcommand{\renm}{{\mbox{\rm\scriptsize ren}}}

\newcommand{\Bobig}{{\mbox{\rm\scriptsize Bo}}}

\newcommand{\Bo}{\Bobo}
\newcommand{\Bobo}{\mathrm{Bo}}

\newcommand{\Bh}{{\mbox{\rm\scriptsize Bh}}}
\newcommand{\Bht}{{\mbox{\rm\tiny Bh}}}

\newcommand{\HM}{{\mbox{\rm\scriptsize HM}}}
\newcommand{\HMt}{{\mbox{\rm\tiny HM}}}

\newcommand{\rbo}{r_{\mathrm{Bo}}}

\newcommand{\betaBondi}{\omega}

\newcommand{\sigmaafAB}{\sigma_A {}^{ B}}
\newcommand{\sigmabo}{\sigma^{\Bo}{}}
\newcommand{\sigmaboAB}{\sigma^{\Bo}{}\!_A {}^{ B}}

\newcommand{\Charmass}{Characteristic mass}
\newcommand{\charmass}{characteristic mass}
\newcommand{\og}{{\overline{g}}}

\newcommand{\tnabla}{\widetilde \nabla}
\newcommand{\zGamma}{{\mathring \Gamma}{}}

\newcommand{\zero}{{\hat 0}}
\newcommand{\one}{{\hat 1}}
\newcommand{\two}{{\hat 2}}
\newcommand{\three}{{\hat 3}}
\newcommand{\hmu}{{\hat \mu}}
\newcommand{\hnu}{{\hat \nu}}

\newcommand{\hB}{{\hat B}}

\newcommand{\localch}{{\check h}}
\newcommand{\mham}{m_{\textrm{\scriptsize Ham}}}
\newcommand{\gK}{generalized Kottler}

\newcommand{\eqref}[1]{\eq{#1}}

\newcommand{\hs}{\cH_{\mbox{\scriptsize sing}}}

\newcommand{\beadl}[1]{\begin{deqarr}\label{#1}}
\newcommand{\eeadl}[1]{\arrlabel{#1}\end{deqarr}}%

\newcommand{\ol}{\overline}

\def\nz{\ifmmode {I\hskip -3pt N} \else {\hbox {$I\hskip -3pt N$}}\fi}
\def\zz{\ifmmode {Z\hskip -4.8pt Z} \else
       {\hbox {$Z\hskip -4.8pt Z$}}\fi}
\def\qz{\ifmmode {Q\hskip -5.0pt\vrule height6.0pt depth 0pt
       \hskip 6pt} \else {\hbox
       {$Q\hskip -5.0pt\vrule height6.0pt depth 0pt\hskip 6pt$}}\fi}
\def\rz{\ifmmode {I\hskip -3pt R} \else {\hbox {$I\hskip -3pt R$}}\fi}
\def\cz{\ifmmode {C\hskip -4.8pt\vrule height5.8pt\hskip 6.3pt} \else
       {\hbox {$C\hskip -4.8pt\vrule height5.8pt\hskip 6.3pt$}}\fi}
\def\au{{\setbox0=\hbox{\lower1.36775ex\hbox{''}\kern-.05em}\dp0=.36775ex\hs
kip0pt\box0}}
\def\ao{{}\kern-.10em\hbox{``}}

\newcommand\Gregbeq{\begin{eqnarray}}
\newcommand\Gregeeq{\end{eqnarray}}

\newcommand{\scri}{{\mycal I}}%
\newcommand{\scrip}{\scri^{+}}%
\newcommand{\Scri}{\scri}

\def\cH{{\cal H}}

\def\h1{{\hat 1}}
\def\h2{{\hat 2}}

\def\hA{{\hat A}}

\def\3f{\frac{3}{2}}

\newcommand{\roscoff}[1]{}

{\catcode `\@=11 \global\let\AddToReset=\@addtoreset}
\AddToReset{figure}{section}

\DeclareFontFamily{OT1}{rsfs}{}
\DeclareFontShape{OT1}{rsfs}{m}{n}{ <-7> rsfs5 <7-10> rsfs7 <10-> rsfs10}{}
\DeclareMathAlphabet{\mycal}{OT1}{rsfs}{m}{n}

{\catcode `\@=11 \global\let\AddToReset=\@addtoreset}
\AddToReset{equation}{section}

\newcounter{mnotecount}[section]

\renewcommand{\themnotecount}{\thesection.\arabic{mnotecount}}

\newcommand{\oversetty}[2]{%
\mathop{#2}\limits^{\vbox to -.1ex{%
\kern -1.5ex\hbox{$\scriptstyle #1$}\vss}}}

\newcommand{\tmcM}{\,\,\,\,\widetilde{\!\!\!\!\mcM}}%

\newcommand{\jlcax}[1]{}
\newcommand{\eean}{\nonumber\end{eqnarray}}

\newcommand{\Mtwo}{{M}{}^2}

\newcommand{\tGamma}{{\widetilde \Gamma}}%
\newcommand{\ztGamma}{\mathring{\tGamma}{}}%

\newcommand{\kk}[1]{}

\newcommand{\eqs}[2]{\eq{#1}-\eq{#2}}

\newcommand{\beq}{\begin{equation}}

\newcommand{\rd}{\,{ d}} 

\newcommand{\T}{\mathbb T}

\newcommand{\zn} {\mathring{\nabla}} 

\newcommand{\FS}       
                  {F}

\newcommand{\HS} 
       {H_{\mbox{\scriptsize volume}}}

{\ptc{this should be removed in the oberwolfach version}}%

\newcommand{\mcT}{{\mycal T}}%

\newcommand{\ourU}{\mathbb U}%
\newcommand{\eel}[1]{\label{#1}\end{equation}}
\newcommand{\eeal}[1]{\label{#1}\end{eqnarray}}
\newcommand{\bed}{\begin{deqarr}}
\newcommand{\eed}{\end{deqarr}}
\newcommand{\bedl}[1]{\begin{deqarr}\label{#1}}
\newcommand{\eedl}[2]{\arrlabel{#1}\label{#2}\end{deqarr}}

\newcommand{\tg}{{\widetilde{g}}}
\newcommand{\ztg}{\,\mathring{\!\widetilde{g}}}

\newcommand{\mcO}{{\mycal O}}

\newcommand{\mcN}{{\mycal N}}

\newcommand{\bel}[1]{\begin{equation}\label{#1}}
\newcommand{\bea}{\begin{eqnarray}}
\newcommand{\bean}{\begin{eqnarray}\nonumber}
\newcommand{\beal}[1]{\begin{eqnarray}\label{#1}}
\newcommand{\eea}{\end{eqnarray}}

\newcommand{\nn}{\nonumber}
\newcommand{\Eq}[1]{Equation~\eq{#1}}

\newcommand{\qed}{\hfill $\Box$}

\newcommand{\be}{\begin{equation}}
\newcommand{\eeq}{\end{equation}}
\newcommand{\ee}{\end{equation}}
\newcommand{\beqa}{\begin{eqnarray}}
\newcommand{\eeqa}{\end{eqnarray}}
\newcommand{\beqan}{\begin{eqnarray*}}
\newcommand{\eeqan}{\end{eqnarray*}}
\newcommand{\ba}{\begin{array}}
\newcommand{\ea}{\end{array}}

\newcommand{\const}{\mbox{\rm const}} 

\newcommand{\hyp}{\mycal S}

\newcommand{\mcM}{{\mycal M}}

\newcommand{\zmcD}{\,\,\mathring{\!\!{\mycal D}}}

\newcommand{\mnote}[1]
{\protect{\stepcounter{mnotecount}}$^{\mbox{\footnotesize
$
\bullet$\themnotecount}}$ \marginpar{
\raggedright\tiny\em
$\!\!\!\!\!\!\,\bullet$\themnotecount: #1} }

\newcommand{\warn}[1]
{\protect{\stepcounter{mnotecount}}$^{\mbox{\footnotesize
$
\bullet$\themnotecount}}$ \marginpar{
\raggedright\tiny\em
$\!\!\!\!\!\!\,\bullet$\themnotecount: {\bf Warning:} #1} }

\newcommand{\R}{\mathbb R}

\newcommand{\backg}{b}
\newcommand{\eq}[1]{(\ref{#1})}

\newcommand{\ptc}[1]{\mnote{{\bf ptc:}#1}}

\newcommand{\mcL}{{\mycal L}}

\newcommand{\beqar}{\begin{deqarr}}
\newcommand{\eeqar}{\end{deqarr}}

\newcommand{\beaa}{\begin{eqnarray*}}
\newcommand{\eeaa}{\end{eqnarray*}}

\newcommand{\znabla}{\mathring{\nabla}}
\newcommand{\zR}{\mathring{R}}

\newcommand{\bmetric}{{b}} 

\newcommand{\tK}{\tilde K} 
\newcommand{\hrho}{\hat\rho}

\newcommand{\bethm}{\begin{theorem}}
\newcommand{\et}{\end{theorem}}
\newcommand{\bl}{\begin{Lemma}}

\newtheorem{Theorem} {\sc  Theorem\rm} [section]
\newtheorem{theorem} [Theorem] {\sc  Theorem\rm}

\newtheorem{Lemma} [Theorem] {\sc  Lemma\rm}

\newtheorem{Proposition} [Theorem] {\sc  Proposition\rm}

\theorembodyfont{\upshape}
\newtheorem{Remark}[Theorem]{\sc Remark\rm}

\theoremsymbol{\ensuremath{\diamondsuit}}
\newcommand{\fcoco}{\small}
\theorembodyfont{\fcoco}\theoremseparator{}\theoremindent0.5cm

\theoremstyle{nonumberplain}\theorembodyfont{\fcoco}
\theoremseparator{}\theoremindent0.5cm

\DeclareFontFamily{OT1}{rsfs}{}
\DeclareFontShape{OT1}{rsfs}{m}{n}{ <-7> rsfs5 <7-10> rsfs7 <10-> rsfs10}{}

\begin{document}
\title{The cosmological constant and the energy of gravitational radiation%
\thanks{Preprint UWThPh-2016-4.  Supported in part by the Austrian Science Fund (FWF): P 23719-N16. PTC is grateful to the Center for Mathematical Sciences and Applications, Harvard, and to Monash University for hospitality and support during part of work on this paper.}
}

\author{Piotr T.\ Chru\'{s}ciel\thanks{Email
\protect\url{piotr.chrusciel@univie.ac.at}; URL \protect\url{http://homepage.univie.ac.at/piotr.chrusciel}}
\
and Lukas Ifsits\thanks{Email
\protect\url{lukas.ifsits@univie.ac.at}}
\\ Erwin Schr\"odinger Institute and
Faculty of Physics \\
University of Vienna
}

\maketitle

\begin{abstract}
 We propose a definition of mass for characteristic hypersurfaces in asymptotically vacuum space-times with non-vanishing cosmological constant $\Lambda \in {\mathbb R}^*$, generalising the definition of Trautman and Bondi for $\Lambda =0$. We show that our definition reduces to some standard definitions in several situations. We establish a balance formula linking the characteristic mass and a suitably defined renormalised volume of the null hypersurface, generalising the positivity identity of one of us (PTC) and Paetz proved when $\Lambda=0$.
\end{abstract}

\noindent
\hspace{2.1em} PACS: 04.20.Cv, 04.20.Ex, 04.20.Ha

\tableofcontents
\section{Introduction}

While the notion of total mass of general relativistic gravitating systems with $\Lambda\le 0$ is well understood by now (cf., eg., \cite{ChEnergy} and references therein), the notion of energy in the radiating regime in the presence of a positive cosmological constant appears to be largely unexplored (see, however, \cite{AshtekarBongaKesavanI,AshtekarBK,TodSzabados}). The object of this work is to contribute to filling this gap.

In this paper we address the question of properties of total mass and energy for radiating systems when $\Lambda \ne 0$. This will be done in the spirit of the pioneering work by Bondi et al.~\cite{BBM,Sachs}, by analysing the asymptotic behavior of the gravitational field on characteristic hypersurfaces extending to asymptotic regions.
Many formal aspects of the problem turn out to be independent of the sign of the cosmological constant, and while we are mainly interested in the case $\Lambda>0$,   we allow $\Lambda<0$ wherever relevant as several results below apply regardless of the sign of $\Lambda$. The case $\Lambda<0$ becomes a useful test-bed for the quantities involved in those aspects thereof which are well understood. It should, however, be emphasised that many of our results, such as e.g.\ the balance equation~\eqref{eq:massfinal}, are new both for $\Lambda<0$ and $\Lambda>0$.

It should be kept in mind that an elegant approach to the definition of energy has been proposed in~\cite{AbbottDeser} for field configurations that asymptotically approach a preferred background with Killing vectors. This provides a widely accepted definition of asymptotic charges in the case where $\Lambda \le 0$. The approach of~\cite{AbbottDeser} does not work for non-trivial radiating fields with $\Lambda>0$, where no natural asymptotic background is known to exist. In retrospect, our work below can be used to provide such a background, namely the metric obtained by keeping only the leading order terms of $g$ in Bondi coordinates, but the decay rates of the metric to this background do not appear to be compatible with what is needed in the Abbott-Deser prescription.

The first issue that one needs to address is that of boundary conditions satisfied by the fields. A popular approach is to assume smooth conformal compactifiability of the space-time, and we develop a framework which covers such fields. We start by deriving in Section~\ref{s10I15.1} below the restrictions on the free characteristic initial data that follow from existence  of smooth conformal compactification. In particular, in Proposition~\ref{P12XII15.1} below  we generalise  to all $\Lambda \in \R$ a result established in~\cite{TamburinoWinicour}   for $\Lambda=0$,  that existence of a smooth conformal compactification guarantees existence of Bondi coordinates in which the metric coefficients have full asymptotic expansions in terms of inverse powers of the Bondi coordinate $r_{\Bobo}$. In Section~\ref{s24XII15.1} we review those aspects of the characteristic Cauchy problem which are relevant for the issues at hand. In Section~\ref{sec:solvingconstraints} we derive the asymptotic expansions of the metric along the characteristic surfaces. Our analysis   is similar in spirit to that of~\cite{ChPaetz3,TimAsymptotics}; however, here the asymptotic expansions have to be carried-out to higher orders because of new $\Lambda  $-dependent nonlinear couplings between some asymptotic expansion coefficients. We also allow matter sources, while vacuum was assumed in~\cite{TimAsymptotics}. In particular in Section~\ref{ss18II16.1} we derive the conditions \eq{18II16.2} on the free initial data in  Bondi-type coordinates which are necessary for absence of log terms in the metric.

The asymptotic expansions of Section~\ref{sec:solvingconstraints}  lead  naturally, in Section~\ref{s28XII15.1}, to the definition of a quantity analogous to the Trautman-Bondi mass. We derive there a key integral identity expressing this mass in terms of the free initial data and the \emph{renormalised volume} of the characteristic surface, Equation~\eqref{eq:massfinal}. This is one of the main results of this work.

\begin{figure}[th]
\begin{center}
 {\includegraphics[width=0.65\textwidth]{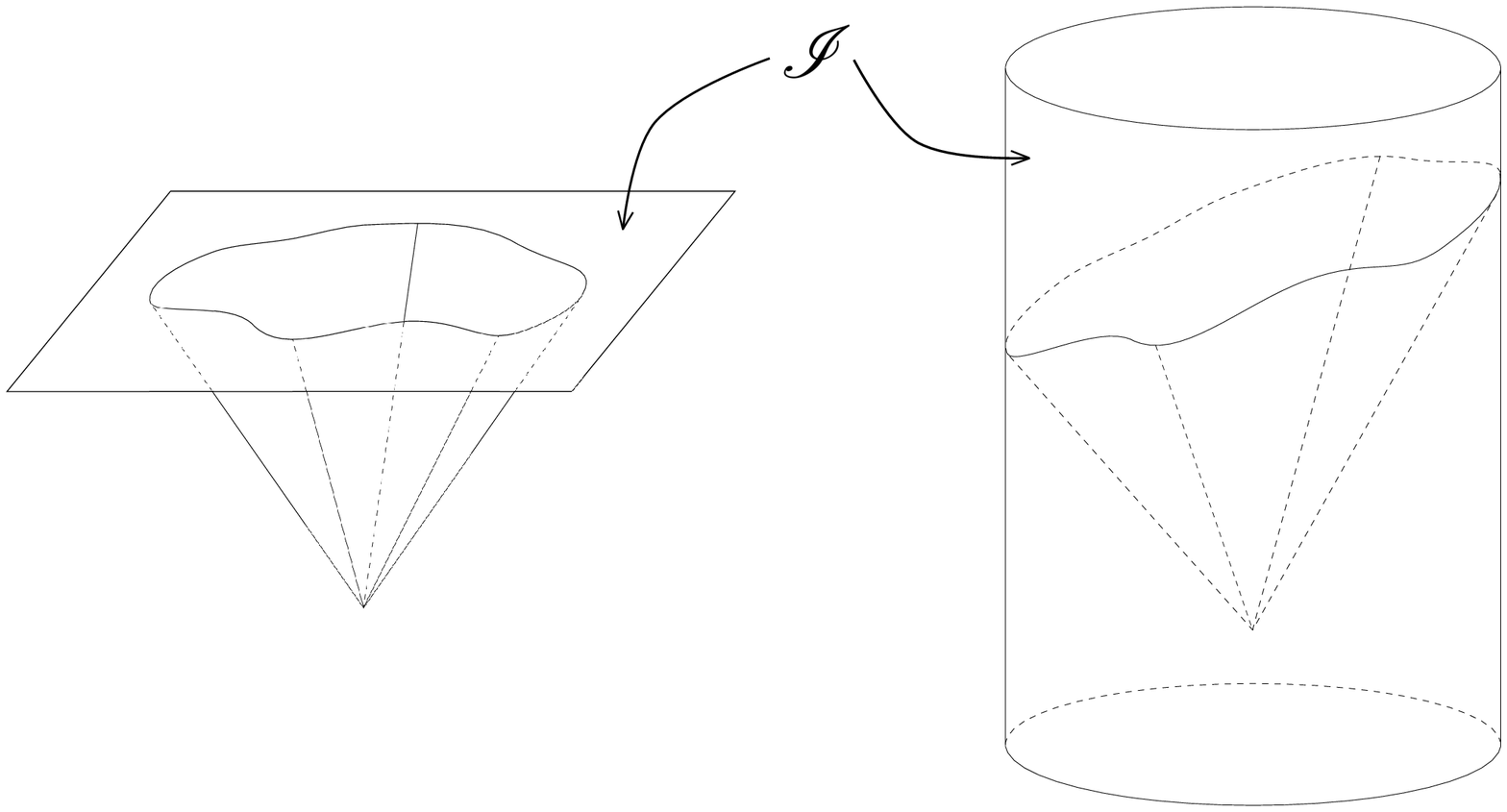}}
\caption{\label{F20I16.1}
Globally smooth light-cones in space-times with a smooth conformal completion at a conformal boundary $\scri$ at timelike infinity ($\Lambda >0$, left) or spacelike infinity ($\Lambda <0$, right). }
\end{center}
\end{figure}

When $\Lambda=0$, our mass identity~\eqref{eq:massfinal} reduces to the one derived in~\cite{ChPaetzBondi} (compare~\cite{TafelBondi2}), giving then an elementary proof of positivity of the Trautman-Bondi energy  for space-times containing globally smooth light-cones extending smoothly to $\scrip$. (As is well known the global structure of such space-times depends crucially upon the sign of $\Lambda$, see Figure~\ref{F20I16.1}.)
In addition to the renormalised volume, boundary terms, and volume integrals involving the free data, the new formula, for asymptotically empty metrics with $\Lambda \ne 0$, involves several terms depending upon coefficients determined by the asymptotic behaviour of the fields multiplied by  $\Lambda$. One can think of this equation as a \emph{balance formula} relating the mass with the remaining quantities at hand.

To get some insight into the formula, in the remaining sections we turn our attention to the case $\Lambda<0$, where energy is much better understood. We review the notion of coordinate mass in Section~\ref{s13I16.3}. We calculate the various quantities appearing in the mass identity~\eqref{eq:massfinal} for the Birmingham metrics and the Horowitz-Myers (HM) metrics in Section~\ref{s13I16.2}.
 \ptctodo{say something about general core geodesics space-times if done}
In Section~\ref{s18II15.1} we derive simple formulae for the Hamiltonian mass for asymptotically Birmingham and asymptotically Horowitz-Myers metrics, in all space-times dimensions $n+1\ge 4$,%
\footnote{%
This extends the analysis in~\cite{ChBamberg} and references therein to higher dimensions with the above boundary conditions.
}
and for smoothly conformally compactifiable four-dimensional space-times with an ultrastatic conformal boundary.
These formulae are used to show that the Hamiltonian mass coincides with the characteristic mass for a family of null hypersurfaces.
 In Appendix~\ref{s15II16.1} we examine separately various contributions to our ``energy balance'' equation for Horowitz-Myers metrics.

\ptctodo{do the renormalised volume identities in the static case?}

Unless explicitly indicated otherwise, we assume throughout that $\Lambda \ne 0$.

\section{Boundary conditions}
 \label{s10I15.1}

Consider an $(n+1)$-dimensional smoothly conformally compactifiable space-time $(\mcM,g)$, $n\ge 2$, solution of the vacuum Einstein equations with cosmological constant $\Lambda\ne 0$.  By definition, there exists a manifold $(\tmcM,\tg)$  with boundary $\ptmcM$ and a defining function $\Omega$ for $\ptmcM$ such that
\bel{10I15.1}
 g=\Omega^{-2}\tg
 \,,
  \quad
   \ptmcM=\{\Omega=0\}\,,
   \quad
   d\Omega(p)\ne 0 \ \mbox{for}\ p\in\ptmcM
   \,.
\ee
Again by definition, both $\Omega $ and $\tg$ are smooth.

\subsection{Bondi coordinates}
 \label{ss5XII15.1}

In the asymptotically flat case, in spacetime dimension $4$ and assuming $\Lambda=0$, Bondi et al.\ have introduced a set of coordinates convenient for analysing gravitational radiation~\cite{BBM}. We will refer to them  as {\em Bondi coordinates}. In these coordinates the metric takes the form
\bel{19II16.1}
	 \og = \og_{00} du^2 -2 e^{2\betaBondi} dr \,du - 2 r^2 U_A dx^A du + r^2  \underbrace{h_{AB}  dx^A dx^B}_{=:h}
	\,,
\ee
where the determinant of $h_{AB}$ is $r$-independent.%
\footnote{We have used the symbol $\betaBondi=\betaBondi(u,r,x^A)$ for a function which is usually denoted by $\beta$ in the literature to avoid a conflict of notation with a constant $\beta$ elsewhere in the paper.}
(One further requires the fields $g_{00}, \, U_A, \, \betaBondi$ and $h_{AB}$ to fulfill appropriate asymptotic conditions.) When using Bondi coordinates, we will decorate all fields and coordinates with a symbol ``$\Bobig$''. Existence of such coordinates in asymptotically vacuum space-times with $\Lambda =0$ and admitting smooth conformal completions has been established in~\cite{TamburinoWinicour}, and in~\cite{ChMS} for polyhomogeneous $\Scri$'s.

We wish to prove existence of such coordinates, and to derive the asymptotic behaviour of smoothly compactifiable metrics in those coordinates in a neighborhood of the conformal boundary, with $\Lambda\in \R^*$. It turns out that, similarly to the $\Lambda=0$ case (cf., e.g.,  \cite[Section~4]{ChMS}, compare~\cite{TimAsymptotics}), smoothness imposes restrictions on some lower-order coefficients in the asymptotic expansion of the free data on the null hypersurfaces meeting the conformal boundary smoothly and transversally.

Let, thus, $\Lambda \in \R$, and let $y^0:\ptmcM \to\R$ be a smooth function defined on an open subset of the conformal boundary $\ptmcM $, with $dy^0$ without zeros, such that the  level sets
$$
 S_c
 : =\{y^0=c\} \subset \ptmcM
$$
of $y^0$ form a smooth foliation by spacelike submanifolds. Passing to a subset of $\ptmcM$ if necessary, we can assume that $y^0$ is defined throughout $\ptmcM$.

So far $y^0$ has only been defined on the conformal boundary. Note that the gradient of $y^0$ within the boundary will be necessary timelike when $\ptmcM$ is timelike, and spacelike when $\ptmcM$ is spacelike. Nevertheless, we will extend $y^0$ to a function in space-time so that $\nabla y^0$ is null regardless of the causal character of the conformal boundary.

Now, at every $p\in S_c$ there exists a unique vector $\mathring X_p$ which is null, future directed, outwards pointing, orthogonal to $T_pS_c$, and normalised to unit length with respect to some smooth auxiliary Riemannian metric. This defines a smooth vector field $\mathring X$ on $\ptmcM$. We choose time-orientation so that $-X_p$ points towards the physical space-time.

Let $\gamma_p$ denote a maximally extended null geodesic with initial tangent $-\mathring X_p$ at $p$.
Standard transversality and injectivity-radius arguments show that there exists a neighborhood $\mcO$ of $\tmcM $ such that for every $c\in\R$ the union of the (images of the) null geodesics
$$
 \mcN_c:=\cup_{p\in S_c}\gamma_p
$$
forms a smooth null hypersurface, with $\cup_c S_c$ foliating $\mcO$.

To obtain Bondi-type coordinates we proceed now as follows:

\begin{enumerate}
  \item Let $x^ A$ denote local coordinates on $S_c$, in $3+1$ space-time dimensions we choose the conformal representative $\ringlambda_{AB}$ of the metric induced on $S_c$ by $\tg$ to take a canonical form. For example, if $S_0$ is diffeomorphic to a two-dimensional sphere, we choose $\ringlambda_{AB}$ to be the canonical metric $s_{AB}$ on $S^2$. In higher dimension one might wish to require instead that the volume element $\sqrt{\det\ringlambda_{AB}}$ takes a convenient form, depending upon the geometry of the conformal boundary.

      \item We extend the local coordinates $x^A$ from $\ptmcM $ to $\mcO$ by requiring the $x^A$'s to be constant along the null geodesics $\gamma_p$.
          \item Let $q\in \mcO$, then $q$ belongs to some null geodesic $\gamma_p$ defined above. We define the function $u$ by letting $u(q)=y^0(p)$. In other words, $u$ is defined to be equal to $c$ on $\mcN_c$.

              \item Set $x:=\Omega$, the conformal compactifying factor as in \eq{10I15.1}. Passing to a subset of $\mcO$ if necessary, the functions $(u,x,x^A)$ form a coordinate system on $\mcO$. By construction the curves $s\mapsto (u,x=s,x^A)$ are null geodesics initially normal to $S_c$:
\bel{29XI15.2}
 \tg(\partial_x,\partial_x) =0
  \,,
  \quad
 \tg(\partial_x,\partial_{x^A})|_{x=0} =0
 \,.
\ee
We recall the usual calculation, which uses the fact that $\partial_x$ is tangent to null geodesics, $\nabla_{\partial_x}\partial_x = \kappa \partial_x$ for some function $\kappa$:
\bean
 \lefteqn{
 \partial_x\big(\tg(\partial_A,\partial_x)\big)
  =\tg(\nabla_{\partial_x}\partial_A,\partial_x)
 + \tg(\partial_A,\nabla_{\partial_x}\partial_x)
 }
 &&
\\
\nn
 && =  \tg(\nabla_{\partial_A}\partial_x,\partial_x)
 + \kappa \tg(\partial_A,\partial_x) =
  \frac 12{\partial_A}\big( \tg(\partial_x,\partial_x)\big)
 + \kappa \tg(\partial_A,\partial_x)
\\
 && =
  \kappa \tg(\partial_A,\partial_x)
 \,.
\eea
Thus
$$
  \partial_x\big(\tg(\partial_A,\partial_x)\big)
  =
  \kappa \tg(\partial_A,\partial_x)
 \,,
$$
which provides a linear homogeneous ODE in $x$ for $\tg(\partial_A,\partial_x)$, with vanishing initial data at $x=0$. We conclude that
\bel{29XI15.1}
 \tg(\partial_x,\partial_x) =0
  \,,
  \quad
 \tg(\partial_x,\partial_{x^A}) =0
 \,.
\ee
Equivalently, the level sets of $u$ are null hypersurfaces generated by the integral curves of $\partial_x$.
\end{enumerate}

The Bondi radial coordinate $r_{\Bobo}$  (the subscript ``$\Bo$'' stands for ``Bondi'') is defined by setting
\bel{9VIII15.1}
                r_{\Bobo}:=
                \left( \frac{ \det g_{AB}}{\det \ringlambda_{AB}}\right) ^{\frac 1{2(n-1)}}=
                \frac 1  x \left( \frac{ \det \tg_{AB}}{\det \ringlambda_{AB}}\right) ^{\frac 1{2(n-1)}}
                \,,
\ee
where $n$ is the space-dimension.

The final coordinate system $(u_{\Bobo}, r_{\Bobo}, x^A_{\Bobo})$ is obtained by setting, in addition to \eq{9VIII15.1},
\bean
 &
 u_{\Bobo} := u
 \,,
 \
 x^A_{\Bobo} := x^ A
 \quad
 \Longrightarrow
 \quad
 \partial_{x} = \frac{\partial r_{\Bobo}} {\partial {x}} \partial_{r_
\Bobo}
 \,,
 \
 \partial_{x^A} = \partial_{x^A_{\Bobo}} + \frac{\partial x} {\partial {x^A_{\Bobo}}} \partial_x
 \,.
 &
\eea
It follows from \eq{29XI15.1} together with the last implication that
\bean
 &
   \tg(\partial_{r_
\Bobo},\partial_{r_
\Bobo}) = 0 =   \tg(\partial_{r_
\Bobo},\partial_{x^ A_
\Bobo})
  \,,
  \quad
   \tg( \partial_{x^A},\partial_{x^B})
   =
    \tg( \partial_{x^A_{\Bobo}},  \partial_{x^B_{\Bobo}} \big)
 \,,
 &
\eeal{9VIII15.3}
which shows that, on $
\mcO$,  the metric satisfies indeed the \emph{Bondi conditions}
\bel{9VIII15.4}
 \tg_{r_{\Bobo}x^A_{\Bobo}} =
  0
  =
 \tg_{r_{\Bobo}r_{\Bobo}}
 \,,
  \quad
 \sqrt{\det g_{x^A_{\Bobo} x^B_{\Bobo}}}
  = r^{n-1}
 \sqrt{\det \ringlambda_{x^A_{\Bobo} x^B_{\Bobo}}}
 \,.
\ee

\Eq{9VIII15.1} implies that $r_{\Bobo}$ has a full asymptotic expansion in terms of powers of $x$:
\bel{29XI15.4}
 r_{\Bobo}= \frac 1 x +
 (r_{\Bobo})_0(u,x^A) +
 (r_{\Bobo})_1(u,x^A) x  + \ldots
  \,,
\ee
where the asymptotic expansion coefficients $(r_{\Bobo})_n$ are functions of $(u,x^A)$.
This can be inverted to give a full asymptotic expansion
$$
  x = \frac 1 {r_{\Bobo}} +
 \frac{(r_{\Bobo})_0(u,x^A)}{r ^2_{\Bo}} +
 \frac{(r_{\Bobo})_0(u,x^A)^2 +
 (r_{\Bobo})_1(u,x^A)}{ r^{3}_{\Bo}}  +
 \ldots
  \,.
$$
(Indeed, if we set $y:=1/ r_{\Bobo}$, then \eq{9VIII15.1} becomes
               \bel{9VIII15.1bis}
               y =   x \left( \frac{\det \ringlambda_{AB}}{ \det \tg_{AB}}\right) ^{\frac 1{2(n-1)}}
                \,,
                \ee
and existence of a smooth function $x=x(y)$ follows from the implicit function theorem.)

Since all metric functions are smooth
in $(u,x^A,x)$, they have complete asymptotic expansions in terms of $1/r_{\Bobo}$, with coefficients depending smoothly upon
$(u_{\Bobo}, x^A_{\Bobo})$.

As a special case of the construction above, we have proved:

\begin{Proposition}
 \label{P12XII15.1}
Let $\mcN$ be a null hypersurface intersecting smoothly and transversally a section $S$ of conformal infinity in a smoothly conformally compactifiable space-time with cosmological constant $\Lambda\in \R$. There exist adapted coordinates $(r,x^ A)$ on $\mcN$ in which the restrictions $\overline g_{AB}$ to $\mcN$ of the metric functions $g_{AB}$ take the form
\bel{24I15.21}
 \overline g_{AB} = r^2 \Big( \ringlambda_{AB}(x^C) + O(r^{-1}) \Big)
 \,,
\ee
with  $r^{-2}\overline g_{AB}$ having full asymptotic expansions in terms of inverse powers of $r$.
These coordinates can be chosen to satisfy the Bondi conditions~\eqref{9VIII15.4} near $\mcN$. In dimension $3+1$ the metric $\ringlambda$ on $S$ can be arbitrarily chosen, in higher dimension $
\sqrt{\det\ringlambda}$ can be arbitrarily chosen.
\end{Proposition}

In what follows we wish to address two questions:
\begin{enumerate}
  \item Assuming vacuum, can the expansion above be made more precise? and
  \item How to read-off the mass from the above expansions?
\end{enumerate}
For this, some preliminary results will be needed.

Since the case $\Lambda=0$ has been satisfactorily covered elsewhere (cf.~\cite{CJL,ChPaetzBondi} and references therein),
from now on we assume
\bel{9VIII15.10}
 \Lambda \ne 0
 \,.
\ee
 \ptctodo{add a discussion of the $n$-th order term, related to conservation laws associated with conformal Killings of the conformal boundary?}
 \ptctodo{lots of stuff went to recycling.tex}

\subsection{Fefferman-Graham expansions}
	\label{s14III16.1}

Recall that \emph{smooth conformal compactifiability} of a metric satisfying the vacuum Einstein equations implies existence of a coordinate system
$$
 (x,x^a)\equiv (x,x^0,x^A)
$$
near $\ptmcM$ in which the metric admits a \emph{Fefferman-Graham} expansion~\cite{FG,GrahamVolume}: We can write the metric as
\bea
 g
 &=& x^{-2}\ell^2\big(\pm dx^2+ \hhere_{ab}(x,x^c)dx^ a dx^ b\big)
 \,,
\eeal{10I15.3}
where $\ell$ is a constant related to the cosmological constant, and where the sign in front of $dx^2$ is the negative of the sign of $\Lambda$. For even values of $n $ we have
\bean
 \hhere_{ab}(x, x^c)
  & = &
 \ringlambda _{ab}(x^c) +   (\hhere_2)_{ab}(x^c) x^2
 + \ldots +   (\hhere_{n-2})_{ab}(x^c) x^{n-2}
\\
 &&
 +   (\hhere_{n})_{ab}(x^c) x^n
 +   (\hhere_{\log})_{ab}(x^c) x^n \log x
 + o( x^{n})
 \,.
\eeal{10I15.2}
Here $\ptmcM$ is the zero-level set of $x$,  the tensor field
$$
 \ringlambda  := \ringlambda_{ab}dx^adx^b
$$
is a representative of the conformal class of metrics induced by $\tg$ on $\ptmcM$ (Riemannian if $\Lambda>0$, Lorentzian if $\Lambda<0$), and for $i=1,\ldots n-1$ the smooth tensor fields
$$
 \mbox{$\hhere_i:= (\hhere_{i})_{ab}dx^adx^b$ and  $\hhere_{\log}:= (\hhere_{\log})_{ab} dx^adx^b$}
$$
on $\ptmcM$ are uniquely determined by $\ringlambda $ and its derivatives, with $\hhere_{2k+1}\equiv 0$ for $2k+1<n$.
We will interchangeably write $(\hhere_{i})_{ab}$ and $(\hhere_{ab})_i$ in what follows.

For odd values of $n $ the expansion reads instead
\bean
 \hhere_{ab}(x, x^c)\textbf{•}
  & = &
 \ringlambda _{ab}(x^c) +   (\hhere_2)_{ab}(x^c) x^2
 + \ldots +   (\hhere_{n-1})_{ab}(x^c) x^{ n-1 }
\nonumber
\\
 &&
 +   (\hhere_{n})_{ab}(x^c) x^n
 + o( x^{n})
 \,,
\eeal{10I15.2x}
with again $\hhere_{2k+1}\equiv 0$ for $2k+1<n$.

We have, both for even and odd  $n\ge 3$,
\bel{11I15.4}
 (\hhere_2)_{ab} = - \frac 1 {n-2 } \left( \mathring R_{ab} - \frac {\mathring R }{2(n-1)}\ringlambda_{ab}
  \right)
  \,,
\ee
where $\mathring R_{ab}$ is the Ricci tensor of the metric $\ringlambda$.

As an example, if $g$ is a Birmingham metric, \eq{6XI12.4} below, with zero mass, we set
\bel{11I15.1}
 \frac {dx} x = - \frac {dr} {\ell \sqrt{\frac {r^2}{\ell^2} +\beta}}
 \quad
 \Longrightarrow
 \quad
  r = \ell \left(\frac{1}{x}-\frac{\beta  x}{4}\right)
  \,,
\ee
where a convenient choice of an integration constant has been made. The metric becomes
\bel{18II16.1}
 g = x^{-2}\ell^2\left( {dx^2}   - \left(1+ \frac{\beta  x^2}4\right)^2 \ell^{-2} dt^2 +
  \left(1- \frac{\beta  x^2}4\right)^2 \mathring h
  \right)
  \,.
\ee

In any case, we are  led to consider metrics of the form
\bea
 g
 &=& x^{-2}\ell^2\big(\pm dx^2+
 \ringlambda
 +   \hhere _2  x^2
 + \redOof x^{p})
 \big)
 \,,
\eeal{10I15.4}
with
\bel{11I15.2}
 \ringlambda (\partial_x,\cdot)=0= \hhere_2(\partial_x,\cdot) = \partial_x\ringlambda=\partial_x \hhere_2
 \,,
\ee
where $p=4$ in dimensions $n\ge 5$, $p=3$ in dimension $n=3$, and $p$ is any number smaller than four when $n=4$ (in that last dimension $O( x^p)$ with $p<4$ can in fact be replaced by $\redOofO x^4\ln x)$).

\subsection{The next term and the geometry of $\sectionofScri$}
 \label{ss29XI15.1}

We will see below that, in a characteristic-Cauchy-problem context, the regularity  properties of the space-time metric are determined by the first three coefficients in the expansion  \eq{24I15.21}. This raises the question, whether or not conformal smoothness implies that some of those coefficients are zero. The aim of this section is to show that
the \emph{next-to-leading term} in the expansion \eq{24I15.21} will \emph{not} vanish in general.
This will be done by relating this term to the trace-free part of the extrinsic curvature, within the conformal boundary, of a section $\sectionofScri$ of $\ptmcM$.

Consider, thus  a null hypersurface $\mcN$ with field of future-directed null tangents $\ellhere $ such that the closure $
\clmcN$ in $\tmcM$ of $\mcN$ intersects $\ptmcM$ transversally in a smooth spacelike submanifold $\sectionofScri$.
Let $B$ denote the ``null extrinsic curvature'' of $\mcN$,
\bel{18III16.31}
 B(X,Y) :=   g(\nabla _X \ellhere , Y)
 \,,
\ee
defined for $X,Y$ tangent to $\mcN$.
We will invoke the Fefferman-Graham expansions,
and the law of conformal transformations of the objects involved.

In what follows we use the notation of \cite[Appendix~A]{CCM2}.
From that last reference we have
\bea
    \overline{\Gamma}^{0}_{AB}
    &=&
    - \frac{1}{2} \nu^{0} \partial_{1} \overline{g}_{AB}
 \,.
\eea
Hence, when $\ellhere =\partial_r$,
 \bean
 B_{AB}
&=&   g(\nabla _A \ellhere ,\partial_B )
 = - g(\nabla _A\partial_B, \ellhere   )
 = - g(\nabla _A\partial_B, \partial_r  )
\\
 \nn
&=&  - g_{\mu r} \Gamma ^\mu _{AB}
 = - g_{0 r} \Gamma ^0 _{AB}
 = - \nu_0 \Gamma ^0 _{AB}
    =  \frac{1}{2}  \partial_{1} \overline{g}_{AB}
\\
&=& \chi_{AB}
\,,
 \label{18III16.32}
\eea
with $\chi$ as in \eq{18III16.21} below.

Let
$$
\tg = \Omega^2 g
$$
be the unphysical conformally rescaled metric. Let $\scri\equiv \ptmcM$ be the conformal boundary, which in vacuum is spacelike if $\Lambda>0$ and timelike if $\Lambda <0$. In what follows we will assume that $\Lambda <0$; the argument applies to the case $\Lambda >0$ after obvious modifications.

Let  $\tilde N$ be the inwards-directed $\tg$-unit normal to $\scri$. Let $\hyp$ be a smooth \emph{spacelike} hypersurface in $\tmcM$ meeting $\scri$ orthogonally at $\sectionofScri$. Let $\tilde T$ denote a future-directed $\tg$-unit normal to $\hyp$. Let $\tilde \ellhere $ be a smooth-up-to-boundary field of tangents of generators of $\clmcN$. There exists a strictly positive function $\omega$ so that
\bel{29XI15.3}
  \tilde \ellhere  =
    \omega
    (\tilde T-\tilde N)
  \
  \mbox{at}
  \
  \sectionofScri
  \,.
\ee
Here $\mcN$ is thought to lie to the past of $\hyp$ and is thus the boundary of the past domain of dependence of $\hyp$ in the unphysical, conformally rescaled space-time (and hence also in the physical space-time).

Let $x$ be a defining function for $\scri$, and let the conformal factor be $\Omega =x$. Using $r_{\Bobo}\approx 1/x$ (compare~\eq{29XI15.4}) as the parameter along the generators in the physical space-time, with $\ellhere =\partial_{r_{\Bobo}}$, we see that the function $\omega$ in \eq{29XI15.3} can be chosen so that
%
\bel{29VII15.2}
  \ellhere = \Omega^2 \tilde \ellhere
  \,,
\ee
and note that with this choice the vector field $\tilde \ellhere $ extends smoothly across the conformal boundary $\{x=0\}$.
Letting $\tilde \chi_{AB}$ denote the corresponding ``unphysical $\chi$-tensor'' of $\mcN$, we have
 \bean
 \tilde \chi_{AB}
  & = &
 \tilde B_{AB}
 = - \tg(\tnabla _A\partial_B, \tilde \ellhere   )
\\
 \nn
 & = &
  - \Omega^2 g\big(\nabla _A\partial_B + \frac 1{  \Omega} (\nabla_A \Omega\partial_B +
 \nabla_B \Omega\partial_A - g_{AB}\nabla \Omega),
  \tilde \ellhere   \big)
\\
 \nn
 & = &
  -  g\big(\nabla _A\partial_B + \frac 1{  \Omega} (\nabla_A \Omega\partial_B +
 \nabla_B \Omega\partial_A - g_{AB}\nabla \Omega),
    \ellhere   \big)
\\
 & = &
  \chi_{AB}  + \frac 1{  \Omega} \ellhere (\Omega) g_{AB}
  \,.
 \eeal{29VII15.3}
On the other hand, on $\sectionofScri$  it holds that
\bea
 \tilde \chi_{AB}|_{\sectionofScri}
  & = &
 - \tg
  \big(\tnabla _A\partial_B, \omega
   (\tilde T - \tilde N  )
    \big)
  = \omega( \tilde K_{AB} -\tilde \HLhere_{AB} )
  %
%
 \,,
\eeal{29VII15.4}
where $\tilde \HLhere$ is the extrinsic curvature tensor of $\scri$ in $(\mcM,\tg)$, and $\tilde K$ is that of $\hyp$.

The Fefferman-Graham expansion shows that the trace-free part of $\tilde \HLhere $ vanishes at $\scri$, so that
 \bea
 \tilde \chi_{AB}|_{\sectionofScri}
  & = &   \omega  \tilde K_{AB} |_{\{x=0\}}
  %
%
 \,.
\eeal{29VII15.4+}
For further reference we note that the trace-free part of $\tilde \HLhere $  is in fact  $\redOofO x^2)$ when a) $\scri$ is locally conformally flat,
or when b)
\bel{30VII15.8}
 \mbox{$\mathring R_{AB}$ is proportional to $\ringlambda_{AB}$.}
\ee

In order to determine  $\tilde K$ we use the coordinates of \eq{10I15.3}-\eq{10I15.2}. In these coordinates let $\hyp$ be given by the equation
\bel{29VII15.5}
 x^0 = f(x,x^A) = f_0(x^A) + x f_1 ( x^A) + \redOof x^2)
  \,,
\ee
with smooth functions $f_0$, $f_1$ (in fact, $f_1$  vanishes if $\hyp$ meets the boundary orthogonally, but this is not needed for our conclusions below), then
$$
 \tilde T = \varepsilon \tilde \alpha^{-2}
  \big(
   dx^0 - \partial_A f_0 dx^A - f_1 dx + \redOof x)
    \big)
 \,,
$$
where $\tilde \alpha$ is determined by the condition $\tg(\tilde T, \tilde T)=\varepsilon \in \{\pm 1\}$:
\bel{29VII15.6}
 \tilde \alpha^{2} = \varepsilon \big(
    \ringlambda ^{00} -2 \ringlambda ^{0A}\partial_A f_0 + \ringlambda ^{AB}\partial_A f_0 \partial_B f_0 + f^2_1 + \redOof x )
        \big)
            \,.
\ee
We emphasise that if the intersection of $\hyp$ with $\scri$ is a smooth spacelike submanifold of $\scri$, as assumed here, then both $\tilde \alpha $ and $\tilde \alpha^{-1}$ are smooth.

We will denote by $e_A$ the tangential lift of $\partial_A$ to the graph of $f $:
$$
 e_A = \partial_A + \partial_A f \partial _0=: e_A{}^\mu \partial_\mu
  \,.
$$

For small $x$ the metric $\tg:= x^2 g$ behaves as
\bea
\tg \to _{x\to0} \ztg {}
 &:=&  \ell^2\big(dx^2+ \ringlambda_{ab}(x^c)dx^ a dx^ b \big)
 \,.
\eeal{30VII15.1}
The Christoffel symbols $ \mathring { \tGamma}{}^\alpha_{\beta \gamma}$ of the asymptotic metric $\ztg {}$ read
\begin{eqnarray*}
 &
 \ztGamma{}{}^c_{ab} \equiv  \Gamma[\ringlambda]{}^c_{ab} =: \mathring \Gamma^c_{ab}
\,,
\quad
\ztGamma{}{}^1_{\mu\nu}=\ztGamma{}{}^\mu_{1\nu}=  0
\,.
&
\end{eqnarray*}
This can be used to determine the asymptotic behaviour of $\tilde K_{AB}$:
\bean
 \tilde K_{AB}
  & = & -
   \tg(\tnabla_{e_A} e_B, \tilde T)
   = - \big(   {e_A}( e_B{}^{\mu})+ \tGamma^\mu_{\alpha\beta} e_A{}^{\alpha} e_B{}^{\beta}\big)\tilde T_\mu
\\
 \nn
  &     =  &
   -\varepsilon \tilde \alpha^{-2}
     \bigg(\underbrace{ \partial_A \partial_B f_0 -
     \mathring \Gamma{}^C_{AB} \partial_C f_0}_{=: \zmcD_A \zmcD_B f_0}
     +\mathring \Gamma{}^0_{AB}
       +  2 (\mathring \Gamma{}^0_{0(A} -\partial_C f_0 \mathring \Gamma{}^C_{0(A}  )\partial_{B)} f_0
\\
 \nn
  &       &
      + ( \mathring \Gamma{}^0_{00} -  \mathring \Gamma{}^C_{00} \partial_C f_0
    ) \partial_{A} f_0\partial_{B} f_0
     \bigg)
     + \redOof x)
\\
 & =  &
  \mathring{\tK}_{AB} +\redOof x)
 \,,
\eeal{30VII15.5}
where
\bel{30VII15.6}
 \mathring{\tK}_{AB}
  := \tilde K_{AB}|_{x=0}
     \,.
\ee
Using \eq{29VII15.3} and \eq{29VII15.4+} we obtain
\bel{30VII15.7}
 \sigma_{AB} =  \omega_0 \big(\mathring{\tK}_{AB} - \frac { \ztg {}{}^{CD} \mathring{\tK}_{CD}} {n-1}  {\ztg}{}_{AB}\big)  + \redOof x)
 \,,
\ee
where $ \omega_0:=\omega|_{x=0}$.

We conclude that on characteristic hypersurfaces smoothly meeting $\scri$ we have $\sigma_{AB} =\redOof 1)$ for large $r$, with
$$
 \mbox{$\sigma|_{x=0}$ being non-zero in general.}
$$

As an example, consider the case $\Lambda<0$ with
\bel{31VII15.3}
 \mbox{
  a metric  $\tg_{ab}|_{x=0}$ which is static  up to a conformal factor.}
\ee
We can then rescale the metric, and adjust the $x$-coordinate accordingly, so that  $\tg_{ab}|_{x=0}$ is in fact $x^0$-independent. A further, $x^0$-independent, rescaling can be done so that $\tg_{00}$ is constant. We can further choose manifestly static local coordinates, where by definition $\tg_{A0}|_{x=0}=0$ (this can be done globally  when $\tg_{A0}dx^A$ is exact, which will certainly be the case if $\sectionofScri$  is simply connected). Setting $X=\partial_0$ and using
   $$
    \partial_0 \tg_{ab} = \mcL_{\partial_0} \tg_{ab}
     = \tnabla_a X_b + \tnabla_b X_a = -2 \tGamma^0_{ab}
     \,,
   $$
we see that all the $\zGamma^0_{ab}$'s vanish, and in fact $\zGamma^a_{0b}=0$. With these choices we will have
\bel{30VII15.8x}
 \mathring{\tK}_{AB} \sim \ztg_{AB}
\ee
if and only if
\begin{equation}
 \zmcD_A \zmcD_B f_0 - \frac{ \zmcD{}^C \zmcD_C f_0}{n-1}\,\ztg_{AB} =0
  \,.
    \label{31VII15.1}
\end{equation}
We conclude that under \eqref{31VII15.3} and \eqref{31VII15.1} we have
\bel{31VII15.2}
 \sigma_{AB} = \redOof r^{-1})
  \qquad
  \Longleftrightarrow
  \qquad
  |\sigma|^2 = \redOof r^{-6})
\ee
for large $r$.

We also see that the \eq{30VII15.8x} is a necessary and sufficient condition for \eq{31VII15.2} in any case.

 \section{Characteristic hypersurfaces}
 \label{s24XII15.1}

Throughout this section we allow arbitrary space-time dimension $n+1\ge 4$.

As a first step towards understanding the mass of characteristic hypersurfaces, a review of the characteristic Cauchy problem for the Einstein equations is in order.

\subsection{Wave-map gauge}
\label{subsec:wavemap}

Let, thus $\mcN$ be a characteristic hypersurface. Following~\cite{CCM2},
we split the Einstein equations along $\mcN$ into constraint and evolution equations using the \emph{generalized wave-map gauge} \cite{CCM2,FriedrichCMP}, which is characterized by the vanishing of the \emph{generalized wave-gauge vector}:
\begin{equation}
\label{eq:wavegauge}
	H^{\lambda} = 0
	\,,
\end{equation}
which is defined as
\begin{equation}
\label{eq:wavegaugedef}
	H^{\lambda} := \Gamma^{\lambda} - V^{\lambda}
	\,, \;\;
	\text{where} \;\;\;
	V^{\lambda} := \hat{\Gamma}^{\lambda} + W^{\lambda}
	\,, \;\;\;
	\Gamma^{\lambda} := g^{\alpha \beta} \Gamma^{\lambda}_{\alpha \beta}
	\,, \;\;\;
	\hat{\Gamma}^{\lambda} := g^{\alpha \beta} \hat{\Gamma}^{\lambda}_{\alpha \beta}
	\,.
\end{equation}
Here $\hat \Gamma$ are the Christoffel symbols of an auxiliary target space metric $
\hat g$, which can be chosen as convenient for the problem at hand.
The  \emph{gauge source functions} $W^{\lambda} = W^{\lambda}(x^{\mu},g_{\mu \nu})$ can be freely specified and are allowed to depend upon the coordinates chosen and the metric itself, but not upon derivatives thereof. In \eqref{eq:wavegaugedef} and in what follows we decorate objects associated with the \emph{target metric} $\hat{g}$ with the hat symbol ``$\,\, \hat{} \,\,$''.

\subsection{Characteristic surfaces, adapted null coordinates and assumptions on the metric}
\label{subsec:nullcoords}

 It is convenient to use coordinates adapted to the characteristic surface, called \emph{adapted null coordinates} ($x^0=u, \, x^1=r, \, x^A$),
{$A\in\{2,\dots,n\}$}.
The coordinate $r>r_0\ge 0$, where we allow a boundary or a vertex at a value $r=r_0$ possibly different from zero, parametrizes the null geodesics issuing from $\{r_0\}$ and generating the null hypersurface, which coincides with the set $\{u=0\}$. The $x^A$'s are local coordinates on the level sets $\{u=0, r=\text{const.}\}$. The trace of the metric on the characteristic surface can then be written as (we will interchangeably use $x^0$ and $u$)
\begin{equation}
\label{eq:metricrestriction}
	\og = \og_{\mu \nu} dx^{\mu} dx^{\nu} = \og_{00} (du)^2 + 2 \nu_0 du dr + 2 \nu_A du dx^A + \coneg
	\,,
\end{equation}
where we use the notation
\begin{eqnarray}
	\nu_0 := \og_{0r}
	\,, \quad
	\nu_A := \og_{0A}
	\,, \quad
	\coneg := \og_{AB} dx^A dx^B
	\,.
\end{eqnarray}
Here and throughout an overline denotes the restriction of a space-time object to $\mcN$.

Under the hypotheses of Proposition~\ref{P12XII15.1},
for $r$ large we can write
\bel{eq:asymptgAB}
	\og_{AB} = \hring_{AB} r^2 + \left( \og_{AB} \right)_{-1} r + \left( \og_{AB} \right)_0 + \wascalOof  r^{-1} )
	\,,
\ee
where we use the symbol $\hring$ to denote the standard metric on the boundary manifold,  and $\left( \og_{AB} \right)_{-i}=\left( \og_{AB} \right)_{-i}(x^C)$, $i \in \mathbb{N}$, are some smooth tensors on that manifold. We also require that $\wascalOof  r^{-1})$ terms remain $\wascalOof  r^{-1})$ under $x^C$-differentiation, and become $\wascalOof  r^{-2})$ under $r$-differentiation; similarly for $O({r}^{-n})$.

The restriction of the inverse metric to $\mcN$ takes the form
\begin{equation}
	\og^{\#} = 2 \nu^0 \partial_u \partial_r + \og^{rr} \partial_r \partial_r + 2 \og^{rA} \partial_r \partial_A + \og^{AB} \partial_A \partial_B
	\,,
\end{equation}
where $\og^{AB}$ is the inverse of $\og_{AB}$ and
\begin{equation}
\label{eq:fielddef2}
	\nu^0 := \og^{0r} = \frac{1}{\nu_0}
	\,, \;\;\;
	\og^{rA} = -\nu^0 \nu^A = -\nu^0 \og^{AB} \nu_B
	\,, \;\;\;
	\og^{rr} = \left(\nu^0\right)^2 \left( \nu^A \nu_A - \og_{00} \right)
	\,.
\end{equation}

The \emph{null second fundamental form} of $\mcN$ is intrinsically defined and does not depend on transverse derivatives of the metric, see \eqref{18III16.31}. In adapted null coordinates it reads (compare \eqref{18III16.32})
\begin{equation}
 \label{18III16.21}
	\chi_{AB} = \frac 12 \partial_r \og_{AB}
	\,.
\end{equation}
The expansion, also called divergence, of the characteristic surface will be denoted by
\begin{equation}
	\tau := \chi_A{}^A = \og^{AB} \chi_{AB}
	\,,
\end{equation}
while the trace-free part of the null second fundamental form
\begin{equation}
	\sigma_A{}^B = \chi_A{}^B - \frac 12 \tau \delta_A{}^B
\end{equation}
is called the \emph{shear} of $\mcN$.

The constraint equations for the characteristic problem  will be referred to as \textsl{Einstein wave-map gauge constraints}. In  space-time dimension $n+1\geq 3$ they read \cite{CCM2}
%
\begin{eqnarray}
	\left( \partial_r - \kappa \right) \tau + \frac{1}{n-1} \tau^2
	&=& - |\sigma|^2 - 8 \pi \ol{T}_{rr}
	\,,   \label{eq:wavegaugeconstraint1}\\
	\left( \partial_r + \frac{1}{2} \tau + \kappa \right) \nu^0
	&=& - \frac{1}{2} \ol{V}^0
	\,,   \label{eq:wavegaugeconstraint2}\\
	\left( \partial_r + \tau \right) \xi_A &=& 2 \conenabla_B \sigma_A^B - 2 \frac{n-2}{n-1} \partial_A \tau - 2 \partial_A \kappa - 16 \pi \ol{T}_{rA}
	\,,
 \phantom{xx}
   \label{eq:wavegaugeconstraint3}
\\
	\left( \partial_r + \frac{1}{2} \nu_0 \ol{V}^0 \right) \nu^A &=& \frac{1}{2} \nu_0 \left(\ol{V}^A - \xi^A - \og^{BC} \coneGamma^A_{BC} \right)
	\,,   \label{eq:wavegaugeconstraint4}
\\
	\left( \partial_r + \tau + \kappa \right) \zeta &=& \frac{1}{2} |\xi|^2 - \conenabla_A \xi^A - \coneR
 \nn
\\
 && + \underbrace{8\pi \left(\og^{AB} \ol{T}_{AB} - \ol{T} \right)}_{=:S} + 2 \Lambda
	\,,   \label{eq:wavegaugeconstraint5}\\
	\left( \partial_r + \frac{1}{2} \tau + \kappa \right) \og^{rr} &=& \frac{1}{2} \zeta - \ol{V}^r
	\,,   \label{eq:wavegaugeconstraint6}
\end{eqnarray}
where
\bel{eq:wavegaugeadditions}
	|\sigma|^2 := \sigma_A{}^B \sigma_B{}^A
	\,, \;\;\;
	|\xi|^2 := \og^{AB} \xi_A \xi_B
	\,, \;\;\;
	\xi^A := \og^{AB} \xi_B
	\,, \;\;\;
	\ol{T} := \og^{\mu \nu} \ol{T}_{\mu \nu}
	\,.
\ee
All objects associated with the one-parameter family of Riemannian metrics $\coneg$ are decorated with the check symbol ``$\,\, \check{} \,\,$''.
The boundary conditions needed to integrate \eqref{eq:wavegaugeconstraint1}-\eqref{eq:wavegaugeconstraint6} starting from a light-cone vertex $r_0=0$ follow from the requirement of smoothness of the metric  there, see \cite[Section~4.5]{CCM2}.
 \ptctodo{mention core geodesics here?}

The function $\kappa$ is defined through the equation
\begin{equation}
	\nabla_{\partial_r} \partial_r = \kappa \partial_r
\end{equation}
and reflects the freedom to choose the coordinate $r$ which parametrizes the null geodesic generators of $\mcN$. The ``auxiliary'' fields $\xi_A$ and $\zeta$ have been introduced to transform \eqref{eq:wavegaugeconstraint1}-\eqref{eq:wavegaugeconstraint6} into first-order equations. The field $\xi_A \equiv -2\ol{\Gamma}^r_{rA}$ represents connection coefficients, while the field $\zeta$ is the divergence of the family of suitably normalized null generators normal to the spheres of constant radius $r$ and transverse to the characteristic surface. In coordinates adapted to the light-cone as in \cite{CCM2} the space-time formula for $\zeta$ reads
(compare~\cite[Equations~(10.32) and (10.36)]{CCM2}; note, however, that there is a term $\tau \overline {g}^{11}/2$ missing at the right-hand side of the second equality in (10.36) there)
%
\bel{eq:zeta_spacetime}
 \zeta := (2\partial_r + 2\kappa +\tau) \ol g^{rr} + 2 \ol\Gamma^r
  \equiv 2 \og^{AB} \ol{\Gamma}^r_{AB} + \tau \og^{rr}
 \,.
\ee

To integrate the wave-map gauge constraints \eqref{eq:wavegaugeconstraint1}-\eqref{eq:wavegaugeconstraint6} one also needs the components $\ol{V}^{\mu}$, which are determined by the wave-map gauge \eqref{eq:wavegauge}-\eqref{eq:wavegaugedef}. We have, in adapted null coordinates \cite[Appendix A, Equations~(A.29)-(A.31)]{CCM2},
\begin{eqnarray}
	\overline{\Gamma}^{0}
	\equiv
	\overline{g}^{\lambda \mu} \overline{\Gamma}_{\lambda \mu}^{0}
	&\equiv&\
	\nu^{0} \left( \nu^{0}\overline{\partial_{0}g_{11}}-\tau \right)
	\,,
	\label{Gamma0}
	\\
	\overline{\Gamma}^{1}
	\equiv
	\overline{g}^{\lambda \mu} \overline{\Gamma}_{\lambda \mu}^{1}
	\nn
	&\equiv&\
	-\partial_{1} \overline{g}^{11}
	+ \overline{g}^{11} \nu^{0} \left(
    \frac{1}{2} \overline{\partial_{0} g_{11}}
  - \partial_{1} \nu_{0}
  - \tau \nu_{0}
	\right)
	\\
	&&	+ \nu^{0} \overline{g}^{AB} \check{\nabla}_{B} \nu_{A}
	- \frac{1}{2} \nu^{0} \overline{g}^{AB} \overline{\partial_{0}g_{AB}}
	\,,
	\label{eq:Gamma1}
	\\
	\overline{\Gamma}^{A}
	\equiv
	\nn
	\overline{g}^{\lambda \mu} \overline{\Gamma}_{\lambda \mu}^{A}
	&\equiv&\
	\nu^{0} \nu^{A} \left( \tau - \nu^{0} \overline{\partial_{0}g_{11}} \right)
	+
\nu^{0}\overline{g}^{AB}(\overline{\partial_{0}g_{1B}}+\partial_{1}\nu_{B}-\partial_{B}\nu_{0}) \nn
	\\
	&&
  - 2\nu^{0}\nu^{B}\chi_{B}{}^{A} +\check{\Gamma}^{A}
  \,.
	\label{GammaA}
\end{eqnarray}
Now, from the restriction to $\mcN$  of \eqref{eq:wavegauge} and together with the first equation of \eqref{eq:wavegaugedef} one finds that the choice of the target metric only redefines the fields $\ol{\hat{\Gamma}}^\mu$ and $\ol{W}^\mu$ entering in the definition of $\ol{V}^\mu = \ol{\hat{\Gamma}}^\mu + \ol{W}^\mu$ without changing $\ol{V}^\mu$ itself. Therefore only $\ol{V}^\mu$ enters in the Einstein wave-map gauge constraint equations, and so only the explicit form of those fields is relevant in the equations of interest to us.

An adapted coordinate system on a characteristic surface $\mcN$ will be called \emph{Bondi type} if the coordinates satisfy Bondi conditions on $\mcN$, but not necessarily away from $\mcN$, reserving the name \emph{Bondi coordinates} for coordinate systems which satisfy Bondi's condition everywhere.
  \label{bonditypecoordsdef}

We will start by deriving the asymptotic expansions of all relevant fields in \emph{Bondi-type coordinates on the characteristic surface}; it appears that the calculations are simplest  in those coordinates.
We have \cite[Equation~(5.5)]{TimDiss}
\bel{eq:bondiimplications}
	\varphi^\Bo = r_{\Bobo}
	\,, \;\;\;
	\ol{\partial_0 g_{rr}^\Bo} = 0
	\,, \;\;\;
	\ol{\partial_0 g_{rA}^\Bo} = 0
	\,, \;\;\;
	\og^{AB}_\Bo \ol{\partial_0 g_{AB}^\Bo} = 0
	\,,
\ee
where $\varphi$ is defined by $\tau = 2 \partial_r \log{\varphi}$, as well as
%
\begin{eqnarray}
	\ol{V}^0_\Bo
	&=& -\tau^\Bo \nu^0_\Bo
	\,,   \label{eq:VBondi0}\\
	\ol{V}^A_\Bo
	&=& \og^{CD}_\Bo \left( \coneGamma^\Bo \right)^A_{CD} - \nu^0_\Bo \conenabla^A \nu_0^\Bo + \nu^0_\Bo \left( \partial_{r_{\Bobo}} + \tau^\Bo \right) \nu^A_\Bo
	\,,  \label{eq:VBondiA}
\\
	\ol{V}^r_{\Bobo}
	&=& \nu^0_\Bo \conenabla_A \nu^A_\Bo - \left( \partial_{r_{\Bobo}} + \tau^\Bo + \nu^0_\Bo \partial_{r_{\Bobo}} \nu_0^\Bo \right) \og^{rr}_\Bo
	\,.   \label{eq:VBondir}
\end{eqnarray}
As mentioned previously the Einstein wave-map gauge constraints form a hierarchical system of ODEs along the null generators of the characteristic surface which can be solved step-by-step.

\section{Asymptotic solutions of the characteristic wave-map gauge constraints, $\Lambda\ne 0$}
\label{sec:solvingconstraints}

Throughout this section we assume that the space-dimension $n=3$.

In \cite{TimAsymptotics} asymptotic solutions of the Einstein wave-map gauge constraints (\ref{eq:wavegaugeconstraint1})-(\ref{eq:wavegaugeconstraint6}) with $\Lambda=0$ have been obtained in the form of polyhomogeneous expansions of the solution at infinity, i.e., expansions in terms of inverse powers of $r$ and of powers of $\log r$. Our aim is to obtain similar expansions when $\Lambda\ne0$, with the goal  to find a formula for the \charmass.

We will assume that for large $r$
\bel{eq:sigmabondiexpansion}
	\sigmaboAB= \left( \sigmaboAB{} \right)_2 r_{\Bobo}^{-2} + \left( \sigmaboAB{} \right)_3 r_{\Bobo}^{-3} +
 {\wascalO }\big(r_{\Bobo}^{-4}\big) \,,
\ee
which is compatible with, and more general than, Proposition~\ref{P12XII15.1}.
Here $\sigmabo$ is the shear of $\mcN$ in Bondi coordinates, with $\sigmaboAB := \og^{BC}\sigmabo( \partial_A, \partial_C)$.
As already mentioned, wherever needed in the calculations that follow we will assume that differentiation of error terms
$O(r^
\alpha)$ with respect to angles preserves the $O(r^\alpha)$ behaviour, while  differentiation   with respect to $r$ produces terms which are $O(r^{\alpha-1})$.

(It follows from our calculations below that the hypothesis \eq{eq:sigmabondiexpansion} is equivalent to
\bel{eq:sigmaaffineexpansion}
	\sigmaafAB= \left( \sigmaafAB{} \right)_2 r ^{-2} + \left( \sigmaafAB{} \right)_3 r ^{-3} + {\wascalO }\big(r ^{-4}\big) \,,
\ee
where $r$ is an affine coordinate along the generators of $\mcN$.)

\subsection{Matter fields}
 \label{ss22I15.1}

We start by analyzing the influence of the matter fields on the asymptotic expansion of the metric in Bondi-type coordinates. Our aim is to determine a decay rate of the energy-momentum tensor which is compatible with finite total mass. The decay rates for various components of the energy-momentum tensor will be  chosen so that they do not affect the leading-order behavior, as arising in the vacuum case, of the solutions of the equations in which they appear.

For the convenience of the reader we repeat here the relevant equations in Bondi-type gauge (see~\cite[Equations~(5.11)-(5.15)]{TimDiss} with the contribution from the cosmological constant $\Lambda$ added here):
\begin{eqnarray}
	\label{24I15.1}
	\kappa^\Bo  - \frac{1}{2} \rnoBo  \left( |\sigma^\Bo|^2 + 8 \pi \ol{T}_{rr}^\Bo \right)
	\hspace{-2mm}&=& \hspace{-2mm}
	0
	\,,
\\
	( \partial_{r_{\Bobo}} + \frac{\rbo}{2} (|\sigma^{\mathrm{Bo}}|^2+ 8\pi \ol T^{\mathrm{Bo}}_{rr} )) \nu^0_\Bo
	 \hspace{-2mm}&=& \hspace{-2mm}
	 0
	 \,,
	\label{constraint_bondi2h}
\\
	(\partial_{r_{\Bobo}} + \tau^{\mathrm{Bo}} )\xi^{\mathrm{Bo}}_A  - 2\conenabla_B \sigma^{\mathrm{Bo}}_A{}^{B} + \partial_A\tau^{\mathrm{Bo}} + \rbo \partial_A(|\sigma^{\mathrm{Bo}}|^2+8\pi \ol T^{\mathrm{Bo}}_{rr})
	\hspace{-2mm}&=& \hspace{-2mm}
	-16\pi \ol T^{\mathrm{Bo}}_{rA}
	\,, \phantom{xxxx}
	\label{24I15.2}
\\
	\partial_{r_{\Bobo}}\nu^A_{\mathrm{Bo}}+(\check\nabla^A+ \xi_{\mathrm{Bo}}^A)\nu^{\mathrm{Bo}}_0
	\hspace{-2mm}&=& \hspace{-2mm}
	0
	\,,
	\label{24I15.3}
\\
	(\partial_{r_{\Bobo}} + \tau^{\mathrm{Bo}} + \frac{\rbo}{2}( |\sigma^{\mathrm{Bo}}|^2 +8\pi \ol T^{\mathrm{Bo}}_{rr}) )\zeta^{\mathrm{Bo}} +  \coneR^{\mathrm{Bo}} - \frac{|\xi^{\mathrm{Bo}}|^2}{2} +\conenabla_A\xi_{\mathrm{Bo}}^A
	\hspace{-2mm}&=& \hspace{-2mm}
	\nonumber \\
	\underbrace{8\pi(\ol g_{\mathrm{Bo}}^{AB}\ol T^{\mathrm{Bo}}_{AB} - \ol T^{\mathrm{Bo}})}_{S^\Bo}
	\hspace{-2mm}&+& \hspace{-2mm} 2 \Lambda
	\,,
	\label{24I15.4}
\\
	\ol g_{\mathrm{Bo}}^{rr} + (\tau^{\mathrm{Bo}})^{-1}(\zeta^{\mathrm{Bo}}  - 2\nu_{\mathrm{Bo}}^0\check\nabla_A\nu_{\mathrm{Bo}}^A)
	\hspace{-2mm}&=& \hspace{-2mm}
	0
	\,.
	\label{constraint_bondi6h}
\end{eqnarray}
%

In Bondi-type coordinates, the relation
$$
 \tau^\Bo = \frac{2}{\rbo}
$$
is independent of the cosmological constant and of matter fields.

It follows from \eq{24I15.1}, which can be solved algebraically for $\kappa^\Bo$, that a term ${\wascalO }(\rbo^{-\alpha_{rr}})$ in $\ol{T}_{rr}^\Bo$ with $ \alpha_{rr} > 2 $ produces an ${\wascalO }(\rbo^{-\alpha_{rr}+1})$ term in $\kappa^\Bo$ (see also the discussion in Section~\ref{sec:solwaveconstraint1} and Equation~\eqref{eq:kappabondicoeffs}):
\beal{14I15.1}
 \ol{T}_{rr}^\Bo =  {\wascalO }(\rbo^{-\alpha_{rr}})\,, \  \alpha_{rr} > 2
  & \quad \Longrightarrow
  \quad
  \kappa^\Bo = (\kappa^\Bo)_{\mathrm{vacuum}} + {\wascalO }(\rbo^{-\alpha_{rr}+1})
 \,.
\eea
Next, from \eq{constraint_bondi2h} we find
\bea
  \nu^0_\Bo = (\nu^0_\Bo)_{\mathrm{vacuum}} + {\wascalO }(\rbo^{-\alpha_{rr}+2})
 \,.
\eea
%
%
%
In the $\xi_A^\Bo$-constraint equation \eq{24I15.2}, the assumption
\bel{22I15.3}
 \ol{T}_{rA}^\Bo =  {\wascalO }(\rbo^{-\alpha_{rA}+1})\,, \ 4\ne  \alpha_{rA}
 \,, \  4\ne \alpha_{rr}
\,,
\ee
leads to
%
\be
  \xi_A^\Bo  = (\xi_A^\Bo)_{\mathrm{vacuum}} + {\wascalO }(\rbo^{-\alpha_{rA}+2})+ {\wascalO }(\rbo^{-\alpha_{rr}+2})
 \,,
\ee
where the values $\alpha_{rA}=4$ and $\alpha_{rr}=4$ have been excluded to avoid here a supplementary annoying discussion of logarithmic terms
(note that the logarithmic terms will be discussed in  detail in the sections that follow):
\bean
 \lefteqn{
  \alpha_{rA}=4 \ \mbox{or} \ \alpha_{rr}=4 \quad \Longrightarrow}
 &&
\\
 &&
  \ \xi_A^\Bo  = (\xi_A^\Bo)_{\mathrm{vacuum}} + {\wascalO }(\rbo^{-\alpha_{rA}+2})+ {\wascalO }(\rbo^{-\alpha_{rr}+2}) + {\wascalO }(\rbo^{-2}\log \rbo)
 \,.
 \phantom{xxxx}
\eeal{18III16.25}
From now on we assume \eq{22I15.3}. To preserve the vacuum asymptotics $\xi_A^\Bo={\wascalO }(\rbo^{-1})$ we will moreover require
%
\bel{24I15.10}
 \alpha_{rr}>3
  \,,
  \quad
  \alpha_{rA}>3
  \,.
\ee
(Anticipating, we have excluded the case  $\alpha_{rr}=3$, which introduces $1/\rbo$ terms in $\nu^0_\Bo$, which lead subsequently to logarithmically divergent terms in $\nu^A_\Bo$. We further note that $\alpha_{rA}=3$ will produce an additional $(\xi^\Bo_A)_1$-term that would not integrate away in $\mTB$ and would remain as a supplementary $(\ol T^\Bo_{rA})_2$-term in our final mass identity \eqref{eq:massfinal} below.)

%
Integration of \eq{24I15.3} gives
%
\bean
 \lefteqn{
 \nu^A_\Bo  = (\nu^A_\Bo)_{\mathrm{vacuum}} + {\wascalO }(\rbo^{-\alpha_{rA}+1})+ {\wascalO }(\rbo^{-\alpha_{rr}+1})
  }
  &&
   \nonumber
\\
  &&  \Longleftrightarrow
  \nu_A^\Bo  = (\nu_A^\Bo)_{\mathrm{vacuum}} + {\wascalO }(\rbo^{-\alpha_{rA}+3})+ {\wascalO }(\rbo^{-\alpha_{rr}+3})
 \,.
 \phantom{xx}
\eea
Finally, the asymptotic behavior $\xi_A^\Bo = {\wascalO }(\rbo^{-1})$ together with \eq{24I15.4} and \eq{constraint_bondi6h} show that:
 a term ${\wascalO }(\rbo^{-\alpha_S})$ in $S^\Bo$ with $\alpha_S<2$ would change the leading order behavior of $\zeta^\Bo$;
	$\alpha_S=2$ would change the leading order term of $\zeta^\Bo$;
	$\alpha_S=3$ would lead to a logarithmic term  in $\zeta^\Bo$.
This leads to
\bean
 \lefteqn{
 S^\Bo =  {\wascalO }(\rbo^{-\alpha_{S}})\,, \  \alpha_{S}>3
 \,, \  \alpha_{rr} \ne 5
  \,,
 } &&
 \\
   &\Longrightarrow
  &
  \zeta^\Bo  = (\zeta^\Bo)_{\mathrm{vacuum}} + {\wascalO }(\rbo^{-\alpha_{S}+1})+ {\wascalO }(\rbo^{-\alpha_{rA}+1})+ {\wascalO }(\rbo^{-\alpha_{rr}+3 })
 \,,
\\
 &&
  \ol{g}_{\Bo}^{rr} = (\ol{g}_{\Bo}^{rr})_{\mathrm{vacuum}} + {\wascalO }(\rbo^{-\alpha_{S}+2})+ {\wascalO }(\rbo^{-\alpha_{rA} +2 })+ {\wascalO }(\rbo^{-\alpha_{rr}+4 })
 \,,
  \phantom{xx}
\eeal{14I15.5}
and note that a factor $r_{\Bobo}^2$ in the $ {\wascalO }(\rbo^{-\alpha_{rr}+3 })$ terms in $\zeta^\Bo$ arises from  the $4\pi \rbo  \ol T^{\mathrm{Bo}}_{rr}\zeta^{\mathrm{Bo}}$
term in \eq{24I15.4}, taking into account the $ 2
\Lambda \rbo/3 $ leading behavior of $\zeta^\Bo$.

We conclude that the leading order of all quantities of interest will be preserved if we assume that
\bel{11II15.21}
 \alpha_{rr} >3\,,
 \
 \alpha_{rA}>3
 \,,
 \
 \alpha_{S}>3\,.
\ee
Keeping in mind our main assumptions, that all fields can be expanded in terms of inverse powers of $\rbo$ to the order needed to perform our expansions, possibly with some logarithmic coefficients, we will allow below matter fields for which \eq{11II15.21} holds.

In what follows we will actually assume
\bel{eq:Tbondiexpansion}
	\ol{T}_{rr}^\Bo = {\wascalO }\big(r_{\Bobo}^{-4}\big)
	\, , \;\;\;
	\ol{T}_{rA}^\Bo = {\wascalO }\big(r_{\Bobo}^{-3}\big)
	\, , \;\;\;
	\ol g^{AB}_\Bo \ol{T}_{AB}^\Bo - \ol{T}^\Bo = {\wascalO }\big(r_{\Bobo}^{-3}\big)
\,.
\ee
Note that the third equation in \eqref{eq:Tbondiexpansion} is less restrictive than \eq{11II15.21}, allowing a logarithmic term in the asymptotic expansion of $\zeta^\Bo$. This term however will be of the order $\log r_{\Bobo}/r_{\Bobo}^2$ and will not influence our result for the \charmass. It is accounted for in the correction term in \eqref{eq:solzetabondinolambda} below.
 An analog statement holds for the fall-off behavior and the correction term in  \eqref{eq:solzetanolambda} below in affine coordinates.

When solving the wave-map gauge constraints we keep in mind that we eventually want to determine the expansion coefficient $\left(\ol g_{00}^\Bo\right)_1$, as needed to calculate the mass. This determines how far the intermediate asymptotic expansions  need to be carried-out.

\subsection{Solving equation (\ref{eq:wavegaugeconstraint1})}
\label{sec:solwaveconstraint1}

The first equation of (\ref{eq:bondiimplications}) implies $\tau^\Bo=2 r^{-1}_\Bo$ and using this one directly finds from (\ref{eq:wavegaugeconstraint1}) in Bondi-type coordinates, cf.~\eqref{24I15.1},
\bel{eq:solwaveconstraint1}
	\kappa^\Bo = \frac{1}{2} r_{\Bobo} \left( |\sigma^\Bo|^2 + 8 \pi \ol{T}_{rr}^\Bo \right)
	\,.
\ee
Note for further reference that this means
\bel{eq:kappabondicoeffs}
	\left(\kappa^\Bo\right)_n
	=
	\frac{1}{2} \left[ |\sigma^\Bo|^2_{n+1} + 8 \pi \big(\ol{T}_{rr}^\Bo\big)_{n+1} \right]
	\,
\ee
for the expansion coefficients of $\kappa^\Bo$, where we have assumed that $n$ is positive.

\subsection{Expansion of $\nu^0_\Bo$}
\label{sec:solwaveconstraint2}

Inserting (\ref{eq:VBondi0}) and (\ref{eq:solwaveconstraint1}) into (\ref{eq:wavegaugeconstraint2}) in Bondi-type coordinates yields, cf.~\eqref{constraint_bondi2h},
\bel{eq:nu0Bondiode}
	\left[ \partial_{r_{\Bobo}} + \frac{r_{\Bobo}}{2} \left( |\sigma^\Bo|^2 + 8 \pi \ol{T}_{rr}^\Bo \right) \right] \nu^0_\Bo = 0
	\,,
\ee
and from (\ref{eq:sigmabondiexpansion}) we have
\bel{eq:sigmabondisquareexpansion}
	|\sigma^\Bo|^2 = \frac{|\sigma^\Bo|^2_4}{r_{\Bobo}^4} + \frac{|\sigma^\Bo|^2_5}{r_{\Bobo}^5} +  \frac{|\sigma^\Bo|^2_6}{r_{\Bobo}^6} + {\wascalO }\big(r_{\Bobo}^{-7}\big)
	\,.
\ee
Using this and (\ref{eq:Tbondiexpansion}) we find the solution of (\ref{eq:nu0Bondiode})
\begin{eqnarray}
\label{eq:solwaveconstraint2}
	\nu^0_\Bo
	&= &  \nuoasymptoticvalue
    \bigg(1
	+
	\frac{1}{4} \Big[ |\sigma^\Bo|^2_4 + 8\pi \big(\ol{T}_{rr}^\Bo\big)_4 \Big] r_{\Bobo}^{-2}
	\nonumber \\
	&&
	+
	\frac{1}{6} \Big[ |\sigma^\Bo|^2_5 + 8\pi \big(\ol{T}_{rr}^\Bo\big)_5 \Big] r_{\Bobo}^{-3}
\bigg)
	+
	{\wascalO }\big(r_{\Bobo}^{-4}\big)
	\,,
\end{eqnarray}
where $\nuoasymptoticvalue$ is a global integration function.

\subsection{Expansion of $\xi^\Bo_A$}
\label{sec:solwaveconstraint3}

Using $\partial_A \tau^\Bo = 0$, (\ref{eq:wavegaugeconstraint3}) in Bondi-type coordinates takes the form
\bel{eq:xibondiode}
	\left( \partial_{r_{\Bobo}} + \tau^\Bo \right) \xi_A^\Bo
	=
	2 \conenabla_B \sigmaboAB{} - 2 \partial_A \kappa^\Bo - 16\pi \ol{T}_{rA}^\Bo
 \,,
\ee
cf.~\eqref{24I15.2}. Using again (\ref{eq:Tbondiexpansion}), (\ref{eq:sigmabondisquareexpansion}) as well as \cite[Equations~(3.24)-(3.26)]{TimAsymptotics} (as revisited to include matter fields)
\bel{eq:sigmader}
	\conenabla_B \sigmaboAB{} = (\Xi^\Bo)^{(2)}_A r_{\Bobo}^2 + (\Xi^\Bo)^{(3)}_A r_{\Bobo}^3 + {\wascalO }\big(r^{-4}_\Bo\big)
	\, ,
\ee
where
\bel{eq:Xi23}
	(\Xi^\Bo)^{(2)}_A := \znabla_B \left(\sigmaboAB{}\right)_2
	\, , \;\;\;
	(\Xi^\Bo)^{(3)}_A := \znabla_B \left(\sigmaboAB{}\right)_3 + \frac{1}{2} \znabla_A |\sigma^\Bo|^2_4
	{ {+ 4\pi \znabla_A \big(\ol{T}_{rr}^\Bobo\big)_4} }
 \,,
\ee
the solution of (\ref{eq:xibondiode}) reads
\begin{eqnarray}
\label{eq:solwaveconstraint3}
	\xi_A^\Bo
	&=& 2 (\Xi^\Bo)^{(2)}_A r^{-1}_\Bo - 2 \left[ \znabla_B \left(\sigmaboAB{}\right)_3 - 8\pi \big(\ol{T}_{rA}^\Bo\big)_3 \right] \frac{\log r_{\Bobo}}{r^2_\Bo}
	\nonumber \\
	&&
	+ C_A^{(\xi_B)} r^{-2}_\Bo + {\wascalO }\big(r^{-3}_\Bo\big)
	\,,
\end{eqnarray}
where the coefficients $C_A^{(\xi_B)} = C_A^{(\xi_B)}(x^C)$ are global integration functions.

It follows from Proposition~\ref{P12XII15.1}  that existence of a smooth conformal completion at infinity requires the relation
\bel{16II16.10}
  \znabla_B \left(\sigmaboAB{}\right)_3  =
	  8\pi \big(\ol{T}_{rA}^\Bo\big)_3
 \,.
\ee

\subsection{Expansion of $\nu_\Bo^A$}
\label{sec:solwaveconstraint4}

Equation~(\ref{eq:wavegaugeconstraint4}) in Bondi-type coordinates does not depend upon $\Lambda$ and reads, cf.~\eqref{24I15.3},
\bel{eq:nuAbondiode}
	\partial_{r_{\Bobo}} \nu^A_\Bo = - \left( \conenabla^A + \xi^A_\Bo \right) \nu_0^\Bo
	\,.
\ee

Now, the transformation from the affine parameter $r$,  described in Section \ref{subsec:nullcoords}, to $r_{\Bobo}$ is given by (see \cite[Equation (51) there]{ChPaetzBondi})
\bel{24XII15.1}
 r_{\Bobo} = r - \tau_2/2 + \wascalOof  r^{-1})
 \,.
\ee
This implies that $\og_\Bo^{AB}$ is of the form
\bel{eq:asymptgbondiAB}
	\og_\Bo^{AB} = \hring^{AB} r^{-2}_\Bo + \left( \og_\Bo^{AB} \right)_3 r^{-3}_\Bo + {\wascalO }\big(r^{-4}_\Bo\big)
	\,,
\ee
Using the form  \eqref{eq:solwaveconstraint2} of $\nu^0_\Bo$,  keeping in mind the relation $\nu^0_\Bo=1/\nu_0^\Bo$ and the form \eqref{eq:solwaveconstraint3} of $\xi_A^\Bo$,
we find the solution of \eqref{eq:nuAbondiode}
\begin{eqnarray}
\label{eq:solwaveconstraint4}
 	\nu^A_\Bo &= &%
  \nuAa
	+
	\hring^{AB} \conenabla_B \nuoa^{-1} r_{\Bobo}^{-1}
	\nonumber \\
	&&
	+
		\left[
			\nuoa^{-1} \hring^{AB} \znabla_C \left(\sigma^{\Bo C}_B\right)_2
			+
			\frac{1}{2} \left(\og_\Bo^{AB}\right)_3 \conenabla_B \nuoa^{-1}
		\right]
	r_{\Bobo}^{-2}
	\nonumber \\
	&&
	+
	\frac{2}{3} \hring^{AB} \left[ \znabla_A \left(\sigma^{\Bo A}_B\right)_3   - 8\pi \big(\ol{T}_{rB}^\Bo\big){}_3 \right] \frac{\log r_{\Bobo}}{r_{\Bobo}^2}
	\nonumber \\
	&&
	-
	\frac{1}{3}
		\Bigg[
			\frac{2}{3} \hring^{AB} \left( - \znabla_A \left(\sigma^{\Bo A}_B\right)_3 + 8\pi \big(\ol{T}_{rB}^\Bo\big){}_3 \right)
			\nonumber \\
			&&
			+
			2 \left(\og_\Bo^{AB}\right)_3 \znabla_A \left(\sigma^{\Bo A}_B\right)_2
			+
			\hring^{AB} C_B^{(\xi_C)}
			-
			\left(\og_\Bo^{AB}\right)_4 \znabla_B \nuoa^{-1}
			\nonumber \\
			&&
			+
			\hring^{AB} \left(\nu^0_\Bo\right)_2 \znabla_B \nuoa^{-1}
			+
			\nuoa^{-1} \hring^{AB} \znabla_B \left(\nu^0_\Bo\right)_2
		\Bigg]
	r_{\Bobo}^{-3}
	\nonumber \\
	&&
	+
	\hbox{o}\big(r_{\Bobo}^{-3}\big)
	\,,
\end{eqnarray}
where $\nuAa$ is a global integration function.

Note that the coefficient of the logarithmic term vanishes when \eq{16II16.10} holds.

\subsection{Expansion of $\zeta^\Bo$}
\label{sec:solwaveconstraint5}

Inserting $\tau^\Bo=2r^{-1}_\Bo$ and (\ref{eq:solwaveconstraint1}) into (\ref{eq:wavegaugeconstraint5}) in Bondi-type coordinates yields, cf.~\eqref{24I15.4},
\begin{eqnarray}
\label{eq:zetabondiode}
 \lefteqn{
	\Big( \partial_{r_{\Bobo}} + \frac{2}{r_{\Bobo}}  + \frac{r_{\Bobo}}{2} \big( |\sigma^{\mathrm{Bo}}|^2 +8\pi \ol T^{\mathrm{Bo}}_{rr} \big) \Big) \zeta^{\mathrm{Bo}} =
}
&&
	\nonumber
\\
	&&-  \coneR^{\mathrm{Bo}} + \frac{|\xi^{\mathrm{Bo}}|^2}{2} - \conenabla_A\xi_{\mathrm{Bo}}^A
	+ 8 \pi \big( \ol g_{\mathrm{Bo}}^{AB}\ol T^{\mathrm{Bo}}_{AB}-\ol T^{\mathrm{Bo}} \big) + 2 \Lambda
\,.
\end{eqnarray}
In order to solve this equation we start by defining
\bel{eq:deltazetabondidef}
	\zeta^\Bo := \zeta^\Bo_{\Lambda=0} + \delta\zeta^\Bo
	\,,
\ee
where $\zeta^\Bo_{\Lambda=0}$ is the solution of (\ref{eq:zetabondiode}) in the case $\Lambda=0$. Its asymptotic expansion is known:   \cite[Equation~(3.40)]{TimAsymptotics}
gives the formula in general coordinates for general $\zR$, while \cite[Equation~(5.23)]{TimDiss}
 the one in Bondi-type coordinates with $\zR=2 $)  and reads
%
%
\bel{eq:solzetabondinolambda}
	\zeta^\Bo_{\Lambda=0} = - \frac{{\zR}}{r_{\Bobo}} + \left(\zeta^\Bo_{\Lambda=0}\right)_2 r^{-2}_\Bo + \hbox{o}\big(r^{-2}_\Bo\big)
	\,,
\ee
where $\left(\zeta^\Bo_{\Lambda=0}\right)_2$ is a global integration function and $\zR$ denotes the leading order coefficient of the asymptotic expansion of $\coneR$ in terms of $r$, which coincides with the Ricci scalar of the boundary metric $\lim_{r\to\infty}r^{-2}\overline g_{AB} dx^A dx^B$.   The expansion of $\delta\zeta^\Bo$ on the other hand can be calculated by subtracting (\ref{eq:zetabondiode}) from the corresponding equation in the case $\Lambda=0$, leading to
\bel{eq:deltazetabondieq}
	\left( \partial_{r_{\Bobo}} + \tau^\Bo + \kappa^\Bo \right) \delta\zeta^\Bo = 2\Lambda
	\,.
\ee
This equation can be solved by using (\ref{eq:Tbondiexpansion}), (\ref{eq:solwaveconstraint1}) as well as \eqref{eq:sigmabondisquareexpansion} and we end up with
\begin{eqnarray}
\label{eq:soldeltazetabondi}
	\delta\zeta^\Bo
&= &   \frac{2 \Lambda}{3} r_{\Bobo}
	-
	\frac{\Lambda}{3} \left[ |\sigma^\Bo|^2_4 + 8\pi \big(\ol{T}_{rr}^\Bo\big)_4 \right] r_{\Bobo}^{-1}
	\nonumber
	+	\frac{\Lambda}{3} \left[ |\sigma^\Bo|^2_5 + 8\pi \big(\ol{T}_{rr}^\Bo\big)_5 \right] \frac{ \log{\left(r_{\Bobo}\right)}}{r_{\Bobo}^2}
	\nonumber \\
	&&
	+	\frac{(\delta\zeta^\Bo)_2}{r_{\Bobo}^2} + \hbox{o}\big(r_{\Bobo}^{-2}\big)
	\,,
\end{eqnarray}
%
where $(\delta\zeta^\Bo)_2$ is again a global integration function. Summing up the solution of (\ref{eq:zetabondiode}) in Bondi-type coordinates reads
\begin{eqnarray}
\label{eq:solwaveconstraint5}
	\zeta^\Bo &=& \frac{2 \Lambda}{3} r_{\Bobo}
	-
	\left( \mathbin{ {\zR}} + \frac{2\Lambda}{3} (\kappa^\Bo)_3 \right) r_{\Bobo}^{-1}
	\nonumber \\
	& &
	+	\frac{2\Lambda}{3} (\kappa^\Bo)_4 \frac{\log r_{\Bobo}}{r_{\Bobo}^2}
	+
	\frac{(\zeta^\Bo)_2}{r_{\Bobo}^2} + \hbox{o}\big(r_{\Bobo}^{-2}\big)
	\nonumber
\\
	&
 =
&
\frac{2 \Lambda}{3} r_{\Bobo}
	-
	\left( \mathbin{ {\zR}} + \frac{\Lambda}{3} \left[ |\sigma^\Bo|^2_4 + 8\pi \big(\ol{T}_{rr}^\Bo\big)_4 \right] \right) r_{\Bobo}^{-1}
	\nonumber \\
	&&
	+	\frac{\Lambda}{3} \left[ |\sigma^\Bo|^2_5 + 8\pi \big(\ol{T}_{rr}^\Bo\big)_5 \right] \frac{\log r_{\Bobo}}{r_{\Bobo}^2}
	+
	\frac{(\zeta^\Bo)_2}{r_{\Bobo}^2} + \hbox{o}\big(r_{\Bobo}^{-2}\big)
	\,,
\end{eqnarray}
and we have combined both integration functions $(\delta\zeta^\Bo)_2$ and $\left(\zeta^\Bo_{\Lambda=0}\right)_2$ into $(\zeta^\Bo)_2$.

In view of the analysis of Section~\ref{s10I15.1}, existence of a smooth conformal completion leads to the condition
\bel{16II16.71}
   \left(\kappa^\Bo\right)_4 =0
   \qquad
    \Longleftrightarrow
    \qquad
       |\sigma^\Bo|^2_5 + 8\pi \big(\ol{T}_{rr}^\Bo\big)_5 =0
  \,.
\ee

\subsection{Analyzing (\ref{eq:wavegaugeconstraint6})}
\label{sec:solwaveconstraint6}

Inserting (\ref{eq:VBondir}) into (\ref{eq:wavegaugeconstraint6}) in Bondi-type coordinates and keeping in mind that, by \eqref{eq:nu0Bondiode}, $\partial_{r_{\Bobo}} \nu^0_\Bo = - \kappa^\Bo$ one finds, cf.~\eqref{constraint_bondi6h},
\bel{eq:grrbondiode}
	\og^{rr}_\Bo + (\tau^\Bo)^{-1} \left( \zeta^\Bo - 2 \nu^0_\Bo \conenabla_A \nu^A_\Bo \right) = 0
\ee
%
for (\ref{eq:wavegaugeconstraint6}) in Bondi-type coordinates, which is an algebraic equation for $\og^{rr}_\Bo$. Inserting the asymptotic expansions (\ref{eq:solwaveconstraint2}), (\ref{eq:solwaveconstraint4}) and (\ref{eq:solwaveconstraint5}) we found for $\nu^0_\Bo$, $\nu^A_\Bo$ and $\zeta^\Bo$ respectively we obtain the asymptotic expansion
%
\begin{eqnarray}
\label{eq:solwaveconstraint6}
	\og^{rr}_\Bo &=& -\frac{\Lambda}{3} r^2_\Bo
	+
	\nuoa \conenabla_A \nuAa r_{\Bobo}
	+
	\left( \frac{{\zR}}{2} + \frac{\Lambda}{3} \left(\kappa^\Bo\right)_3 \right)
	-
	\frac{\Lambda}{3} (\kappa^\Bo)_4  \frac{\log r_{\Bobo}}{r_{\Bobo}}
	\nonumber \\
	&&
	+	
  \left( \nuoa \left[
		\znabla_A \left(\nu^A_\Bo\right)_2 + \left(\nu^0_\Bo\right)_2 \znabla_A \nuAa
		\right]
		- \frac{1}{2} (\zeta^\Bo)_2
	\right) r^{-1}_\Bo
	\nonumber \\
	&&
	+ \hbox{o} \big(r^{-1}_\Bo\big)
	\,,
\end{eqnarray}
%
%
where, as before, $\zR$ is the Ricci scalar of the boundary metric $\lim_{r\to\infty}r^{-2}\overline g_{AB} dx^A dx^B$ and $\znabla_A$ is the associated covariant derivative.

Note that the coefficient of the logarithmic term in \eqref{eq:solwaveconstraint5} vanishes if \eq{16II16.71} holds.

\subsection{No-logs}
 \label{ss18II16.1}

Consider characteristic initial data on $\mcN$ such that the functions  $r^{-2}_\Bobo g^\Bo_{AB}$ have a full asymptotic expansion in terms of inverse powers of $r_{\Bobo}$. From what has been said it follows that the equations
\bel{18II16.2}
  \znabla_B \left(\sigmaboAB{}\right)_3  - 8\pi \big(\ol{T}_{rA}^\Bo\big)_3  = 0 =
       |\sigma^\Bo|^2_5 + 8\pi \big(\ol{T}_{rr}^\Bo\big)_5
  \,,
\ee
(see \eq{16II16.10} and \eq{16II16.71})
provide a necessary condition for conformal smoothness of the associated space-time.
It is likely that an analysis along the lines of~\cite{ChPaetz3}, using~\cite{friedrich:JDG,Friedrich,TimConformal}, will prove that these equations are also sufficient in space-times with conformally well-behaved matter fields (cf.~\cite{friedrich:JDG,FriedrichDust,FriedrichMassive}), but we have not investigated this.
 \ptctodo{to be done?}

When one, or both, of Equations \eq{18II16.2} fails, the characteristic initial data set will have a full polyhomogeneous expansion at infinity. One expects that the evolved metric will similarly have a polyhomogeneous expansion, but no  evolution theorems guaranteeing this are available so far even in vacuum.

As such, the no-logs conditions \eq{18II16.2} require the data to be transformed to Bondi coordinates, if not already so given. When $\Lambda=0$, a coordinate-invariant version of the no-logs conditions has been established by Paetz in~\cite{TimAsymptotics}. It would be of interest to find the equivalent of his conditions for $\Lambda\in \R^*$.
 \ptctodo{to be done?}

\section{\Charmass}
 \label{s28XII15.1}

 Throughout this section we assume that the space-time dimension is $n+1=4$.

\subsection{The Trautman-Bondi mass}
\label{subsec:mTB}

In \cite{BBM,Sachs,T}, assuming $\Lambda =0$ and in space-dimension $n=3$,  it was proposed how to define the mass of a null hypersurface $\mcN$ at a given moment of ``retarded time'' $u$ at the cross-section $\sectionofscri$ where it intersects null infinity $\scrip$.
This mass, usually referred to as the \emph{Trautman-Bondi mass}, is defined as
\bel{eq:mTB}
	 \mTB = \frac{1}{4\pi} \int_{\sectionofScri } M \, \hringmeasure
 \,,
\ee
where%
\footnote{Bondi et al. introduced this formalism in the asymptotically flat case, where $\hring_{AB} \equiv s_{AB}$, the standard metric on $S^2$. In anticipation of other boundary topologies, e.g. a torus, we will use the symbol $\hring$ to denote the chosen metric on the relevant manifold.}
 $\hringmeasure=\sqrt{\det \hring_{AB}}dx^2dx^3$, and where $M$ denotes the \emph{mass aspect function} $M:\sectionofScri\to \R$ (compare, e.g., \cite{BBM,Sachs}),
\bel{eq:massaspectdef}
	M := \frac{1}{2} \left(\ol g_{00}^\Bo \right)_1
	\,.
\ee
The definition uses Bondi coordinates, as seen in Section~\ref{ss5XII15.1}, and recall that $(g^\Bo_{00})_k$ denotes the coefficient in front of $1/r^k_\Bobo$ in an asymptotic expansion of $g^\Bo_{00}$ for large $r_{\Bobo}$, in particular
$$
 g^\Bo_{00} = -1 +   \frac{\left(g^\Bo_{00}\right)_1} {r_{\Bobo}} + o(r_{\Bobo}^{-1})
  \,.
$$

Our aim is to obtain an analogue of the Trautman-Bondi mass in  space-times with $\Lambda\ne 0$. We seek to derive a formula which applies to a class of space-times which includes  vacuum space-times with a smooth conformal completion at null infinity $\scrip$, such that the characteristic surface intersects $\scrip$ in a smooth cross-section $\sectionofScri$.

From our point of view, the key justification of \eq{eq:massaspectdef} as providing a good candidate for the integrand for \emph{a total mass} is the fact that $M$ is one of the non-local integration function which arise  when solving the characteristic equations when $\Lambda=0$. It turns out that this remains true for $\Lambda\ne 0$.

Consider, thus, characteristic data in Bondi-type
coordinates, defined perhaps only for large values of $r_{\Bobo}$. The space-time metric on $\mcN=\{u ^\Bo=0\}$ can then be written as
\bel{eq:metriconN}
 \ol g = \ol g_{00}^\Bo {d}u_\Bo^2 + 2\nu_0^\Bo{d}u_\Bo {d}r_{\Bobo} + 2\nu_A^\Bo{d}u _\Bo {d}x^A_\Bo + \coneg^\Bo
 \,.
\ee
Now, Bondi \emph{et al.} assume
\bel{eq:bc}
	\lim_{r_{\Bobo} \to \infty} \nu^A_\Bo = 0
	\,, \quad
	\lim_{r_{\Bobo} \to \infty} \nu^0_\Bo =   1
	\,, \quad
	\lim_{r_{\Bobo} \to \infty} \left( r_{\Bobo}^{-2} \og^\Bo_{AB} \right) = \hring_{AB}
	\,.
\ee
It follows from Proposition~\ref{P12XII15.1} that the last equation in \eq{eq:bc} is justified under the hypotheses there. However, it is not clear at all whether the first two can be assumed to hold \emph{for all retarded times} in general: When $\Lambda<0$ this is part of asymptotic conditions which are usually imposed in this context, but which one might  \emph{not} want to impose in \emph{some} situations. However, when $\Lambda>0$  there is little doubt that all three conditions in \eq{eq:bc} can be simultaneously satisfied for all times by a restricted class of metrics only.
For this reason we have allowed general fields $\left(\nu^A_\Bo\right)_0 $ and $\left(\nu^0_\Bo\right)_0(x^B)$ when solving the constraint equations so far.

Nevertheless, it is easy to see that the first two equations \eq{eq:bc} are determined by the propagation of the coordinates $u$ and $x^A$  away from the initial data surface $\mcN$, and can always be imposed on the $\mcN$  as long as one does not assume that they  hold at later times. In particular, the first two equations in \eq{eq:bc} imply no loss of generality \emph{as long as no evolution equations are used.} Since we only work at $\mcN$, and use only the constraint equations, we will assume \eq{eq:bc} from now on.


\subsection{The \charmass\  in Bondi-type coordinates}
\label{sec:massbondicoords}

The asymptotic expansion of $\og^\Bo_{00}$ needed to obtain the mass aspect function can be calculated using the third equation in \eqref{eq:fielddef2} in Bondi-type coordinates
\bel{g00Bondidef}
	\og_{00}^\Bo = \og^\Bo_{AB}\nu^A_\Bo\nu^B_\Bo - (\nu_0^\Bo)^2 \og_\Bo^{rr}
	\,.
\ee
and we note again (cf. \eqref{eq:asymptgbondiAB}) that $\og_\Bo^{AB}$ is of the form
\be
	\og_\Bo^{AB} = \hring^{AB} r^{-2}_\Bo + \left( \og_\Bo^{AB} \right)_3 r^{-3}_\Bo + {\wascalO }\big(r^{-4}_\Bo\big)
	\,.
\ee
Using this and (\ref{eq:solwaveconstraint6}) leads us to
%
\begin{eqnarray}
	\og^\Bo_{00} &=& \bigg( \hring_{AB} \nuAa \nuBa - \nuoa^{-2} \left( \og^{rr}_\Bo \right)_{-2} \bigg) r^2_\Bo
 \nn
\\
 \nn
 &&
	+
	\left(\og^\Bo_{AB}\right)_{-1} \nuAa \nuBa r_{\Bobo}
	\nonumber \\
	&&
	+
		\nuAa
		\Big[
			2 \hring_{AB} \left(\nu^B_\Bo\right)_2 + \left(\og^\Bo_{AB}\right)_0 \nuBa
		\Big]
	\nonumber \\
	&&
		-
		\nuoa^{-1}
		\Big[
			\nuoa^{-1} \left( \og^{rr}_\Bo\right)_0 + 2 \left( \og^{rr}_\Bo \right)_{-2} \left(\nu_0^\Bo\right)_2
		\Big]
	\nonumber \\
	&&
	-
	\nuoa^{-2} \left(\og^{rr}_\Bo\right)_{\text{log},1} \frac{\log r_{\Bobo}}{r _\Bo}
	\nonumber \\
	&&
	+
	\bigg(
		\nuAa
		\Big[
			2 \hring_{AB} \left(\nu^B_\Bo\right)_3
			+ 2 \left(\og^\Bo_{AB}\right)_{-1} \left(\nu^B_\Bo\right)_2
			+ \left(\og^\Bo_{AB}\right)_1 \nuBa
		\Big]
	\nonumber \\
	&&
		- \nuoa^{-1}
		\Big[
			\nuoa^{-1} \left(\og^{rr}_\Bo\right)_1
			+ 2 \left(\og^{rr}_\Bo\right)_{-1} \left(\nu_0^\Bo\right)_2
			+ 2 \left(\og^{rr}_\Bo\right)_{-2} \left(\nu_0^\Bo\right)_3
		\Big]
	\bigg) r^{-1}_\Bo
	\nonumber \\
	&&
	+
	\hbox{o}\big(r^{-1}_\Bo\big)
	\,,
\label{eq:g00bondiexpansion}
\end{eqnarray}
where we can directly read off an expression for the mass aspect function $M$:
\begin{eqnarray}
	M & =& \frac{1}{2} \left(\og^\Bo_{00}\right)_1
	\nonumber \\
	&
	= &
	\frac{1}{2}
	\bigg(
		\nuAa
		\Big[
			2 \hring_{AB} \left(\nu^B_\Bo\right)_3
			+ 2 \left(\og^\Bo_{AB}\right)_{-1} \left(\nu^B_\Bo\right)_2
			+ \left(\og^\Bo_{AB}\right)_1 \nuBa
		\Big]
	\nonumber \\
	&&
		- \frac{
			\nuoa^{-1} \left(\og^{rr}_\Bo\right)_1
			+ 2 \left(\og^{rr}_\Bo\right)_{-1} \left(\nu_0^\Bo\right)_2
			+ 2 \left(\og^{rr}_\Bo\right)_{-2} \left(\nu_0^\Bo\right)_3}
		{\nuoa}
	\bigg)
 \,.
\end{eqnarray}
Now, using that $\nu_0^\Bo = 1 / \nu^0_\Bo$ (cf. \eqref{eq:fielddef2}) and \eqref{eq:solwaveconstraint2}, we have
\begin{eqnarray}
\label{eq:nu0downbondiexpansion}
	\nu_0^\Bo
	&= & \nuoasymptoticvalue^{-1}
    \bigg(1
	-
	\left(\nu^0_\Bo\right)_2 r_{\Bobo}^{-2}
	-
	\left(\nu^0_\Bo\right)_3 r_{\Bobo}^{-3}
\bigg)
	+
	{\wascalO }\big(r_{\Bobo}^{-4}\big)
	\nonumber \\
	&=  &
 \nuoasymptoticvalue^{-1}
    \bigg(1
	-
	\frac{1}{4} \Big[ |\sigma^\Bo|^2_4 + 8\pi \big(\ol{T}_{rr}^\Bo\big)_4 \Big] r_{\Bobo}^{-2}
	\nonumber \\
	&&
	-
	\frac{1}{6} \Big[ |\sigma^\Bo|^2_5 + 8\pi \big(\ol{T}_{rr}^\Bo\big)_5 \Big] r_{\Bobo}^{-3}
\bigg)
	+
	{\wascalO }\big(r_{\Bobo}^{-4}\big)
	\,.
\end{eqnarray}
Inserting this and the expansion coefficients of $\og^{rr}_\Bo$ and $\nu^A_\Bo$ we calculated before $M$ reads
\begin{eqnarray}
\label{eq:solMgeneral}
	M &=&
		\nuAa
		\Big[
			 \hring_{AB} \left(\nu^B_\Bo\right)_3
			+ \left(\og^\Bo_{AB}\right)_{-1} \left(\nu^B_\Bo\right)_2
			+ \frac{1}{2} \left(\og^\Bo_{AB}\right)_1 \nuBa
		\Big]
	\nonumber \\
	& &
		 + \nuoa^{-1} \frac{1}{2}
		\bigg(
		 \frac{1}{2} \left( (\zeta^\Bo)_2 + \conenabla_A \nuAa \Big[ |\sigma^\Bo|^2_4 + 8\pi \big(\ol{T}_{rr}^\Bo\big)_4 \Big] \right)
		\nonumber \\
	&&
    -		
		\znabla_A \left(\nu^A_\Bo\right)_2
- \nuoa^{-1} \frac{\Lambda}{4} \Big[ |\sigma^\Bo|^2_5 + 8\pi \big(\ol{T}_{rr}^\Bo\big)_5 \Big]
		\bigg)
 \nn
\\
 &&
	- \frac{1}{8} \Big[ |\sigma^\Bo|^2_4 + 8\pi \big(\ol{T}_{rr}^\Bo\big)_4 \Big] \znabla_A \nuAa
 \,.
\end{eqnarray}

We return, now, to the definition of the \charmass, \eqref{eq:mTB}, and assume in the remainder of the present work that the boundary conditions on $\nu^0_\Bo$ and $\nu^A_\Bo$, introduced in \eqref{eq:bc}, hold.
With these boundary conditions, and using the fact that the  divergence terms
 in \eqref{eq:solMgeneral} will integrate out to zero, we find
\begin{eqnarray}
\label{eq:solamssbondi}
	\mTB &=& \frac{1}{16\pi}\int_{\sectionofScri} (\zeta^\Bo)_2 \, \hringmeasure
	+
	\frac{\Lambda}{12\pi}\int_{\sectionofScri} \left(\nu^\Bo_0\right)_3 \, \hringmeasure
	\nonumber \\
	&=& \frac{1}{16\pi}\int_{\sectionofScri} (\zeta^\Bo)_2 \, \hringmeasure
	-
	\frac{\Lambda}{72\pi}\int_{\sectionofScri} \left[ |\sigma^\Bo|^2_5 + 8\pi \big(\ol{T}_{rr}^\Bo\big)_5 \right] \, \hringmeasure
	\,.
\end{eqnarray}

\subsection{The \charmass\ in terms of characteristic data}
    \label{sec:masscharacteristic}

To continue, we want to relate the fields occurring in Bondi-type coordinates to their representation in coordinates where $r$ is an affine parameter along the radial null outgoing geodesics of $\ol{g}$.
We start with $(\zeta^\Bo)_2$ and follow the argumentation in \cite[leading to Equation~(51) there]{ChPaetzBondi}, which we repeat here for the convenience of the reader.

First, we have
\bel{eq:rrBotransf}
	r_{\Bobo} = r - \frac{\tau_2}{2} + {\wascalO }(r^{-1})
	\,.
\ee

Next, the  transformation formulae for $\tau$ and  $\zeta$ (compare (\ref{eq:zeta_spacetime})) read:
%
\bea
%
%
\label{eq:tautransformation}
  \tau^\Bo(r^\Bo)
  &=&
  \frac{\partial r}{\partial r^\Bo} \tau(r( r^\Bo)) \,=\, \frac{2}{r_{\Bobo}} 
\,,
\\
 \zeta^\Bo
  &=&
 2(\og^\Bo)^{AB}{(\overline{\Gamma}^\Bo)}{}^{r_{\Bobo}}_{AB} +  \tau^\Bo (\og^\Bo)^{r_{\Bobo} r_{\Bobo}}
 \nonumber
\\
%
%
 &=&
   2(\og^\Bo)^{AB}\Big( \frac{\partial r_{\Bobo} }{\partial  x^k} \frac{\partial  x^i}{\partial x_\Bo ^A}\frac{\partial  x^j}{\partial x_\Bo ^B}{\overline{\Gamma}}{}^k_{ij} +  \frac{\partial r_{\Bobo} }{\partial r} \frac{\partial^2 r}{\partial x_\Bo ^{A}\partial x_\Bo ^B}\Big)
  \nn
\\
 &&
 +   \tau\frac{\partial r}{\partial r_{\Bobo} } \frac{\partial r_{\Bobo} }{\partial x^{i}}\frac{\partial r_{\Bobo} }{\partial x^{j}}{\og}{}^{ij}
  \nn
\\
&=&
 2\og^{AB} \frac{\partial  r_{\Bobo} }{\partial r} \frac{\partial r}{\partial x_\Bo^A}\frac{\partial r}{\partial x_\Bo ^B} \kappa
 +  2\og^{AB} \frac{\partial  r_{\Bobo} }{\partial  x^C} \frac{\partial r}{\partial x_\Bo^A}\frac{\partial  r}{\partial x_\Bo ^B}\underbrace{{\ol\Gamma}{}^C_{11}}_{=0}
  \nn
\\
 &&
 +
  \frac{\partial  r_{\Bobo} }{\partial  r}\zeta
 +  2 {\og}^{AB} \frac{\partial  r_{\Bobo} }{\partial  x^C}\tilde\Gamma{}^C_{AB}
 +  2 \underbrace{{\og}^{AB} \chi_{AB}}_{=\tau}\frac{\partial  r_{\Bobo} }{\partial  x^C}\nu^0\nu^C
  \nn
\\
  &&
  - 2 \og^{AB} \frac{\partial  r_{\Bobo} }{\partial  r} \frac{\partial r}{\partial x_\Bo ^B}\xi_A
  +  2\tau \og^{AB} \frac{\partial  r_{\Bobo} }{\partial  x^A} \frac{\partial r}{\partial x_\Bo^B}
+  4\og^{AB} \frac{\partial  r_{\Bobo} }{\partial  x^C} \frac{\partial r}{\partial x_\Bo^B}\sigma_A{}^C
  \nn
\\
&&
 +  2\og^{AB}  \frac{\partial r_{\Bobo} }{\partial r} \frac{\partial^2 r}{\partial x_\Bo^{A}\partial x_\Bo ^B}
 +   \tau\frac{\partial r}{\partial r_{\Bobo} } \frac{\partial r_{\Bobo} }{\partial x^{A}}\frac{\partial r_{\Bobo} }{\partial x^{B}}{\og}{}^{AB}
  \nn
\\
\hspace{-2mm}&=&
  \frac{\partial  r_{\Bobo} }{\partial  r}\zeta
%
%
%
%
 +  2  \frac{\partial r_{\Bobo} }{\partial r}\Delta_{{\check g}}r
 + O\big(r_{\Bobo}^{-3}\big)
\label{eq:zetatransformation}
\,,
\eea
%
where $\Delta_{{\check g}}$ is the Laplace operator of the two-dimensional metric $\check g_{AB}dx_\Bo^A dx_\Bo^B$.

To continue we need the asymptotic expansion of $\zeta$ and therefore solve the respective constraint equation (\ref{eq:wavegaugeconstraint5}). Note that we have already done this in Bondi-type coordinates, but we also need the result in affine coordinates.

 We begin with the same procedure as in Section~\ref{sec:solwaveconstraint5} and define
\bel{eq:deltazetadef}
	\zeta := \zeta_{\Lambda=0} + \delta\zeta
	\,,
\ee
where $\zeta_{\Lambda=0}$ is the solution of (\ref{eq:wavegaugeconstraint5}) in the case $\Lambda=0$. Its asymptotic expansion is known and reads \cite[Equation~(3.40)]{TimAsymptotics}
%
\bel{eq:solzetanolambda}
	\zeta_{\Lambda=0} = - \frac{\zR}{r} + \left(\zeta_{\Lambda=0}\right)_2 r^{-2} + \hbox{o}\big(r^{-2}\big)
	\,,
\ee
with $\left(\zeta_{\Lambda=0}\right)_2$ being a global integration function. We assume that the relevant fields satisfy analog fall-off behavior in affine coordinate $r$ as we assumed in Bondi-type coordinate $r_{\Bobo}$ (cf. Equations~\eqref{eq:sigmabondiexpansion} and \eqref{eq:Tbondiexpansion}).
The equation for $\delta\zeta$ reads
\bel{eq:deltazetaeq}
	\left( \partial_r + \tau + \kappa \right) \delta\zeta = 2\Lambda
 \,.
\ee
From now on we choose the coordinate $r$ so that $\kappa=0$. We start by solving the Raychaudhuri equation (\ref{eq:wavegaugeconstraint1}) in this gauge and obtain the expansion of $\tau$
%
\beal{eq:tauexpansion}
	  \tau  &=&  \frac{2}{r} + \frac{\tau_2}{r^2}
	+ \frac{2 \left[ |\sigma|^2_4 + 8 \pi \big(\ol T_{rr}\big)_4 \right] + \tau_2^2}{2 r^3}
	\nonumber \\
	&& + \frac{2 \left[ |\sigma|^2_5 + 8 \pi \big(\ol T_{rr}\big)_5 \right] + 2 \tau_2 \left[ |\sigma|^2_4 + 8 \pi \big(\ol T_{rr}\big)_4 \right] + \tau_2^3}{4 r^4}
	+ {\wascalO }\big(r^{-5}\big)
	\,,
	\phantom{xxx}
\eea
where $\tau_2$ is a global integration function and $|\sigma|^2_n$ are the expansion coefficients of $|\sigma|^2$:
\beal{eq:firstsigmasquareexpansion}
	|\sigma|^2 = \frac{|\sigma|^2_4}{r^4} + \frac{|\sigma|^2_5}{r^5} + {\wascalO }\big(r^{-6}\big)
	\,.
\eea
Using (\ref{eq:tauexpansion}) we find from (\ref{eq:deltazetaeq})
\begin{eqnarray}
\label{eq:deltazetasolution}
	\delta\zeta
 &=&
  \Lambda \left( \frac{2r}{3} - \frac{\tau_2}{3} + \frac{\tau_2^2-2\tau_3}{3r} + \frac{3 \tau_2 \tau_3 - \tau_2^3 - 2 \tau_4}{3} \frac{\log r}{r^2} \right)
\nonumber \\
 &&
 + \frac{\delta\zeta_2}{r^2}
  + \hbox{o}\big(r^{-2}\big)
	\,,
 \phantom{xx}
\end{eqnarray}
where $\delta\zeta_2$ is again a global integration function. 
Summing, and combining the two integration functions $(\zeta_{\Lambda=0})_2$ and $\delta\zeta_2$ into $\zeta_2$ the solution of (\ref{eq:wavegaugeconstraint5}) gives
%
\begin{eqnarray}
\label{eq:zetasolution}
	\zeta
	&=&
	\frac{2\Lambda}{3} r
	- \frac{\Lambda \tau_2}{3}
	- \left( \zR + \frac{\Lambda \big(2\tau_3-\tau_2^2\big)}{3} \right)
    r^{-1}
	\nonumber \\
	&&
	+ \frac{ \Lambda \left( 3 \tau_2 \tau_3 - \tau_2^3 - 2 \tau_4 \right)}{3} \frac{\log r}{r^2}
	+ \frac{\zeta_2}{r^2}
	+ \hbox{o}\big(r^{-2}\big)
	\,.
\end{eqnarray}
Using this and the asymptotic expansion of $\Delta_{\coneg}r$
(compare~\cite[Equation~(51)]{ChPaetzBondi})
\bel{eq:deltarexpansion}
	\Delta_{\coneg}r = \frac{\Delta_{\hring}\tau_2}{2 r^2} + {\wascalO }\big(r^{-3}\big)
\ee
and expressing (\ref{eq:zetatransformation}) in terms of $r_{\Bobo}$ one obtains
%
\begin{eqnarray}
\label{eq:zetabondi2transformation}
	(\zeta^\Bo)_2 &=& \zeta_2 + \frac{\zR}{2} \tau_2 + \Delta_{\hring} \tau_2
	\nonumber \\
	&  &
	+ \frac{\Lambda}{3} \left( -\frac{|\sigma^\Bo|^2_5 + 8 \pi \left(\ol T^\Bo_{rr}\right)_5}{3} + \tau_2 \Big[ |\sigma|^2_4 + 8 \pi \big(\ol T_{rr}\big)_4 \Big] \right)
	\,,
	\\
	 \zeta^\Bo_{\log,2}
	 & = &
	  \frac{\Lambda}{3} \Big( 2 \tau_2 \Big[ |\sigma|^2_4 + 8 \pi \big(\ol T_{rr}\big)_4 \Big] - |\sigma|^2_5 - 8 \pi \big(\ol T_{rr}\big)_5 \Big)
	\,.
\end{eqnarray}
Inserting \eqref{eq:zetabondi2transformation} into (\ref{eq:solMgeneral}) and using the boundary conditions on $\nu^0_\Bo$ and $\nu^A_\Bo$, introduced in \eqref{eq:bc}, we find
\begin{eqnarray}
\label{eq:intermediateMcharacteristic}
	M &=&
  \frac{1}{4} \left( \zeta_2 + \frac{\zR}{2} \tau_2 + \Delta_{\hring} \tau_2 \right)
	-
	\frac{1}{2} \znabla^A \left(\nu_A^\Bo\right)_0
	\nonumber \\
	&&
	-	\frac{\Lambda}{12} \bigg( |\sigma^\Bo|^2_5 + 8\pi \big(\ol{T}_{rr}^\Bo\big)_5 - \tau_2 \Big[ |\sigma|^2_4 + 8 \pi \big(\ol T_{rr}\big)_4 \Big] \bigg)
	\,.
\end{eqnarray}

We now calculate the expansions of $|\sigma|^2$ and $|\sigma^\Bo|^2$ to insert explicit expressions for the coefficients occurring in $M$. Further we want to relate the relevant coefficients of the energy-momentum tensor in Bondi-type coordinates to their representation in affine coordinates. For $|\sigma|^2$ one obviously has
\begin{eqnarray}
\label{eq:sigmasquareexpansion}
	|\sigma|^2 &=& \frac{ |\sigma|^2_4}{r^4} + \frac{ |\sigma|^2_5}{r^5} + {\wascalO }\big(r^{-6}\big)
	\nonumber \\
	&=& \frac{ \left(\sigma_A{}^B\right)_2 \left(\sigma_B{}^A\right)_2}{r^4}
	+
	2 \frac{ \left(\sigma_A{}^B\right) _2 \left(\sigma_B{}^A\right)_3}{r^5}
	+
	{\wascalO }\big(r^{-6}\big)
	\,,
\end{eqnarray}
and performing a coordinate transformation and replacing the dependence on $r$ with $r_{\Bobo}$ we obtain
\begin{eqnarray}
\label{eq:sigmabondisquareexpansionexplicit}
	|\sigma^\Bo|^2
  &=&
  \frac{ \left(\sigma_A{}^B\right)_2 \left(\sigma_B{}^A\right)_2}{r_{\Bobo}^4}
	+
	2 \frac{ \left(\sigma_A{}^B\right)_2 \left(\sigma_B{}^A\right)_3 - \left(\sigma_A{}^B\right)_2 \left(\sigma_B{}^A\right)_2 \tau_2}{r_{\Bobo}^5}
	\nonumber
\\
	&&
	+	{\wascalO }\big(r_{\Bobo}^{-6}\big)
	\nonumber
\\
	&=&	\frac{|\sigma|^2_4}{r_{\Bobo}^4} + \frac{ |\sigma|^2_5 - 2|\sigma|^2_4 \tau_2 }{r_{\Bobo}^5} +
	{\wascalO }\big(r_{\Bobo}^{-6}\big)
	\,.
\end{eqnarray}

By assumption, or by smooth conformal compactifiability we can write $\ol{T}_{rr}^\Bo$ in the form
\beal{Trrbondiexpansion}
	\ol{T}_{rr}^\Bo = \frac{\big( \ol{T}_{rr}^\Bo \big)_4}{r^4_\Bo}
	+ \frac{\big( \ol{T}_{rr}^\Bo \big)_5}{r^5_\Bo}
	+ {\wascalO }\big(r_{\Bobo}^{-6}\big)
	\,.
\eea
Performing a coordinate transformation we find an analog expansion for $\ol{T}_{rr}$ in affine coordinates and replacing  again the dependence on $r$ with $r_{\Bobo}$ we obtain
\beal{Trrbondiexpansion+}
	\ol{T}_{rr}^\Bo = \frac{\left( \ol{T}_{rr}  \right)_4}{r^4_\Bo}
	+ \frac{\left( \ol{T}_{rr}  \right)_5 - 2 \left( \ol{T}_{rr}  \right)_4 \tau_2}{r^5_\Bo}
	+ {\wascalO }\big(r_{\Bobo}^{-6}\big)
	\,.
\eea
Therefore we end up with the following formula for the mass aspect expressed through characteristic data%
%
%
\begin{eqnarray}
\label{eq:solMcharacteristic}
	M &=& \frac{1}{4} \left( \zeta_2 + \frac{\zR}{2} \tau_2 + \Delta_{\hring} \tau_2 \right)
	-
	\frac{1}{2} \znabla^A \left(\nu_A^\Bo\right)_0
	\nonumber \\
	&&
	+	\frac{\Lambda}{12} \bigg( 3 \tau_2 \Big[ |\sigma|^2_4 + 8 \pi \big(\ol T_{rr}\big)_4 \Big] - \Big[ |\sigma|^2_5 + 8\pi \left( \ol{T}_{rr}  \right)_5 \Big] \bigg)
	\,.
\end{eqnarray}
Using again the definition of the \charmass\  and bearing in mind that the divergence terms will vanish after integration over $\sectionofScri$ we find
\begin{eqnarray}
\label{eq:solmassbondicharecteristic}
	\mTB &=& \frac{1}{16\pi}\int_{\sectionofScri} \left( \zeta_2 + \frac{\zR}{2} \tau_2 \right) \, \hringmeasure
	\nonumber \\
	&&
	+ \frac{\Lambda}{48\pi} \int_{\sectionofScri} \bigg( 3 \tau_2 \Big[ |\sigma|^2_4 + 8 \pi \big(\ol T_{rr}\big)_4 \Big] - \Big[ |\sigma|^2_5 + 8\pi \left( \ol{T}_{rr}  \right)_5 \Big] \bigg) \, \hringmeasure
	\,.
 \phantom{xxxx}
\end{eqnarray}
%

\subsection{The \charmass\  and the renormalized volume}
\label{sec:massgeometric}

We are ready to prove our final formula for the \charmass, which will be in terms of geometric fields defined on a characteristic surface parametrized by an affine parameter $r$ ranging from $r_0$ to infinity. In the case of a light-cone we take $r_0=0$, but we allow non-zero $r_0$ to cover other situations of interest.

We first note the asymptotic expansion of $\sqrt{\det \ol g_{AB}}$ for large $r$, which is obtained by using the considerations in \cite[leading to equation (3.13) there]{TimAsymptotics}
 and our result for the expansion of $\tau$, \eqref{eq:tauexpansion}:
%
\beal{eq:detgdets}
 \lefteqn{
	\sqrt{\det \ol g_{AB}}
	=
	r^2 \sqrt{\det \hring_{AB}}
	\Bigg(
	1
	-
	\frac{\tau_2}{r}
	+
	\frac{\tau_2^2 - 2 \Big[ |\sigma|^2_4 + 8 \pi \big(\ol T_{rr}\big)_4 \Big] }{4r^2}
}
&&
	\nonumber \\
	&&
\phantom{xxxx}
	+
	\frac{2 \tau_2 \Big[ |\sigma|^2_4 + 8 \pi \big(\ol T_{rr}\big)_4 \Big] -  \Big[ |\sigma|^2_5 + 8 \pi \big(\ol T_{rr}\big)_5 \Big] }{6r^3}
	+ \,
	{\wascalO }(r^{-4})
	\Bigg)
	\,.
\eea
Using this, $d\mu_{\coneg}=\sqrt{\det \ol g_{AB}}dx^2dx^3$ and the expansion (\ref{eq:zetasolution}) of $\zeta$ we find
\begin{eqnarray}
\label{eq:zetaintegral}
 \lefteqn{
	\int_{\sectionofscri}{\zeta}d\mu_{\coneg} = \frac{2\Lambda}{3} r^3 \underbrace{\int_{\sectionofscri}\hringmeasure}_{=:\hringvolume}
%
	- \Lambda r^2 \int_{\sectionofscri}{\tau_2}\hringmeasure
	- r \int_{\sectionofscri} \zR \hringmeasure
%
 }
 &&
	\nonumber
\\
	&&
	- \Lambda r \int_{\sectionofscri}{\left( \Big[ |\sigma|^2_4 + 8 \pi \big(\ol T_{rr}\big)_4 \Big] - \frac{1}{2} \tau_2^2 \right)}\hringmeasure
	\nonumber \\
	&&
	- \frac{\Lambda}{3} \log r \int_{\sectionofscri}{ \bigg( \Big[ |\sigma|^2_5 + 8 \pi \big(\ol T_{rr}\big)_5 \Big] - 2 \Big[ |\sigma|^2_4 + 8 \pi \big(\ol T_{rr}\big)_4 \Big] \tau_2 \bigg) }\hringmeasure
	\nonumber \\
	&&
	+ \int_{\sectionofscri}{\left[ \zeta_2 + \zR \tau_2 \right]} \hringmeasure
	- \frac{\Lambda}{12} \int_{\sectionofscri}{ \tau_2^3 } \hringmeasure
	\nonumber \\
	&&
	+ \frac{\Lambda}{18} \int_{\sectionofscri}{ \bigg( 19 \tau_2 \Big[ |\sigma|^2_4 + 8 \pi \big(\ol T_{rr}\big)_4 \Big] - 2 \Big[ |\sigma|^2_5 + 8 \pi \big(\ol T_{rr}\big)_5 \Big] \bigg) } \hringmeasure
	+ \hbox{o}(1)
	\,.
 \phantom{xxxxx}
\end{eqnarray}
%
%
%
%
%
From (\ref{eq:wavegaugeconstraint5}) with $\kappa=0$ and the Gauss--Bonnet theorem we have,
\bel{eq:integderivationzeta}
	\int_{\sectionofscri}(\partial_r + \tau) \zeta d\mu_{\coneg} = -4\pi \chi(\sectionofScri) + \int_{\sectionofscri}\left(\frac{1}{2}|\xi|^2+S\right) d\mu_{\coneg} + 2\Lambda \int_{\sectionofscri} d\mu_{\coneg}
	\,,
\ee
where $\chi(\sectionofScri)$ is the Euler characteristic of $\sectionofScri$. The integral in the last term of that equation is the area of the constant-$r$ sections of $\mcN$, and we define the volume function $V(r)$ to be its integral
\bel{eq:volume}
	V(r) := \tintfromrzero^{r} \frac{dV(\tilde{r})}{d\tilde{r}} d\tilde{r}  = \tintfromrzero^{r} \int_{\sectionofscri} d\mu_{\coneg} d\tilde{r}
	\,.
\ee
\begin{Remark}
  \label{R12II16.1}
We note that $V(r)$ is uniquely defined up to the choice $r_0$ of the origin of $r$ and up to scaling on each generator.

When cross-sections of $\scri$ are negatively curved compact manifolds, the asymptotic conditions imposed in our construction define  the scaling uniquely.

When cross-sections of $\scri$ are flat compact manifolds, the asymptotic conditions imposed in our construction define  the scaling up to a constant. This freedom can be gotten rid of by requiring the $\mathring h$-volume of the cross-section to take some convenient value, e.g.\ one or $(2\pi)^{2}$.

When cross-sections of $\scri$ are two-dimensional spheres, the asymptotic conditions imposed in our construction define  the scaling uniquely up to the action of the group of conformal transformations of $S^2$.
This freedom reflects the fact that in this case $\mTB$ is not a mass but the time-component of a covector.

A redefinition of $r_0$ affects the explicit formula for $V$ as a function of $r$, and hence the numerical value of the ``renormalized volume'', to be defined shortly.
When $\mcN$ is a globally smooth light-cone, or is a smooth hypersurface emitted from a submanifold of codimension larger than one, then the origin of the affine parameter $r_0=0$ is determined by the location of the ``emitting'' submanifold, which gets rid of the last ambiguity.
\qed
\end{Remark}

Using $\partial_r\sqrt{\det \ol g_{AB}}=\tau\sqrt{\det \ol g_{AB}}$ we find
\bel{eq:derivationzetainteg}
	\partial_r \int_{\sectionofscri} \zeta d\mu_{\coneg} = -4\pi \chi(\sectionofScri) + 2\Lambda \frac{dV(r)}{dr} + \int_{\sectionofscri}\left(\frac{1}{2}|\xi|^2+S\right) d\mu_{\coneg}
	\,,
\ee
which we can integrate in $r$ starting from $r=r_0$
%
%
%
\bean
 \lefteqn{
	\lim_{r \rightarrow \infty} \left( \int_{\sectionofscri} \zeta d\mu_{\coneg} + 4\pi \chi(\sectionofScri) r - 2\Lambda V(r) \right)
 }
 &&
\\
 &&  = \limatrzero \int_{\sectionofscri} \zeta d\mu_{\coneg}
	+
	4 \pi \chi(\sectionofScri) r_0
	+
	\intfromrzero^{\infty}\int_{\sectionofscri} \left( \frac{1}{2} |\xi|^2 + S \right) d\mu_{\coneg} dr
	\,.
	\phantom{xxxxxx}
\eeal{eq:limitzetaintegral}
We leave the symbol $\limatrzero$ in the last equation to accommodate a vertex at $r=r_0$, where $\zeta$ is singular, but note that   light-surfaces emanating from  smooth space co-dimension-two submanifolds will also be of interest to us. One needs to make sure to use appropriate boundary conditions for the lower bound of the integration depending on what kind of characteristic surface is studied. In the case of a light-cone, i.e. a null-hypersurface emanating from a point at $r_0=0$, the necessary boundary conditions follow from regularity at the tip of the cone as has been discussed in \cite[Section~4.5]{CCM2}.

When the first term in the last line vanishes, we can infer non-negativity of the left-hand side by assuming the \textit{dominant energy condition} for non-vanishing matter fields. This condition implies then \cite{TimDiss}
\bel{eq:dominantenergy}
	S := 8\pi (\ol g^{AB} \ol T_{AB} - \ol T )\geq 0
\ee
which means that the right-hand side of (\ref{eq:limitzetaintegral}) is manifestly non-negative.
Assuming that  the right-hand side of (\ref{eq:limitzetaintegral}) is finite, we see
  that the divergent terms  in $2\Lambda V(r)$ and $4\pi\chi(\sectionofScri)r$
need to cancel those in the expression on the right-hand side of (\ref{eq:zetaintegral}) exactly. To make this precise we continue by calculating an explicit expression for the volume function $V(r)$.
We start by using again (\ref{eq:detgdets}) and find
\begin{eqnarray}
\label{eq:areaexplicit}
	\frac{dV(r)}{dr}
& = &
  \int_{\sectionofscri} d\mu_{\coneg}  = r^2 \hringvolume
	- r \int_{\sectionofscri} \tau_2 \hringmeasure
	\nonumber \\
	&&
	+ \frac{1}{2} \int_{\sectionofscri} \left( \frac{1}{2} \tau_2^2 - \Big[ |\sigma|^2_4 + 8 \pi \big(\ol T_{rr}\big)_4 \Big] \right) \hringmeasure
	\nonumber \\
	&&
	+ \frac{1}{6r} \int_{\sectionofscri} \bigg( 2 \Big[ |\sigma|^2_4 + 8 \pi \big(\ol T_{rr}\big)_4 \Big] \tau_2 - \Big[ |\sigma|^2_5 + 8 \pi \big(\ol T_{rr}\big)_5 \Big] \bigg) \hringmeasure
	\nonumber \\
	&&
	+ {\wascalO }\big(r^{-2}\big)
	\,.
\end{eqnarray}
%
It follows that there exist constants so that the function $V(r)$ has an asymptotic expansion of the form
$$
V(r) =  \frac{1}{3} r^3 \hringvolume + V_{-2} r^2 +   V_{-1} r  +  V_{\log} \log r  +  V_0 + V_{1}r^{-1} + {\color{blue}o }\big(r^{-1}\big)
 \,.
$$
We define the \emph{renormalized volume} $V_{\renm}$ as ``the finite left-over in the expansion'':
$$
V_{\renm} := V_0
\,.
$$
One can think of $V_{\renm}$ as the global integration function arising from integrating the equation for $dV/dr$. The numerical value of $V_{\renm}$ is defined up to the ambiguities pointed out in Remark~\ref{R12II16.1}.

Integrating \eq{eq:areaexplicit} we obtain in fact
\begin{eqnarray}
\label{eq:2lambdavolume}
	-2\Lambda V(r) &=&%
\Lambda \Bigg[ - \frac{2}{3} \left(\left(r- \frac {\tau_2} 2 \right)^3
  +\left(\frac{\tau_2} 2\right)^3\right) \hringvolume
	-2 V_\renm
	\nonumber \\
	&&
+ r \int_{\sectionofscri} \left( \Big[ |\sigma|^2_4 + 8 \pi \big(\ol T_{rr}\big)_4 \Big]  \right) \hringmeasure
	\nonumber \\
	&&
	+ \frac{1}{3} \log r \int_{\sectionofscri} \bigg( \Big[ |\sigma|^2_5 + 8 \pi \big(\ol T_{rr}\big)_5 \Big] - 2 \Big[ |\sigma|^2_4 + 8 \pi \big(\ol T_{rr}\big)_4 \Big] \tau_2 \bigg) \hringmeasure \Bigg]
	\nonumber \\
	&&
	+ {\wascalO }\big(r^{-1}\big)
	\,,
\end{eqnarray}
thus
\begin{eqnarray}
\label{8I15.1}
 \lefteqn{
  V_\renm =
\lim_{r\to\infty} \Bigg[   V(r)  - \frac{r^3}{3}  \hringvolume
	+ \frac {r^2} 2  \int_{\sectionofscri} \tau_2 \hringmeasure
 }
 &&
	\nonumber \\
	&&
	+ \frac r2    \int_{\sectionofscri} \left( \Big[ |\sigma|^2_4 + 8 \pi \big(\ol T_{rr}\big)_4 \Big] - \frac{1}{2} \tau_2^2 \right) \hringmeasure
	\nonumber \\
	&&
	+ \frac{1}{6} \log r \int_{\sectionofscri} \bigg( \Big[ |\sigma|^2_5 + 8 \pi \big(\ol T_{rr}\big)_5 \Big] - 2 \Big[ |\sigma|^2_4 + 8 \pi \big(\ol T_{rr}\big)_4 \Big] \tau_2 \bigg) \hringmeasure \Bigg]
 \,.
 \phantom{xxx}
\end{eqnarray}
Now, by (\ref{eq:limitzetaintegral}) and using (\ref{eq:zetaintegral}) and (\ref{eq:2lambdavolume}),
\beal{eq:limitzetaintegralrewritten}
	\lefteqn{
	\lim_{r \rightarrow \infty} \left( \int_{\sectionofscri} \zeta d\mu_{\coneg} + 4\pi \chi(\sectionofScri) r - 2\Lambda V(r) \right)
}
&&
	\nonumber \\
	&
	=
 &
	\lim_{r \rightarrow \infty} \bigg(- r \int_{\sectionofscri} \zR \hringmeasure + 4\pi \chi(\sectionofScri) r \bigg)
	- \frac{\Lambda}{12} \int_{\sectionofscri}{ \tau_2^3 } \hringmeasure
	\nonumber \\
	&&
	+ \frac{\Lambda}{18} \int_{\sectionofscri}{ \bigg( 19 \tau_2 \Big[ |\sigma|^2_4 + 8 \pi \big(\ol T_{rr}\big)_4 \Big] - 2 \Big[ |\sigma|^2_5 + 8 \pi \big(\ol T_{rr}\big)_5 \Big] \bigg) } \hringmeasure
	\nonumber \\
	&&
	+ \int_{\sectionofscri} \left( \zeta_2 + \zR \tau_2 \right) \hringmeasure
	- 2\Lambda V_\renm
	\nonumber \\
	&=&
    \limatrzero   \int_{\sectionofscri} \zeta d\mu_{\coneg}
  +
  4 \pi \chi(\sectionofScri) r_0
	+
  \intfromrzero^{\infty}\int_{\sectionofscri} \left( \frac{1}{2}|\xi|^2+S \right) d\mu_{\coneg} dr
	\,.
	\phantom{xxxxx}
\eea
Next we rewrite (\ref{eq:solmassbondicharecteristic}) as
\beal{eq:massexpressionlimitrewritten}
	16\pi \mTB \hspace{-2mm}&=&
	\int_{\sectionofscri}{\left( \zeta_2 + \zR \tau_2 \right)} \hringmeasure
	- \int_{\sectionofscri} \frac{\zR}{2} \tau_2 \hringmeasure
	\nonumber \\
	&&
	+ \frac{\Lambda}{3} \int_{\sectionofscri} \bigg( 3 \tau_2 \Big[ |\sigma|^2_4 + 8 \pi \big(\ol T_{rr}\big)_4 \Big] - \Big[ |\sigma|^2_5 + 8\pi \left( \ol{T}_{rr}  \right)_5 \Big] \bigg)
    \hringmeasure
    \,,
\phantom{xxxx}
\eea
and find, by (\ref{eq:limitzetaintegralrewritten}) and (\ref{eq:massexpressionlimitrewritten}), using $\int_{\sectionofscri} \zR \hringmeasure = 4\pi \chi(\sectionofScri)$,
\bea
	16\pi \mTB
	\label{eq:massalmostfinal}
	 &=&
	\limatrzero   \int_{\sectionofscri} \zeta d\mu_{\coneg}
	+	4 \pi \chi(\sectionofScri) r_0
  + \intfromrzero^{\infty}\int_{\sectionofscri} \left( \frac{1}{2}|\xi|^2+S \right) d\mu_{\coneg} dr
	\nonumber \\
	&&
	- \int_{\sectionofscri} \frac{\zR}{2} \tau_2 \hringmeasure
	+	2\Lambda V_\renm
	+ \frac{\Lambda}{12} \int_{\sectionofscri}{ \tau_2^3 } \hringmeasure
	\nonumber \\
	&&
	- \frac{\Lambda}{18} \int_{\sectionofscri} \bigg( \tau_2 \Big[ |\sigma|^2_4 + 8 \pi \big(\ol T_{rr}\big)_4 \Big]
	+ 4 \Big[ |\sigma|^2_5 + 8 \pi \big(\ol T_{rr}\big)_5 \Big]
	\bigg) \hringmeasure
	\,.
	\phantom{xxxxi}
\eea

We continue with a generalisation of the arguments leading to equation (43) in \cite{ChPaetzBondi}. Indeed, we
allow the case $r_0 \neq 0$. Next, for further reference, we allow an asymptotic behaviour for small $r$ for light-cones emanating from a submanifold of general space
\newcommand{\codims}{d}%
co-dimension $\codims$,
 and not only a light-cone. Finally, for future reference the following calculations, up to the resulting expansion of $\tau$, \eqref{eq:tauexpressionnew}, are performed for arbitrary space-time dimensions $n+1 \geq 3$.

Keeping in mind the expansion \eqref{24I15.21} for large $r$, we note that
\begin{equation}
 {\tau   \label{eq:tauexpansionsmalllargex} }
 = \left\{
     \begin{array}{ll}
	\frac{n-1}{r} + \frac{\tau_2}{r^2} + {\wascalO }\big(r^{-3}\big)
	\,, & \hbox{for large $r$},\label{eq:tauexpansionlargex}\\
       \frac{\codims-1}{r} + {{\wascalO }}(1)
	\,, &  \hbox{ for small $r$ .}\label{eq:tauexpansionsmallx}
     \end{array}
   \right.
\end{equation}
Here the behaviour for small $r$ is the one which occurs
when the set $\{r=0\}$  has space co-dimension $\codims$ (e.g., $\codims=n$ for a light-cone emanating from a point). If $r_0>0$ we assume that $\tau$  is smooth up-to-boundary when the boundary $r=r_0$ is approached.

Next, let
 \ptctodo{ the next correction to tau encodes $r_0$ if solving the asymptotic Raychaudhuri equation, maybe needs including in another life}
\bel{eq:tau1def}
	\tau_1 :=	\frac{n-1}{r}
\,.
\ee
This is the value of $\tau$ for a light-cone in Minkowski space-time, and it follows from \eqref{24I15.21}
that this is the value approached asymptotically along null hypersurfaces meeting $\scri$ smoothly and transversally.
Let
$$
 \delta \tau := \tau - \tau_1
$$
denote the deviation of $\tau$ from its asymptotic value for large $r$, then
\be  \label{eq:deltatauexpansionsmalllarge}
 {\delta \tau  }
 = \left\{
     \begin{array}{ll}
	\frac{\tau_2}{r^2} + {\wascalO }\big(r^{-3}\big)
	\,, & \hbox{for large $r$\,;}
	\label{eq:deltatauexpansionlarge}
	\\
	\frac{\codims-n}{r} + {\wascalO }(1)
	\,, & \hbox{for small $r$}
	\label{eq:deltatauexpansionsmall} \,.
     \end{array}
   \right.
\ee
(Note that $\delta \tau$ is diverging at the same rate as $\tau$ for small $r$  when $\codims \ne n$.)
From the Raychaudhuri equation  \eqref{eq:wavegaugeconstraint1}  with $\kappa=0$ one finds that $\delta \tau$ satisfies the equation
\bel{eq:deltatauequation}
	\frac{d \delta \tau}{dr} + \left( \frac{\delta \tau}{n-1} + \frac{2}{r} \right) \delta \tau
    = - |\sigma|^2 - 8 \pi \ol{T}_{rr}
	\,.
\ee
Define
$$
 \Psi \equiv r^{-2} \Phi := r^{-2}  \exp\bigg(  \int_{r_*}^r \left( \frac{\delta \tau}{n-1} + \frac{2}{\tilde r} \right) \,d\tilde r \bigg)
  \,,
$$
for some $r_*$ (possibly depending upon $x^A$)   which will be irrelevant for our final formula \eq{eq:tauexpressionnew} below except for the requirement that the integral  converges. Thus
%
%
\bel{eq:Phideltatauequation}
	\frac{1}{\Phi} \frac{d \Phi}{dr} = \frac{\delta \tau}{n-1} + \frac{2}{r}
	\,,
\ee
so that
\eq{eq:deltatauequation} is equivalent to
\bel{eq:Phiequation}
	\frac{1}{\Phi} \frac{d \left( \Phi \, \delta \tau \right)}{dr} = - |\sigma|^2 - 8 \pi \ol{T}_{rr}
	\,,
\ee
Using \eqref{eq:deltatauexpansionsmalllarge}, we are led to the following three equivalent expressions for the function $\Psi$:
\begin{subnumcases}{\Psi(r,x^A) = \label{11II15.11}}
	\exp \left( -\int_r^\infty \frac{\delta \tau}{n-1}  (s,x^ A) ds +C_1(x^A) \right)
	\,;
	\label{11II15.11a} \\
	\exp \left(  \int_{r_0}^r \frac{\delta \tau}{n-1}  (s,x^ A)   ds +C_2(x^A) \right)
	\,;
	\label{11II15.11b} \\
	r^{\frac{\codims-n}{n-1}} \exp \left( \frac{1}{n-1} \int_{0}^r \left[ \delta \tau  (s,x^ A)  - \frac{\codims-n}{s} \right] ds +C_3(x^A) \right)
  \,,
	\label{11II15.11c}
	\phantom{xxxxxx}
\end{subnumcases}
for some functions $C_i(x^A)$, depending upon the choice of $r_*$.
In \eq{11II15.11b} we have assumed that $r_0>0$, while
\eq{11II15.11c} holds when $\delta\tau(r,x^A) \sim   (\codims-n) r^{-1}$ for small $r$, compare \eq{eq:deltatauexpansionsmall}.

We emphasise that both $\Phi$ and $\Psi$ are auxiliary functions which are only needed to derive \eq{eq:tauexpressionnew} below, and there is some freedom in their definition. In particular either of the functions $C_i(x^A)$, $i=1,2,3$, can be chosen to be zero if convenient for a specific problem at hand, and we note that the  $C_i(x^A)$'s cancel out in the final expression for $\tau$ in any case.

We further stress that in the special case of a light-cone  we have $\codims=n$ and treating the case for small $r$ separately is not necessary. In this case  \eqref{11II15.11b} coincides with \eqref{11II15.11c}.

Thus, using $\delta \tau = \tau - \tau_1$ and \eqref{eq:tau1def},
\begin{subnumcases}{\frac{ \Psi(r_0,x^A)}{ \Psi(r,x^A)} = \label{11II15.12}}
	\exp \left( - \frac{1}{n-1}\int_{r_0}^r \left ( \tau  (s,x^ A) - \frac {n-1}s\right) ds  \right)
	\,;
	\label{11II15.12a}
	\\
	\left(\frac{r_0}r\right)^{\frac{\codims-n}{n-1}} \exp \left( \frac{1}{n-1} \int_r ^{r_0} \left ( \delta \tau  (s,x^ A) - \frac {\codims-n}s\right)ds  \right)
  \,,
	\label{11II15.12b}
  \phantom{xxxxx}
\end{subnumcases}
with both  \eqref{11II15.11a} and \eqref{11II15.11b} leading to \eqref{11II15.12a} as long as the right-hand-side of  \eqref{11II15.12a} converges, and
with \eq{11II15.12b} holding with $r_0=0$ when $\delta\tau(r,x^A) \sim (\codims-n) r^{-1}$ for small $r$.

Integrating \eq{eq:Phiequation} and using \eq{11II15.12a},
 without denoting the dependence on coordinates $x^A$ explicitly in what follows,
\bean
\label{eq:tauexpressionnew}
	\tau &=&
	 \frac{n-1}{r}
	-   r^{-2}\bigg[ \Psi(r)^{-1} \tintfromrzero^r \Big( |\sigma(\tilde{r})|^2 + 8\pi \ol T_{rr}(\tilde{r}) \Big) \Psi(\tilde{r}) \tilde{r}^2 d\tilde{r}
	\\
	&&
	 - \lim_{s\to r_0}\frac{\Psi(s)}{ \Psi(r)} 	  \bigg( \tau(s) - \frac{n-1}{s}  \bigg) s^2 \bigg]
	\,.
\eea
We can directly read off the expression for $\tau_2$
from this:
%
\bean
	\tau_2  &=&
	- \lim_{r \rightarrow \infty} \bigg\{ \Psi(r)^{-1} \tintfromrzero^r \Big(
    |\sigma(\tilde{r})|^2 + 8\pi \ol T_{rr}(\tilde{r}) \Big) \Psi(\tilde{r}) \tilde{r}^2 d\tilde{r} \bigg\}
\\
	&&
	- \bigg[\lim_{r\to\infty}\Psi(r )^{-1}	\bigg]\times \lim_{r \to r_0} \bigg[ \Psi(r) \bigg( \frac{n-1}{r} - \tau \bigg) r^2 \bigg]
	\label{eq:tau2expressionnew2}
	\,.
\eea
%

From now on we return to space-time dimension four:
$$
 n+1 =4
 \,.
$$
Returning  to (\ref{eq:massalmostfinal}), inserting the result for $\tau_2$ we just found,
and using further
%
\bel{eq:dmugdmusexponential}
	d\mu_{\coneg} = \text{e}^{ -\int_r^{\infty} \frac{\tilde{r}\tau-2}{\tilde{r}} d\tilde{r} } r^2 \hringmeasure
\ee
we obtain our final formula for the \charmass\  $\mTB$ of a null hypersurface $\mcN=[r_0,\infty)\times \sectionofscri $:
\bean
	\mTB \hspace{-2mm}&=&
	\frac{1}{16\pi}
	\intfromrzero^{\infty} \int_{\sectionofscri} \Bigg( \frac{1}{2} |\xi|^2 + S
	\nonumber \\
	&&
	\;\;\;\;\;\;
	+ \left[ \frac{\zR}{2}
	+
	\frac{\Lambda}{18} \Big( |\sigma|^2_4 + 8 \pi \big(\ol T_{rr}\big)_4 \Big)
	\right] \Big( |\sigma|^2 + 8\pi \ol T_{rr} \Big) \text{e}^{ \int_r^{\infty} \frac{\tilde{r}\tau-2}{2\tilde{r}} d\tilde{r} } \Bigg) d\mu_{\coneg} dr
	\nonumber \\
	&&
	+ \frac{1}{16\pi} \Bigg[
		4 \pi \chi(\sectionofScri) r_0
	\nonumber \\
	&&
		+ \limatrzero \bigg( \int_{\sectionofscri} \bigg[ \zeta
		+  \left( \frac{\zR}{2}
	+
	\frac{\Lambda}{18} \Big( |\sigma|^2_4 + 8 \pi \big(\ol T_{rr}\big)_4 \Big)
	\right) \left( \frac{2}{r} - \tau \right) \text{e}^{\int^{\infty}_{r_0} \frac{r\tau-2}{2r} dr} \bigg]  d\mu_{\coneg} \bigg)
		\Bigg]
	\nonumber \\
	&&
	+ \frac{\Lambda}{192 \pi} \int_{\sectionofscri}{ \tau_2^3 } \hringmeasure
	+	\frac{\Lambda V_\renm}{8\pi}
	-
	\frac{\Lambda}{72\pi} \int_{\sectionofscri} \left( |\sigma|^2_5 + 8\pi \left(\ol T_{rr}\right)_5 \right) \hringmeasure
 \,.
\eeal{eq:massfinal}
To obtain this equation, it is irrelevant which form of $\Psi$ in \eq{11II15.11} we take, provided that the same formula is consistently used throughout. For example, if $\Psi(r)$ is given by \eqref{11II15.11a} with $C_1(x^A)=0$, then $\lim_{r\to \infty} \Psi(r)$  equals one, independently of whether $r_0=0$ (so that the null hypersurface  is singular at $r_0$) or $r_0 \neq 0$ (in which case the set $\{r=r_0\}$ has space co-dimension one).

In the special case of a light-cone, where $\codims=n=3$, $\zR=2$ and
\ptctodo{make contact with the core geodesics section in the footnote }
$r_0=0$,%
\footnote{
Recall that $\zR=2$ when $\sectionofscri$ is a two-sphere, $\zR=0$ for a torus, and $\zR<0$ for higher genus topologies of $\scri\approx \R\times\sectionofscri$. In the case of a smooth-light cone the cross-sections are spherical for small $r$, and therefore everywhere, so $\zR=2$.}
\eqref{eq:massfinal} simplifies to
%
\beal{eq:massfinalCone}
	\mTB \hspace{-2mm}&=&
	\frac{1}{16\pi}
	\int_0^{\infty} \int_{\sectionofscri} \bigg( \frac{1}{2} |\xi|^2 + S
	\nonumber \\
	&&
	\;\;\;\;\;\;
	+ \left[ 1
	+
	\frac{\Lambda}{18} \Big( |\sigma|^2_4 + 8 \pi \big(\ol T_{rr}\big)_4 \Big)
	\right] \Big( |\sigma|^2 + 8\pi \ol T_{rr} \Big) \text{e}^{ \int_r^{\infty} \frac{\tilde{r}\tau-2}{2\tilde{r}} d\tilde{r} } \bigg) d\mu_{\coneg} dr
	\nonumber   \\
	&&
	+ \frac{\Lambda}{192 \pi} \int_{\sectionofscri}{ \tau_2^3 } \hringmeasure
	+	\frac{\Lambda V_\renm}{8\pi}
	-
	\frac{\Lambda}{72\pi} \int_{\sectionofscri} \left( |\sigma|^2_5 + 8\pi \left(\ol T_{rr}\right)_5 \right) \hringmeasure
	\,.
	\phantom{xxxxi}
\eea
Assuming further a conformally smooth compactification and vacuum we have $|\sigma|^2_5=0$, and after some rearrangements we obtain the striking identity:
\beal{eq:massfinalConeVac}
	\mTB
	&=&
	\frac{1}{16\pi}	\int_0^{\infty} \int_{\sectionofscri} \bigg( \frac{1}{2} |\xi|^2
	+
	 |\sigma|^2  \text{e}^{ \int_r^{\infty} \frac{\tilde{r}\tau-2}{2\tilde{r}} d\tilde{r} } \bigg) d\mu_{\coneg} dr
	\nonumber   \\
	&&
	+ \frac \Lambda{8\pi}\left(  	{ V_\renm}+ \frac{1}{12}\int_{\sectionofscri}{
    \tau_2\bigg(\frac{ \tau_2^2}{2 }  - \frac{|\sigma|_4^2}{3}\bigg) } \hringmeasure
 \right)
	\,,
\eea
with $\tau_2\le 0 $ given by
\bea
	\tau_2  &=&
	-
    \int_0^\infty
    |\sigma|^2   \text{e}^{- \int_r^{\infty} \frac{\tilde{r}\tau-2}{2\tilde{r}} d\tilde{r} }   {r}^2 d{r}
	\label{9II16.1}
	\,.
\eea
(Recall that $\tau_2=0$ if and only if the metric to the future of $\mcN$ is, at least locally, the de Sitter or anti-de Sitter metric~\cite{CCG}.)

\section{Coordinate mass}
 \label{s13I16.3}

In this section we assume that $\Lambda <0$ and we allow arbitrary space-time dimension $n+1\ge 4$.

There exist several well-defined notions of mass for asymptotically hyperbolic initial data sets (cf., e.g., \cite{Wang,ChruscielSimon,AbbottDeser,ChNagyATMP,ChHerzlich}), which typically coincide whenever simultaneously defined, some of them defined so forth only in dimension $3+1$. Our aim, in this and in the next section, is to show that the characteristic mass coincides with those alternative definitions in some cases of interest. To set the stage, in this section we introduce the notion of ``coordinate mass'' for two classes of metrics. (Compare~\cite[Section~V]{ChruscielSimon} for a similar treatment in dimension $3+1$.)
 \ptclevoca{to be reworded if the case arises}

\subsection{Birmingham metrics}
 \label{s19VIII15.1}

 \ptclevoca{this is completely duplicated with the Birmingham section, included here only for the mass on the light-cones purposes, needs discarding in the black hole notes; $\varepsilon$ should be added}


Consider an $(n+1)$-dimensional metric, $n\ge 3$,  of the form
\bel{6XI12.4}
 g = - f(r) dt^2 + \frac {dr^2} {f(r)} + r^2  \underbrace{\mathring h_{AB}(x^C) dx^A dx^B}_{=:\mathring h}
 \,,
\ee
where $\mathring h$ is a \emph{Riemannian Einstein metric} on the compact manifold which, to avoid a proliferation of notation, we will denote as $\sectionofScri $; we denote by $x^A$ the local coordinates on $\sectionofScri $.  As discussed in \cite{Birmingham}, for any $m\in \R$ and $\ell>0$ the function
%
\bel{6XI12.5}
 f= \frac{\zR}{(n-1)(n-2)} - \frac {2{ m}}{r^{n-2}}   - \varepsilon \frac{r^2}{\ell^2}
 \,,
 \quad
 \varepsilon \in \{0,
 \pm 1\}
 \,,
\ee
where $\mathring R$ is the (constant) scalar curvature of $\mathring h$,
leads to a vacuum metric,
\bel{4I13.1}
 R_{\mu\nu} = \varepsilon \frac{n}{\ell^2} g_{\mu\nu}
 \,,
\ee
where the positive constant $\ell $ is  related to the cosmological constant as
\begin{equation}\label{17XII12.21}
 \frac 1 {\ell^2} = \varepsilon\frac {2\Lambda}{n(n-1)}
 \,.
\end{equation}

Clearly, $n$ is not allowed to equal two in \eq{6XI12.5}, and we therefore exclude this dimension in what follows.

The multiplicative factor two in front of $m$ is convenient in dimension three when $\mathring h$ is a unit round metric
on $S^2$, and we will keep this form regardless of topology and dimension of $\sectionofScri $.

There is a rescaling of the coordinate $r=b  \bar r$, with $b\in \R^*$, which leaves
\eq{6XI12.4}-\eq{6XI12.5} unchanged if moreover
\bel{4I13.2}
 \overline{\mathring h} = b^2\mathring h
 \,,
 \quad
 \bar m =b^{-n} m
  \,,
   \quad
    \bar t=  b t
    \,.
\ee
We can use this to achieve
\bel{4I13.4}
 \beta:= \frac{\zR}{(n-1)(n-2)} \in \{0, \pm 1\}
 \,,
\ee
which will be assumed from now on. The set $\{r=0\}$ corresponds to a singularity when $m\ne 0$. Except in the case $m=0$ and $\beta = -1$,
by an appropriate choice of the sign of $b$ we can always achieve $r>0$ in the regions of interest. This will also be assumed from now on.

\bigskip

%
We  {define}
$$
 \mbox{\emph{the coordinate mass of the metric  \eq{6XI12.4} with $f$ given by \eq{6XI12.5} to be $m$}.}
$$
Similarly, we  {define}
$$
  \mbox{
   \emph{the coordinate mass of any metric  which asymptotes to \eq{6XI12.4}-\eq{6XI12.5} to be $m$}.
    }
$$
Here, \emph{``asymptotes to''} can e.g.\ be understood as
\bean
 g
  & = &
    - \big(f_m(R) +\redOofo R^{2-n })\big)dT^2 + \frac {dR^2} {\big(f_m(R) +\redOofo R^{2-n})\big)}
\\
 &&
     + R^2\big(   \mathring h_{AB}(x^C) +\redOofo 1)\big)dx^A dx^B
 \,,
\eeal{6XI12.4xyz}
for large $R$, at fixed $T$, with $f_m=f$ given  by \eq{6XI12.5}.

\subsection{Horowitz-Myers-type metrics}
 \label{s9VII14.1}

\subsubsection{The metric}

 \ptctodo{are these metrics globally regular when $\beta \ne 0$?}
Consider an $(n+1)$-dimensional metric, $n\ge 3$,  of the form
\bel{6XI12.4x}
 g =  f(r) d\psi^2 + \frac {dr^2} {f(r)} + r^2  \underbrace{\mathring h_{AB}(x^C) dx^A dx^B}_{=:\mathring h}
\ee
where now $\mathring h$ is a Riemannian \emph{or} pseudo-Riemannian Einstein metric on an $(n-1)$-dimensional manifold ${\mathring N} $ with constant scalar curvature $\mathring R$ and, similarly to the last section, the $x^A$'s are local coordinates on ${\mathring N} $.%
\footnote{To avoid a proliferation of notation we use the symbol $\mathring h$ both for the metric on $\sectionofScri $ appearing in \eq{6XI12.4} and for the metric on the manifold $\mathring N$ relevant for \eq{6XI12.4x}. Typically $(\sectionofScri ,\mathring h)$  is a compact Riemannian manifold, while $( \mathring N,\mathring h)$ in \eq{6XI12.4x} will be Lorentzian with $\mathring N$ non-compact.}
This metric can be formally obtained from \eq{6XI12.4} by changing $t$ to $i \psi$. It therefore follows from the discussion of Section~\ref{s19VIII15.1}
that for $m\in \R$ and $\ell\in \R^*$ the function
\bel{6XI12.5x}
 f= \beta - \frac {2{ m}}{r^{n-2}} - \varepsilon \frac{r^2}{\ell^2}
 \,,
 \quad
 \varepsilon \in \{0,
 \pm 1\}
 \,,
 \quad
 \beta=\frac{\zR}{(n-1)(n-2)}
 \,,
\ee
leads to a metric satisfying \eq{4I13.1}.
 \ptclevoca{material repetitive with equation \eq{24III13.22x}, so watch out if this is just next to it, and remove when the case arises}
Rescaling the coordinate $r$ and the metric $
\mathring h$ by a suitable constant if necessary we can without loss of generality assume that
$$
 \beta \in \{0,\pm 1\}
 \,.
$$

Suppose that $f$ has zeros, and let us denote by $r_0$ the largest zero of $f$. We assume that $r_0$ is of first order, and we restrict attention to $r\ge r_0$. Imposing a suitable $\psi_0$-periodicity condition on $\psi \in [0,
\psi_0]$, the usual arguments imply that the set $\{r=r_0\}$ is a rotation axis in a plane on which $\sqrt{r-r_0}$ and $\psi$ are coordinates of polar type: Indeed, if we set
$$
 \rho = F(r)\,,
 \ \mbox{with} \ F = \int_{r_0}^r \frac 1 {\sqrt{f(r)}} dr = \frac { \sqrt{r-r_0}} {2 \sqrt{f'(r_0)}}
  \big(1+\redOof r-r_0)\big)
 \,,
$$
we find
$$
 \frac{dr^2}{f} + f d\psi^2 = d\rho^2 +  f(F^{-1}(\rho))  d\psi^2
  =  d\rho^2 +  (2f'(r_0))^2 \big(1+\redOof \rho^2)\big) \rho^2 d\psi^2
 \,,
$$
which defines a smooth metric near $\rho=0$ if and only if
\bel{22XII14.11}
 \psi=\lambda \ell \chicoordinate
  \,,
\ee
where $\chicoordinate$ is a new $2\pi$-periodic coordinate, and
\bel{29XI14.1}
 \lambda = \frac 1 { 2\ell f'(r_0)}
 \,.
 \ee
%
 \ptclevoca{check for repetitivity, CROSSREF when relevant}

In the case where
 $$
  \varepsilon = -1
  \,,
 $$
 one obtains Einstein metrics with a negative cosmological constant.

Whatever $\varepsilon$, a conformal completion at spacelike infinity can be obtained by introducing a new coordinate $x=\ell/r$, bringing $g$ to the form
\bean
 g &=&  f(\ell x^{-1})  \ell^2 \lambda^2 d\chicoordinate^2 + \frac {\ell^2 dx^2} {x^4f(\ell x^{-1})} + \ell^2 x^{-2}  \mathring h
\\
 & = &
  x^{-2}\ell^2 \big( - ( \varepsilon-\beta x^2 + \redOof x^n))  \lambda^2  d\chicoordinate^2 -  (\varepsilon+\beta x^2 + \redOof x^n))dx^2 + \mathring h\big)
 \,.
 \qquad
\eeal{9VII14.1}
We see explicitly that the conformal class of metrics
induced by $x^2 g$ on   the boundary at infinity,
$$
 \scri =\{x=0\} \approx S^1 \times {\mathring N}
  \,,
$$
is Lorentzian if $\mathring h$ is Lorentzian and if $\varepsilon=-1$.
\subsubsection{ $\beta=0$, $n=3$}
 \label{ss9VII14.1}

In~\cite{HorowitzMyers} Horowitz and Myers consider the case $n+1=4$, $\varepsilon=-1$,%
\footnote{\label{f8II16.1}The case $\beta=0$ and $\varepsilon =1$ leads to a signature $(+---)$ for large $r$; our signature $(-+++)$ is recovered by multiplying the metric by minus one, but then one is back in the case $\varepsilon =-1$ after renaming $m$ to $-m$.}
and choose  $\mathring h = -\ell^{-2} dt^2 + d\varphi^2$, with $\varphi$ being a $2\pi$-periodic coordinate on $S^1$. Thus
\bel{6XI12.4xy}
 g = -\frac{r^2}{\ell^2} dt^2 +  f(r) \ell^2 \lambda^2   d\chicoordinate^2 + \frac {dr^2} {f(r)} + r^2 d\varphi^2
 \,.
\ee
\Eq{9VII14.1} shows that timelike infinity $\scri\approx \R\times S^1\times S^1$
is conformally flat:
 \ptclevoca{mention or synchronize with \eq{Kot4HM}}
\be
 x^2 g \to_{r \to
 \infty}  - dt^2+    \ell^2( \lambda^2   d\chicoordinate^2 +   dx^2 + d\varphi^2)
 \,.
\eel{9VII14.2}

Some comments about factors of $\ell$ are in order: if we think of $r$ as having dimension of length, then $\ell$,  $t$ and $\psi$ also have dimension of length, $m$ has dimension $\text{length}^{n-1}$, while $f$, $x$, and the $x^A$'s (and thus $\varphi$) are dimensionless.

A uniqueness theorem for the metrics \eq{6XI12.4xy} has been established in \cite{WoolgarRigidityHM}.

\subsubsection{$\beta=\pm 1$, $n=3$}

We consider the metric \eq{6XI12.4x} with$^{\ref{f8II16.1}}$ $\varepsilon=-1$ and $\mathring h$ of the form
\bel{14IV15.1}
 \mathring h = \left\{
                 \begin{array}{ll}
                   d\theta^2 + \sin^2 (\theta)\, d\varphi^2, & \hbox{$\beta=1$;} \\
                    d\theta^2 + \sinh^2 (\theta)\, d\varphi^2, & \hbox{$\beta=-1$.}
                 \end{array}
               \right.
\ee
In regions where $f$ is positive, one obtains a Lorentzian metric after a ``double Wick rotation''
$$
 \theta = i \ell^{-1} t
\,,
 \ \varphi = i \phi
 \,,
$$
resulting in
\bel{14IV15.2}
 g =   - \frac{r^2}{\ell^2} d t^2 + \frac {dr^2} {f(r)}
     +  f(r)  \ell^2 \lambda^2
        d\chicoordinate^2 + r^2   \left\{
                 \begin{array}{ll}
                    \sinh^2( \ell^{-1}t)\, d\phi^2, & \hbox{$\beta=1$;} \\
                   \sin^2 (\ell^{-1}t) \, d\phi^2, & \hbox{$\beta=-1$.}
                 \end{array}
               \right.
\ee
Taking $\chicoordinate $ and $\phi$ periodic one obtains again a conformal infinity diffeomorphic to $\R\times \T^2$. Note that the conformal metric at the conformal boundary is \emph{not} conformally stationary anymore, as opposed to the Horowitz-Myers metrics \eq{9VII14.2}. We have not attempted to study the nature of the singularities of $g$ at $\sinh (\ell^{-1}t)=0$ or at $\sin (\ell^{-1}t)=0$.
	\ptctodo{might be interesting, should be horizons? Killing horizons for $\partial_\varphi$? maybe already done in the literature? all curvature invariants should all be regular there, by analytic continuation arguments from those of the metric before the change of variables?}

\subsubsection{Negative coordinate mass}
 \label{ss18II15.1}

For completeness,
 \ptclevoca{reword}
we show that the metric \eq{6XI12.4x} has the striking property that its total coordinate mass is negative when $m$ is positive; the latter is needed for regularity of the metric. This has already been observed in~\cite{HorowitzMyers} in space-dimension three with a toroidal Scri. Here we check that this remains correct in higher dimensions, for a large class of topologies of Scri.

Before continuing, we note that Lorentzian Horowitz-Myers-type metrics with a smooth conformal compactification at infinity exist only with negative $\Lambda$: Indeed, to obtain the right signature for large $r$  when $
\epsilon >0$ one needs to multiply the metric by minus one. But then the resulting  metric has negative Ricci scalar, and hence solves Einstein equations with a negative cosmological constant.

Somewhat more generally, consider those metrics of the form \eq{6XI12.4x} for which
$$
  \mathring N = \R_t\times \Mcheckhere
    \,,
$$
where $(\Mcheckhere  ,\localch)$ is a compact Riemannian manifold, and where
\bel{29XI14.2}
 \mathring h =  -\ell^{-2}
 dt^2 + \localch
 \,,
\ee
so that
\bel{23II15.3}
 g =  f(r) d\psi^2 + \frac {dr^2} {f(r)} + r^2 \left( -\ell^{-2}dt^2 + \localch
 \right)
 \,.
\ee
The question arises, how to define the mass of such a metric.

To avoid ambiguities, let us write $f_m$ for the function $f$ of \eq{6XI12.5x}.

To assign a coordinate mass to a metric \eq{23II15.3}, we need to check whether  metrics satisfying \eq{6XI12.4x}-\eq{6XI12.5x} and \eq{29XI14.2} can be written in the form \eq{6XI12.4xyz} by setting  $r=r(\Rlocal)$: 
\bean
 g & = &   f_m(r) \ell^2 \lambda^2   d\chicoordinate^2 + \frac {dr^2} {f_m(r)} + \frac{r^2}{\ell^2} (- dt^2 + \ell^2\localch)
\\
 & = &
  -\frac{r^2}{\ell^2}dt^2 + \left(\frac {dr}{d\Rlocal } \right)^2 \frac{ d\Rlocal^2}{f_m(r)}
 \nonumber
\\
 && +  r^2\big ((1+\redOof \beta r^{-2}) + \redOof m r^{-n})) \lambda^2  d\chicoordinate^2  +    \localch\big)
 \,,
\eeal{6XI12.4a}
where the error terms have to be understood for large $r$.
We will have
\bean
 g
 & \approx &
  -  f_M(\Rlocal )dt^2 +  \frac{ d\Rlocal^2}{  f_M(\Rlocal)}+  \Rlocal^2\big (  \lambda^2   d\chicoordinate^2  +    \localch\big)
 \,,
\eeal{6XI12.4a+}
for some parameter $M$ possibly different from $m$, provided that
\bel{20XII14.2}
 r^2 = \ell^2 f_M(\Rlocal)\big(1+\redOofo \Rlocal^{-n})\big)
 \,,
 \quad
  \left(\frac {d\Rlocal}{dr } \right)^2  {f_m(r)} = f_M(\Rlocal)\big(1+\redOofo \Rlocal^{-n})\big)
 \,,
\ee
The first equation  determines $r$ as a function of $\Rlocal$ up to  correction terms $\redOofo \Rlocal^{-n})$. Inserting the result into the second equation determines $M$, provided that the asymptotic expansion of the left-hand side is compatible with that of the right-hand side. However, it is straightforward to check that these equations are compatible if and only if
\bel{20XII14.1}
 \beta =0
 \,.
\ee
We conclude that for metrics  satisfying \eq{6XI12.4x}-\eq{6XI12.5x} and \eq{29XI14.2}
$$
 \mbox{the coordinate mass is only defined if $\beta=0$.}
$$
Assuming \eq{20XII14.1}, after asymptotically solving the first equation in \eq{20XII14.2} and inserting the result into the second one, we find that
\bel{19XII14.11}
\Rlocal = r + \frac {\ell^2 M}{r^{n-1}} + O (r^{-(2n-1)})
\,,
\ee
and that the coordinate mass equals
\bel{19XII14.12}
M = - \frac{m}{n-1}
\,.
\ee
In particular $M$ is negative for positive $m$.

\section{Examples}
 \label{s13I16.2}

Throughout this section we allow arbitrary space-time dimension $n+1\ge 4$.
We show that the numerical value of the Trautman-Bondi mass, as generalised to higher dimensions below, and which coincides with the characteristic mass defined in Section~\ref{s28XII15.1} in dimension $3+1$, is proportional to the ``coordinate mass'' for the metrics considered in Section~\ref{s13I16.3}. This, in itself, is not surprising, since these metrics have only the mass parameter $m$ as free parameter, so whatever we will calculate must be a function of $m$. The main conclusion here appears to be that $\mTB$ is a linear function of $m$, with a strictly positive proportionality factor.   A full agreement will be obtained in the analysis of the Hamiltonian mass in Section~\ref{s18II15.1} below, where the proportionality factors will also be matched.

In what follows, we seek to write the metrics under consideration in the form \eq{19II16.1},
\bel{21XII14.3}
 g = g_{uu} du^2 -2 e^{2\betaBondi} dr \,du - 2 r^2 U_A dx^A du + r^2  \underbrace{h_{AB}  dx^A dx^B}_{=:h}
 \,,
\ee
where the determinant of $h_{AB}$ is $r$-independent.
By analogy with \eqref{eq:mTB}-\eqref{eq:massaspectdef}, in space-time dimension $n+1$ we set
\bel{21XII14.5}
	 \mTB = \frac{1}{8\pi}\lim_{r\to\infty}\int_{\sectionofScri }(g_{uu})_{n-2}\,d\mu_h
 \,.
\ee
This definition is motivated by the fact  that, when solving the characteristic constraint equations on a null hypersurface, the $(g_{uu})_{n-2}$-coefficient in the expansion of $ g_{uu}$ arises as a global integration function.

\subsection{Birmingham metrics}
 \label{ss17II16.1}

Consider, first, the original Birmingham metrics \eq{6XI12.4},
\bel{21XII14.1}
 g = - f(r) dt^2 + \frac {dr^2} {f(r)} + r^2   \mathring h
 \,,
\ee
with $f$ given by \eq{6XI12.5}.
Introducing a new coordinate $u=t-\int_{r*}^r f^{-1}(s)ds$, for some conveniently chosen $r_*$, brings $g$ to the desired form
\bean
 g  &= & - f(r) \left(du+ \frac {dr } {f(r)}\right) ^2 + \frac {dr^2} {f(r)} + r^2   \mathring h
 \\
   &&
    = - \left(\frac{\zR}{(n-1)(n-2)} - \frac {2{ m}}{r^{n-2}} - \bvarepsilon \frac{r^2}{\ell^2}\right)  du ^2- 2 du \, dr + r^2   \mathring h
 \,,
  \phantom{xxxx}
\eeal{21XII14.4}
where $\bvarepsilon \in\{\pm 1\}$ is the sign of the cosmological constant $\Lambda = \bvarepsilon |\Lambda|$, which we allow to be either positive or negative.

\subsubsection{Mass and  volume}
The coordinate $r$ provides obviously a radial Bondi coordinate. Moreover, the equality
$\partial_r = - \nabla u$ implies that $r$ is also an affine parameter along the radial
null outgoing geodesics  of $g$. When $m=0$ we have the explicit formulae
$$
 u = t - \left\{
 \begin{array}{ll}
		\left.
		 \begin{array}{ll}
			 \ell \tan^{-1}\left(\frac{r}{\ell}\right)
				 \,, & \hbox{$\bvarepsilon=-1$,} \\
			 \ell \tanh^{-1}\left(\frac{r}{\ell}\right)
				 \,, & \hbox{$\bvarepsilon=1$,} \\
		 \end{array}
		\right\}
			 \, & \hbox{$\beta=1$;}
	\\
	- \bvarepsilon \frac{\ell^2}{r}
		\,, & \hbox{$\beta=0$;}
	\\
		\left.
		 \begin{array}{ll}
			 -\ell \tanh^{-1}\left(\frac{r}{\ell}\right)
				 \,, & \hbox{$\bvarepsilon=-1$,} \\
			 -\ell \tan^{-1}\left(\frac{r}{\ell}\right)
				 \,, & \hbox{$\bvarepsilon=1$,} \\
		 \end{array}
		\right\}
			 \, & \hbox{$\beta=-1$.}
 \end{array}
      \right.
$$

\Eq{21XII14.5} leads to a Bondi-Trautman--type mass
\bel{21XII14.5b}
	 \mTB = \frac{m}{4\pi}\int_{\sectionofScri } \,d\mu_{\mathring h}
	 \equiv
	   m \frac{  \mu_{\mathring h}(\sectionofScri)}{4\pi}
     \,.
\ee
(Here the normalisation factor $8\pi$ in \eq{21XII14.5} is clearly convenient only when $\sectionofScri$ is a unit round two-dimensional sphere,  but   this issue will be of no concern to us here.)
We conclude that the characteristic mass of null hypersurfaces asymptotic to the level sets of $t$ is indeed proportional to the coordinate mass, with a positive proportionality factor.
We will see in Section~\ref{ss18II15.3} that the proportionality factor is the same as the one occurring in the Hamiltonian definition of mass, see \eq{19II16.3} below.
\subsubsection{The balance equation for Birmingham metrics}
 \label{Bss9VII14.2}

 \ptctodo{how does our volume compare with the Brendle-Chodoch one?  }

Consider  metrics of the form
\bel{B21XII14.1}
 g = - f(r) dt^2 + \frac {dr^2} {f(r)} + r^2  \underbrace{\mathring h_{AB}dx^A dx^B }_{=: \mathring h}
 \,.
\ee
 Recall \eq{21XII14.4}: setting $u=t-\int_{r*}^r f^{-1}(s)ds$, for some conveniently chosen $r_*$, brings   $g$ to a Bondi form provided that $\det \mathring h$ is $r$-independent:
$$
 g  = - f  du ^2- 2 du \, dr + r^2   \mathring h
 \,.
$$
The inverse metric reads
$$
 g^\sharp  \equiv  g^{\mu\nu}\partial_\mu\partial_\nu
 =
    f\partial_r^2 -  2 \partial_u \partial_r    + r^{-2}
    \mathring h^\sharp
 \,,
$$
where $\mathring h^\sharp=\mathring h^{AB}\partial_A\partial_B$ is the metric inverse to $\mathring h$.

Similarly to the previous section, the integral curves of the vector field
\bel{B10VII14.1}
 -\nabla u =     \partial_r
\ee
are affinely parameterized geodesics. Whenever $\det \mathring h$ is $r$-independent, the function  $r$ is therefore both an \emph{area coordinate} and
an affine parameter along the generators of the null hypersurfaces $\{u=\const\}$.

Recall  that $V_\renm$ is defined as the limit, as $r$ approaches infinity, of the volume $V(r)$ of the light-cone minus all diverging terms in an asymptotic expansion of $V(r)$ :
\bean
	V_\renm &=&
\lim_{r\to\infty} \Bigg\{V(r)+ \frac 12  \Bigg[ - \frac{2}{3} r^3 \int_{\sectionofscri} d\mu_{\mathring h}
	+ r^2 \int_{\sectionofscri} \tau_2 d\mu_{\mathring h}
	+ r \int_{\sectionofscri} \left( |\sigma|^2_4 - \frac{1}{2} \tau_2^2 \right) d\mu_{\mathring h}
	\nonumber \\
	&&
	+ \frac{1}{3} \log r \int_{\sectionofscri} \left( |\sigma|^2_5 - 2 |\sigma|^2_4 \tau_2 \right) d\mu_{\mathring h} \Bigg]
 \Bigg\}
 < \infty
	\phantom{xxxx}
\eeal{B22XII14.2}
(for simplicity a metric vacuum to sufficiently high order has been assumed in \eqref{B22XII14.2}).

For the Birmingham metrics  \eq{6XI12.4xy} we have $T_{rr}\equiv 0 \equiv S\equiv\sigma\equiv  \xi \equiv \tau_2$, $\zR =2$, and the volume function is straightforward:
\beaa
 V(r) &= & \int_{r_*}^{r } \int_\sectionofScri \sqrt{\det g_{AB}}\, d^2x \, ds
\\
 &= &
\frac{ \mu_{\mathring h}(\sectionofScri)}3 (r^3-r_*^3)
\quad
\Longrightarrow
\quad
V_\renm =
- \frac{ \mu_{\mathring h}(\sectionofScri)r_*^3}3
 \,.
\eeaa

The mass formula \eq{eq:solMgeneral}  reduces to
\bean
	\mTB &=&
	\frac{1}{16\pi} \bigg(  {4 \pi \chi(\sectionofScri) r_*
	+} \lim_{r \to r_*} \int_{\sectionofscri} \zeta d\mu_{\coneg}
	\bigg)
	+	\frac{\Lambda V_\renm}{8\pi}
	\, .
\eeal{B14XII14.1}
Note that this holds for any  value of $r_*$. A natural choice would be to choose $r_*$ to be the location of the outermost past horizon, but we allow $r_*$ to be arbitrary.

Specialising to the Birmingham metrics we find, in space-time dimension $n+1=4$,
\bean
	16 \pi \mTB &=&
 {4 \pi \chi(\sectionofScri) r_*
	+} \lim_{r\to r_*}\mu_{\mathring h}(\sectionofScri) \left(-2r^2\left( \frac \beta r - \frac {2{ m}}{r^{2}} + \frac{r }{\ell^2}  \right) - \frac{2 \Lambda r ^3}3 \right)
\\
 &=&
 {4 \pi \chi(\sectionofScri) r_*
	+} \ \mu_{\mathring h}(\sectionofScri) \left(-2r_*  \beta  + {4 { m}}  \right)
 \nonumber
\\
 &=& 4 \mu_{\mathring h}(\sectionofScri)  m
	\, ,
\eeal{B14XII14.2}
where we have used the Gauss-Bonnet theorem to cancel the term containing the Euler characteristic $\chi(\sectionofScri)$    of $ \sectionofScri$ with the term involving $\beta$.

 \ptctodo{some remarks on renormalized volume in arXiv:1512.07109}

\subsection{Horowitz-Myers type metrics}
 \label{ss18II16.2}

We pass now to the metrics \eq{6XI12.4x} with $ \mathring h $ given by
\eq{29XI14.2} and   $\psi$ replaced by $\lambda \ell \chicoordinate$,
 with $\lambda$ given by \eq{29XI14.1}, and where $\chicoordinate$ is  $2\pi$-periodic:
  \ptctodo{one should look at an ADM type mass when $ \beta$ is non zero, trying to make the fall-off as similar as possible in all three components? though this should be settled by the Hamiltonian mass if it works with beta non-zero there?}
\bean
 g
  &  = &
   - \frac{r^2}{\ell^2} dt^2  + \frac {dr^2} {f} + f \ell^2 \lambda^2 d\chicoordinate^2
     + r^2 \localch
\\
& =  &   - \frac{r^2}{\ell^2} dt^2  + \frac {dr^2} {f} + \underbrace{r^{\frac{2(n-2)}{n-1}}(f \ell^2 )^{\frac{1}{n-1}}}_{=:r^2_\Bobo}
 \underbrace{ \left( \frac{f \ell^2 \lambda^2}{r^2_\Bobo}  d\chicoordinate^2  + \frac{ r^2}{r^2_\Bobo}
 \localch
  \right)}_{\mbox{\scriptsize \rm $r$-independent determinant}}
 \,.
\eeal{21XII14.8}
We want $f$ to be positive for large $r$, and hence we need to assume that $\Lambda<0$.

\subsubsection{Bondi coordinates, \charmass}

Setting $du = dt - \ell \,f^{-1/2} r^{-1} dr$ we obtain
\bean
 g  & =  &
   -  \frac{r^2}{\ell^2} du^2  - \frac{ 2 r }{f^{1/2} \ell} du\, dr + r^2_\Bobo
   \left( \frac{f\ell^2 \lambda^2}{r^2_\Bobo}  d\chicoordinate^2  + \frac{ r^2}{r^2_\Bobo}
 \localch
  \right)
   \nonumber
\\
 & =  &
   -  \frac{r^2}{\ell^2} du^2  - \frac{2 r }{f^{1/2} \ell} \left(\frac {dr }{dr_{\Bobo}}\right) du\, dr_{\Bobo} + r^2_\Bobo
    \left( \frac{f\ell^2 \lambda^2}{r^2_\Bobo}  d\chicoordinate^2  + \frac{ r^2}{r^2_\Bobo}
 \localch
  \right)
 \,.
 \phantom{xxx}
\eeal{21XII14.8b}
Note that the we have obtained Bondi coordinates \emph{only if the determinant $\localch$ is $t$-independent}, as otherwise the replacement of $t$ by its expression in terms of $u$ and $r$ introduces back $r$-dependence in the   determinant of $g_{AB}$. \Eq{14IV15.2} clearly shows that this requires $\beta=0$ in dimension $n+1=4$. Nevertheless we continue our calculations without assuming the vanishing of $\beta$.

In space-time dimension $n+1=4$ we find
%
%
\bel{21XII14.11}
 r_{\Bobo} = r+\frac{\beta  \ell ^2}{4 r}-\frac{\ell ^2 m}{2
   r^2}+O\left( {r} ^{-3}\right)
  \quad
 \Longleftrightarrow
  \quad
 r =r_{\Bobo} -\frac{\beta  \ell ^2}{4 r_{\Bobo} }+\frac{\ell ^2 m}{2 r_{\Bobo} ^2}+O\left(  r_{\Bobo}  ^{-3}\right)
 \,,
\ee
leading to
%
\bel{21XIII14.12}
 g_{uu} = -\left(\frac{r_{\Bobo} ^2}{\ell ^2}-\frac{\beta }{2}+\frac{m}{r_{\Bobo} }+\frac{\beta ^2 \ell ^2}{16
   r_{\Bobo} ^2}\right)
   +O\left( r^{-3}_\Bobo\right)
   \,.
\ee
\Eq{21XII14.5} gives
\bel{22XII14.1}
	 \mTB = -\frac{m \mu_{\hat h}(\HMsection)}{8\pi}
     \,,
\ee
where $\mu_{\hat h}$ is the measure induced on $\HMsection$ by the metric
\bel{24II15.11}
 \hat h := \lim_{r\to\infty}\left( \frac{f \ell^2 \lambda^2}{r^2_\Bobo}  d\chicoordinate^2  + \frac{ r^2}{r^2_\Bobo}
 \localch\right) =   \lambda^2   d\chicoordinate^2  +
 \localch
 \,.
\ee

In all dimensions, when $\beta$ vanishes we find
\bel{23II15.1}
	 \mTB = -\frac{m \mu_{\hat h}(\HMsection)}{4 (n-1)\pi}
     \,,
\ee
and the above remains true whether or not $\beta$ vanishes in odd space-dimensions $n$. We see that in these case the \charmass\ coincides with the coordinate mass, up to a volume normalisation factor related to the integrals involved. We will see in Section~\ref{ss18II15.4} that, similarly to the Birmingham metrics, in space-time dimension four the proportionality coefficient is the same as that for the Hamiltonian mass, see \eq{hammassHM+} below.

In even space-dimensions $n=2k$, when $\beta$ does not vanish, a calculation shows that the definition \eq{21XII14.5} gives instead the curious formula
\bel{23II15.2}
	 \mTB = -\frac{(m  + c_k \beta^k \ell^{n-2})\mu_{\hat h}(\HMsection)}{4 (n-1)\pi}
     \,,
\ee
where $c_k\in \R^*$ is a numerical coefficient depending upon $k$. For example, we have
\bel{24II15.6}
c_2= \frac 1{6}\,,
\
c_3 = -\frac{2}{25}\,,
\
c_4 =  \frac{33}{686}\,,
\
c_5 = -\frac{644  }{19683}\,,
\
c_6 =  \frac{7735  }{322102}
 \,.
\ee

\subsubsection{Renormalized volume}
 \label{ss9VII14.2}

\ptctodo{ this calculation
    should be generalised to any central manifold ?}
With the choice
\bel{26I15.2}
\psi_0 = 2 \pi \ell \lambda
\,,
\ee
where $\lambda$ is given by \eq{29XI14.1},
the curves obtained by letting  $\psi$ vary from zero to $\psi_0$
while keeping $t$ fixed and $r=r_0$, where $f(r_0)=0$ with  $f$ given by \eq{6XI12.5x},
are closed geodesics for the metric \eq{6XI12.4x}:  This follows from the fact that the manifold $\{r=r_0\}$ is the fixed-point set of the group of isometries generated by the Killing vector field $\partial_\psi$, and is therefore totally geodesic. Those geodesics will be referred to as \emph{core geodesics}, or \emph{emission curves}.

From the definition of $r_0$ we have
\bel{28I15.2}
 \frac{r_0^2}{\ell^2} = \frac {2 m}{r_0 }
\qquad
\Longleftrightarrow
\qquad
r_0 = (2m \ell^2)^{\frac 13}
 \,.
\ee

 \ptctodo{something more should be said here, add later when the dust has settled; at this stage I concentrate on outgoing light cones}
It is remarkable that the  null surfaces issuing normally from those geodesics are smooth away from the emission curves, and their union covers the whole space-time.
%

The contravariant metric $g^\sharp$ associated to \eq{21XII14.8b} equals
%
\beaa
 g^\sharp &= &
    f\partial_r^2 -  \frac {2\ell \sqrt{f(r)}}  {r} \partial_u \partial_r   + f^{-1}\ell^{-2} \lambda^{-2} \partial_\chicoordinate^2 + r^{-2} \partial_\varphi^2
 \,.
\eeaa

The vector field
\bel{10VII14.1}
 -\nabla u =     \frac   {\ell \sqrt{f(r)}} {r} \partial_r
\ee
has vanishing Lorentzian length, and a standard argument shows that its integral curves are affinely parameterized geodesics. Hence the parameter
$s$ defined as
\bel{10VII14.2}
 \frac{ds}{dr} =   \frac   {r}{\ell \sqrt{f(r)}}
 \quad
 \Longleftrightarrow
 \quad
 \partial_s =    \frac   {\ell \sqrt{f(r)}} {r} \partial_r
\ee
is an affine parameter along the generators of the null hypersurfaces $\{u=\const\}$.
(An explicit expression for $s$ in terms of elliptic integrals in space-time dimension $n+1=4$ can be given,  which again does not appear to be very useful.)

We are ready to calculate  the renormalized volume
$ V_\renm$. We have
\bean
 V(s') &= & \sintfromrzero^{s'} \int_\sectionofScri \sqrt{\det g_{AB}}\, d^2x \, ds
 = \sintfromrzero^{s'} \int_\sectionofScri\ell \lambda \sqrt{f}r  \, d\chicoordinate \, d\varphi \, \underbrace{\frac{ds}{dr}}_{ \frac   {r}{\ell \sqrt{f(r)}}} dr
\\
 &= &
 \mu_{\hat h}(\sectionofScri)   \int_{r(0)}^{r(s')}  r^2   dr
 =
\frac{1}3
 \mu_{\hat h}(\sectionofScri)    \big(r^3(s') - r_0^3\big)
 \,.
\eeal{12II15.1}
Here one should keep in mind that $r^3(s')$ needs to be reexpressed in terms of the affine parameter $s'$ before removing the singular part of $V(s')$. For this, integration of \eq{10VII14.2} gives, for large $r$,
%
%
\beal{10VII14.6c}
 s & = &
  r-r_0 +  \underbrace{\int_{r_0}^\infty \left( \frac r {\ell \sqrt f} -1 \right) dr}_{=: s_*}
  -\frac{\ell ^2 m}{2
   r^2} -\frac{3\ell ^4 m^2}{10    r^5}  +O\left(r^{-8}\right)
 \,,
\eea
with $0<s_*<\infty$  for $m>0$.

It is convenient to introduce a dimensionless variable $x $ through the formula $s = r_0 x  = (2m \ell^2)^{1/3} x $; set $s_*=   (2m \ell^2)^{1/3} x_*$. After
inverting \eq{10VII14.6c}  one obtains
\bel{29I15.12}
 r^3-r_0^3 = \frac{1}{2} \ell^2 m
   \left(4 x^3-12 x^2
   (x_*-1)+12 x
   (x_*-1)^2-4
   x_*^3+12
   x_*^2-12
   x_*+3\right)
 \,.
\ee
Inserting into \eq{12II15.1} leads to
\bean
 V_\renm
 &= &  \frac{1}{6} \ell^2 m
   \left(-4 x_*^3+12
   x_*^2-12
   x_*+3\right)
  \mu_{\hat h}(\sectionofScri)
\\ &= & \frac{1}{6} \ell^2 m
 \underbrace{\left(4  (1-x_*)^3- 1 \right)}_{\approx -0.6793 }
  \mu_{\hat h}(\sectionofScri)
	\,,
\eeal{29I15.13}
where we have used
\bel{29I15.14}
x_*=\int_1^\infty\left(\frac 1 {\sqrt{1-\frac 1 {x^3}}} - 1\right) dx \approx
 0.568815
 \,.
\ee

More information on the null geometry of Horowitz-Myers metrics, as well as a term-by-term analysis of the balance equation, can be found in Appendix~\ref{s15II16.1}.

\section{Hamiltonian mass, $\Lambda <0$}
 \label{s18II15.1}

Until specified otherwise, we allow arbitrary space-time dimension $n+1\ge 4$.

%
\newcommand{\fullgg}{g}%
\renewcommand{\gK}{Birmingham}%
\renewcommand{\Mtwo}{\sectionofScri}%
The calculations of the mass so far might appear to be \emph{ad-hoc}. In particular one wonders, why the coordinates approach of Section~\ref{ss18II15.1} appears to allow only the $\beta= 0$ case for HM-type metrics.
As such, a systematic way of obtaining an expression for the energy of a field configuration is to use a Hamiltonian approach.
Now, both families of metrics \eq{6XI12.4} and \eq{23II15.3}, with $f$ given by \eq{6XI12.5}, are asymptotic, as $r\to\infty$, to a background metric $\backg$ obtained by setting $m=0$ in $f$ (with different backgrounds for each family). When $\Lambda<0$ one can therefore use,%
\footnote{When $\Lambda>0$ a Hamiltonian definition of mass requires somewhat different considerations, see~\cite{CJK2}.}
in each case,  the formalism of~\cite{ChAIHP} (as already done in~\cite{ChruscielSimon} for $(3+1)$-dimensional asymptotically Kottler metrics),
to define the mass of $g$ relative to $\backg$.
Indeed, the Hamiltonian analysis in \cite{ChAIHP}
shows that to every spacelike hypersuface $\hyp $ and $\backg$-Killing vector $X$ one can associate a Hamiltonian mass $H(X,\hyp)$ through the formula
 \ptclevoca{this is likely to be repetitive when the energy chapters are added}
\begin{eqnarray}
  H(X,\hyp  )&= &\frac 12 \int_{\partial\hyp}
 \ourU^{\alpha\beta}dS_{\alpha\beta}\,,
\label{toto_new}
\end{eqnarray}
where the integral over $\partial \hyp$ is understood as the limit of integrals over a family of well behaved boundaries of sets which exhaust $\hyp$. Here $d
S_{\alpha\beta}$ is defined as
$\frac{\partial}{\partial x^\alpha}\lrcorner \frac{\partial}{\partial
  x^\beta}\lrcorner \rd x^0 \wedge\cdots \wedge\rd x^{n} $, with
$\lrcorner$ denoting contraction, and $\ourU^{\alpha\beta}$ is given by
\begin{eqnarray}
  \ourU^{\nu\lambda}&= &
{\ourU^{\nu\lambda}}_{\beta}X^\beta
  + \frac 1{8\pi} \Delta^{\alpha[\nu}
  {X^{\lambda]}}_{;\alpha}
\ ,
 \phantom{xxx}
 \label{Fsup2new_new}
\\ {\ourU^{\nu\lambda}}_\beta &= & \displaystyle{\frac{2|\det
  \bmetric_{\mu\nu}|}{ 16\pi\sqrt{|\det g_{\rho\sigma}|}}}
g_{\beta\gamma}(e^2 g^{\gamma[\nu}g^{\lambda]\kappa})_{;\kappa}
\,,\label{Freud2.0_new}
\end{eqnarray}
where a
semicolon  denotes covariant
differentiation \emph{with respect to the background metric $b$},
while
\bea
  \label{mas2_new}
   &
    e := \frac{\sqrt{|\det g_{\rho\sigma}|}}{\sqrt{|\det\bmetric_{\mu\nu}|}}
     \;,
&
\\
 &
 \label{18III16.10}
 \Delta^{{  \alpha}\nu}:=\frac{\sqrt{|\det g_{\rho{ \sigma}}|}}{\sqrt{|\det b_{\alpha\beta}|}} ~ g^{{ \alpha}\nu}- b^{{ \alpha}\nu}
    \,.
   &
\eea

\subsection{Asymptotically Birmingham metrics}
 \label{ss18II15.3}
  \ptctodo{watch out several renewcommands}
\renewcommand{\Bh}{{}}
\renewcommand{\Bht}{{}}
\renewcommand{\HM}{{}}
\renewcommand{\HMt}{{}}

We wish, first, to calculate \eq{toto_new} for $(n+1)$ dimensional metrics with  Birmingham asymptotics,
%
%
with a negative cosmological constant $\Lambda$
(equivalently, in \eq{6XI12.5} we take $\varepsilon=-1$),
with the  Killing vector $\partial_0$ and with $\Sigma=\{t=\const\}$.
For this, it is useful to introduce the following $\backg^\Bh$--orthonormal frame:
\begin{equation}
  \label{frame_new}
  e^\Bh_{\zero}= \frac{1}{\sqrt{\beta + \frac{r^2}{\ell^2}}}\partial_0\,, \quad
  e^\Bh_{\one}=  {\sqrt{\beta + \frac{r^2}{\ell^2}}} \partial_r\,, \quad
  e^\Bh_{\hat A} = \frac 1r \chitwo ^\Bh_{\hat A}\,,
\end{equation}
where $\chitwo ^\Bh_{\hat A}$ is an orthonormal (ON) frame for the metric $\mathring h$. To avoid ambiguities: the contravariant form of the background metric is,
by definition
\begin{equation}
  \label{frame_new+}
 \backg^{\mu\nu}\partial_\mu \otimes \partial_\nu:= - e^\Bh_{\zero}\otimes e^\Bh_{\zero} + e^\Bh_{\one}\otimes e^\Bh_{\one} + \sum_{A=2}^{n}e^\Bh_{\hat A}\otimes e^\Bh_{\hat A}
  \,.
\ee
Here and in what follows in the current section we use
$$
 A \in \{2,3,...,n\}
 \,,
$$
and we shall use hatted indices to denote the components of a tensor field
in the frame $e_{{\hat \mu}}$ defined in \eq{frame_new}. The connection coefficients, defined as
$\nabla_{e^\Bht_{{\hat \mu}}}e^\Bh_{{\hat \nu}} = {\omega^\Bh{}^{\hat \rho}}_{{\hat \nu}{\hat \mu}}e^\Bh_{\hat \rho}$ with $\nabla$ associated with $b$, read
%
\begin{eqnarray}
	& \omega^\Bh_{\zero \one \zero} =	- \frac r {\ell^2 \sqrt{\beta + \frac{r^2}{\ell^2}}} = - \frac{1}{\sqrt{\beta\frac{\ell^4}{r^2}+\ell^2}} = -\frac 1 \ell + \redOof r^{-2})
	\,, &
	\nonumber \\
	& \omega^\Bh_{\one \hat A \hat B} =	- \frac{ \sqrt{\beta + \frac{r^2}{\ell^2}} } r
 b_{\hat A \hat B}
  = - \sqrt{\frac{\beta}{r^2}+\frac{1}{\ell^2}}
 b_{\hat A \hat B}
  = \left(-\frac 1 \ell + \redOof r^{-2})
 \right)
 b_{\hat A \hat B}
	\,. &
	\label{29XI14.5_new}
\end{eqnarray}
The remaining  possibly non-vanishing connection coefficients, not obtained from the above by permutations of indices, are the  $\omega^\Bh_{\hat A \hat B \hat C}$'s, with $\hat A\ne \hat B$.
For example, in space-time dimension $3+1$, if we  use a coordinate system
$\theta,\varphi$ on $\Mtwo$ in which ${\mathring h}$
 takes, locally, the form
$d\theta^2 + \sinh^2 \theta\; d\varphi^2$ for $k=-1$, $d\theta^2 +
d\varphi^2$ for $k=0$, and $d\theta^2 + \sin^2 \theta \;d\varphi^2$ for
$k=1$, we find
\begin{eqnarray}
  & \omega^\Bh_{ \two \three \three} = \cases{ -\frac{\coth \theta}r \,, &
    $k=-1\,,$ \cr 0 \,, & $k=0\,,$ \cr -\frac{\cot \theta}r \,, &
    $k=1\,.$} & \label{29XI14.6again_new}
\end{eqnarray}
However, the exact form above, and the one of $\omega^\Bh_{\hat A \hat B \hat C}$ in general, is not needed for what follows.

We further have
%
\begin{eqnarray}
	\label{Killing_new}
	& X_\Bh^{\zero} = \sqrt{\beta + \frac{r^2}{\ell^2}} = \frac r {\ell} + \redOof r^{-1})
	\,, &
	\\
	\label{Killing1_new}
	& e^\Bh_{\one}\big(X_\Bh^{\zero}\big) = \big(X_\Bh^{\zero}\big)_{;\one} = - X^\Bh_{\zero;\one} = X^\Bh_{\one;\zero}
= \frac r {\ell^2} 
	\,, &
\end{eqnarray}
with the third equality in \eq{Killing1_new} following from the Killing equations $X_{\mu;\nu}+ X_{\nu;\mu}=0$; all the remaining $X_\Bh^{\hmu}$'s and $X^\Bh_{\hmu;\hnu}$'s are zero.

Let the tensor field $e^{\mu\nu}$ be defined by the formula
\begin{equation}
  \label{mas3_new}
  e^{\mu\nu}
   := g^{\mu\nu}-b^{\mu\nu}
	\; .
\end{equation}
As already mentioned, we use hatted indices to denote the components of a tensor field
in the frame $e_{{\hat \mu}}$, \emph{e.g.} $e^{{\hat \mu}{\hat \nu}}$
denotes the coefficients of $e^{\mu\nu}$ with respect to that
frame:
$$
 e^{\mu\nu}\partial_\mu\otimes \partial_\nu = e^{{\hat \mu}{\hat \nu}}e_{{\hat \mu}}\otimes
    e_{{\hat \nu}}
 \,.
$$
%


Let the $\theta^{\hat A}$'s form a coframe dual to the $e_{\hat A}$'s. Then
$$
\theta^{\hat 0}\wedge \ldots
\wedge  \theta^{\hat n }
 = \sqrt{|\det b_{\alpha\beta}|} dx^0 \wedge \ldots \wedge dx^n
 \,,
$$
and so on the level sets of $t$ intersected with those of $r$ we have
\beaa
\ourU^{\alpha\beta} dS_{\alpha\beta}\big|_{r=R}
 & = & \ourU^{\hat \alpha\hat \beta} e_{\hat \alpha} \rfloor e_{\hat \beta} \rfloor ( dx^0 \wedge \ldots \wedge dx^n)\big|_{r=R}
\\
 & = &  \frac{ \ourU^{\hat \alpha\hat \beta}}{\sqrt{|\det b_{\alpha\beta}|}}  e_{\hat \alpha} \rfloor e_{\hat \beta} \rfloor ( \theta^{\hat 0}\wedge \ldots
\wedge  \theta^{\hat n })\big|_{r=R}
\\
 & = & \frac{ 2 \ourU^{\hat 1 \hat 0 }}{\sqrt{|\det b_{\alpha\beta}|}}  \, \theta^{\hat 2}\wedge \ldots
\wedge  \theta^{\hat n } \big|_{r=R}
  \,.
\eeaa
{}From \eq{toto_new} we thus find
\begin{eqnarray}
\label{toto1_new}
  H(X,\hyp  )&= &\lim_{R\to\infty}  \int_{\hyp \cap\{r=R\}}  \frac{\ourU^{\one\zero}}{\sqrt{|\det b_{\alpha\beta}|}} \, \theta^{\hat 2}\wedge \ldots \wedge \theta^{\hat n}
	\,.
\end{eqnarray}
%
We wish to analyze when the above limit exists.  Since every $\theta^{\hat A}$ comes with a multiplicative factor of $r$ in local coordinates on the level sets of $R$ within $\hyp$, again in local coordinates the integrand in \eq{toto1_new}  behaves as $r^{n-1} \ourU^{\one\zero}$. Now,
$$
 r^{n-1} {\ourU^{\one\zero}}_{\beta} X^\beta = r^{n-1} {\ourU^{\one\zero}}_{\zero}X^{\zero}\approx \frac{r^n}\ell{\ourU^{\one\zero}}_{\zero}
 \,,
$$
hence in the calculations we only need to keep track of those terms in ${\ourU^{\one\zero}}_{\zero}/\sqrt{|\det b_{\alpha\beta}|} $ which decay  slower than $r^{-n}$, or at that rate. Similarly one sees from \eq{Killing_new}--\eq{Killing1_new} that only those terms in %
$$
 \Delta^{{\hat \alpha}\hnu}
  =  {\sqrt{|\det g_{\hrho{\hat \sigma}}|}} ~ g^{{\hat \alpha}\hnu}- b^{{\hat \alpha}\hnu}
 $$
(compare \eq{18III16.10})
which are $\redOof r^{-n})$, or which are decaying slower, will give a non-vanishing contribution to the term involving the derivatives of $X$ in the integral \eq{toto1_new}.

We will say that a metric is \emph{asymptotically Birmingham} if  there exists $\epsilon>0$ such that in the frame \eq{frame_new} it holds
\begin{equation}
  \label{hamfalof_new1}
  e^{\hmu\hnu}= \redOof r^{-n/2-\epsilon})\,, \quad e_{\hrho}(e^{\hmu\hnu})= \redOof r^{-n/2-\epsilon})
  \,,
   \quad
   \det (e^{\hmu \hnu}) - 1 = \redOof r^{-n-\epsilon})\,.
\end{equation}
We note that we have imposed the volume-element condition to guarantee convergence of mass integrals, see \eq{26XII15.1}-\eq{C3} below.

Recall that we
only consider vector fields $X$ which are $b$-Killing vector
fields, and therefore their tetrad components satisfy
\be \label{A1.}|X^\hnu|+ |\zn_\hmu X^\hnu| \le C r
 \,.
\ee

We claim that \eqref{hamfalof_new1} guarantee a finite total energy in vacuum. Indeed, this follows from the standard integral identity (cf., e.g., \cite{ChAIHP}),
\begin{equation}
   \int_{\{x^0=0,r=R\}}
 \ourU^{\alpha\beta}dS_{\alpha\beta}=  2\int_{\{x^0=0,R_0\le r\le R\}}
 \znabla_\beta \ourU^{\alpha\beta}dS_{\alpha}  + \int_{\{x^0=0,r=R_0\}}
 \ourU^{\alpha\beta}dS_{\alpha\beta}\,,
\label{26XII15.1}
\end{equation}
with
\begin{eqnarray}
\nonumber
 \lefteqn{
    16\pi\znabla_\beta \ourU^{\alpha\beta} =   \left( \sqrt{|\det g|}
    g^{\alpha \gamma} - \sqrt{|\det b|}    b^{\alpha \gamma}\right) b_{\gamma \beta} X^\beta
    }
    &&
\\
 &&
     +
    \left(\mathring{\mcT}^\alpha{}_\kappa-\mcT^\alpha{}_\kappa\right)X^\kappa
+2 \Lambda \left(\sqrt{|\det b| }- \sqrt{|\det g| }\right)X^\beta
\nonumber \\ & &
 + \sqrt{|\det    b|} \;\left( Q^\alpha{}_{\beta}
    X^\beta + Q^\alpha{}_{\beta \gamma}
    \znabla^\beta X^\gamma\right)\;, \label{C3}
\end{eqnarray}
where
    $Q^\alpha{}_{\beta}$ is a quadratic form in $e_a(e^{bc})$, and
    $Q^\alpha{}_{\beta \gamma}$ is bilinear in $e_a(e^{bc})$ and
    $e^{ab}$, both with bounded coefficients.
    Finally,
 \be\label{N.5g2}
 8\pi
 \mcT^\lambda{}_\kappa:=  \sqrt{|\det g|} \bigg( R^\lambda{}_\kappa -\frac 12
 g^{\alpha\beta}R_{\alpha\beta} \delta ^\lambda_\kappa +\Lambda
 \delta ^\lambda_\kappa
  \bigg)
  \;,
\ee
with $\mathring{\mcT}^\lambda{}_\kappa$  defined as in \eq{N.5g2} with
$g$ replaced by $b$.

Passing with $R$ to infinity in \eq{26XII15.1}, under \eq{hamfalof_new1} the right-hand side converges to a finite limit in vacuum, and
 one finds indeed that the resulting Hamiltonians are finite.

If the metric is not vacuum, the same argument applies if one moreover assumes that there exists $\epsilon>0$ such that
\bel{P10.0}  |\mcT^{\mu}{}_{\nu}- \mathring{\mcT}^{\mu}{}_{\nu} |\le C (1+r)^{-1 -\epsilon}
  \,.
\ee

We note that for the calculations of the boundary term the following, slightly weaker, conditions suffice:
\begin{equation}
  \label{hamfalof_new}
  e^{\hmu\hnu}= \redOofo r^{-n/2})\,, \quad e_{\hrho}(e^{\hmu\hnu})= \redOofo r^{-n/2})\,.
\end{equation}
The boundary conditions \eq{hamfalof_new} ensure that one needs to keep track only of those terms in $\ourU^{\one\zero}$ which are linear in $e^{\hmu\hnu}$ and $e_{\hrho}(e^{\hmu\hnu})$, when $\ourU^{\one\zero}$ is Taylor-expanded around $\backg$.

 For example, if $g$ has the same leading order terms as a \gK\ metric \eq{6XI12.4}-\eq{6XI12.5} we find, writing $f^\Bh_0$ for $f^\Bh\big|_{m=0}$, using \eq{frame_new}
\bel{29XI14.8_new}
 g_\Bh^\sharp :=g_\Bh^{\hat \mu \hat \nu}e^\Bh_{\hat \mu} e^\Bh_{\hat \nu} = - \frac{f^\Bh_0}{f^\Bh} \left(e^\Bh_{\zero}\right)^2 + \frac{f^\Bh}{f^\Bh_0} \left(e^\Bh_{\hat 1}\right)^2 + \sum_{\hat A=2}^{n-1} \left(e^\Bh_{\hat A}\right)^2
  \,,
\ee
which yields
\begin{eqnarray}
	& e_\Bh^{\zero\zero}
	=	-	\frac{f^\Bh_0}{f^\Bh} + 1
	= \frac{f^\Bh - f^\Bh_0}{f^\Bh}
	=	- \frac{2m\ell^2}{r^n} \Big( 1 + \redOof r^{-2}) \Big)
	\,, &
	\nonumber \\
	& e_\Bh^{\one\one}
	= \frac{f^\Bh_0 - f^\Bh}{f^\Bh_0}
	=	- \frac{2m\ell^2}{r^n} \Big( 1 + \redOof r^{-2}) \Big)
	\,, &
	\nonumber \\
	& e^\Bh_{\one}(e_\Bh^{\zero\zero})
	= f^\Bh_0 \, \partial_r \left( \frac{f^\Bh - f^\Bh_0}{f^\Bh_0} \right)
	=	\frac{2nm\ell}{r^n} \Big( 1 + \redOof r^{-2}) \Big)
	\,, &
	\nonumber \\
	& e^\Bh_{\one}(e_\Bh^{\one\one})
	= f^\Bh_0 \, \partial_r \left( \frac{f^\Bh - f^\Bh_0}{f^\Bh_0} \right)
	=	\frac{2nm\ell}{r^n} \Big( 1 + \redOof r^{-2}) \Big)
	\,, &
	\label{egk_new}
\end{eqnarray}
with the remaining $e_\Bh^{\hmu\hnu}$'s and $e^\Bh_{{\hat \sigma}}(e_\Bh^{\hmu\hnu})$'s vanishing, so that Equations \eq{hamfalof_new} are satisfied for metrics with leading Birmingham asymptotics.

Rather generally, under \eq{hamfalof_new} one obtains, using $b_{\hat \mu \hat \nu}  =\mathrm{diag}(-1,+1,\cdots,+1)$,
%
%
\begin{eqnarray}
	g_{{\hat \mu}{\hat \nu}} &=&
	b_{{\hat \mu}{\hat \nu}}-b_{{\hat \mu}{\hat \alpha}}b_{{\hat \nu}{\hat \beta}} e^{{\hat \alpha}{\hat \beta}}+\redOofo r^{-n})
	\,,
	\label{metricdown_new}
\\
	\sqrt{|\det g^\Bh_{\mu\nu}|} &=&
	\sqrt{|\det \backg^\Bh_{\mu\nu}|}\left(1 + \frac 12 \Big( e_\Bh^{\zero\zero} - e_\Bh^{\one\one} - b_{{\hat A}{\hat B}}e_\Bh^{{\hat A}{\hat B}} \Big) + \redOofo r^{-n})\right)
	\,,
\\
	\frac{{16\pi}{\ourU^\HM{}^{\one\zero}}_{\zero}}{\sqrt{|\det b_{\alpha\beta}|}}
	&=&
   \displaystyle{\frac{2\sqrt{|\det
  \bmetric_{\mu\nu}|}}{ \sqrt{|\det g_{\rho\sigma}|}}}
g_{\hat 0 \hat \gamma}(e^2 g^{\hat \gamma[\hat 1}g^{\zero]\hat \kappa})_{;\hat \kappa}
	\nonumber
\\
	&=&
  4  g_{\hat 0 \hat \gamma} g^{\hat \gamma[\hat 1}g^{\zero]\hat \kappa}   e_{;\hat \kappa}
 +
   \displaystyle 2e
g_{\hat 0 \hat \gamma}( g^{\hat \gamma[\hat 1}g^{\zero]\hat \kappa})_{;\hat \kappa}
	\nonumber
\\
	&=&
  -2   g^{ 1 \hat \kappa}   e_{;\hat \kappa}
 +
   \displaystyle 2e
g_{\hat 0 \hat \gamma}((b^{\hat \gamma[\hat 1}+e^{\hat \gamma[\hat 1})(b^{\zero]\hat \kappa}+e^{\zero]\hat \kappa}))_{;\hat \kappa}
	\nonumber
\\
	&=&
  -2    e_{;\hat 1}
 +
   \displaystyle 2
 b_{\hat 0 \hat \gamma}( b^{\hat \gamma[\hat 1}b^{\zero]\hat \kappa} +
 b^{\hat \gamma[\hat 1}e^{\zero]\hat \kappa} +
 e^{\hat \gamma[\hat 1}  b^{\zero]\hat \kappa}
 + e^{\hat \gamma[\hat 1}  e^{\zero]\hat \kappa} )_{;\hat \kappa}
 +\redOofo r^{-n})
	\nonumber
\\
	&=&
  -2    e_{;\hat 1}
 +
   \displaystyle 2
 b_{\hat 0 \hat 0}(
 b^{\hat 0[\hat 1}e^{\zero]\hat \kappa} +
 e^{\hat0[\hat 1}  b^{\zero]\hat \kappa} )_{;\hat \kappa}
 +\redOofo r^{-n})
	\nonumber
\\
	&=&
  -2    e_{;\hat 1}
 -
   \displaystyle
 (
 - b^{\hat 0 \hat 0}e^{\hat 1 \hat \kappa} +
 e^{\hat0 \hat 1}  b^{\zero \hat \kappa}  -
 e^{\hat 0 \hat 0}  b^{\hat 1 \hat \kappa} )_{;\hat \kappa}
 +\redOofo r^{-n})
	\nonumber
\\
	&=&
  -2    e_{;\hat 1}
 -
   \displaystyle
  e^{\hat 1 \hat \kappa}{}_{;\hat \kappa}
  +
 e^{\hat0 \hat 1}{} _{;\hat0}
 +
 e^{\hat 0 \hat 0} {}_{;\hat 1}
 +\redOofo r^{-n})
 \,.
  \label{23II15.10}
\end{eqnarray}
This can be further rewritten as 
\begin{eqnarray}
	\frac{{ 16\pi \ourU^\Bh{}^{\one\zero}}_{\zero}}
{\sqrt{|\det b_{\alpha\beta}|}} &=&
	 -2e^\Bh{}_{;\one}
- e^{\one\one}{}_{;\one}
 - e^{\one\hat A }{}_{;\hat A}
 + e^{\zero\zero}{}_{;\one}  +\redOofo r^{-n})
	\nonumber \\
	\phantom{xxx}
	\label{uo1o_new}
	\nonumber \\
	&=&
	  e^\Bh_{\one} \big( b_{{\hat A}{\hat B}}e_\Bh^{{\hat A}{\hat B}} \big)  +
  \omega^\Bh_{\hat A \one \hat B}   e_\Bh^{{\hat A}{\hat B}} -
  \omega^\Bh{}^{\hat A}{}_{ \one \hat A}   e_\Bh^{\one\one}
   - \frac1 r{\mathring\mathcal{ D}}^\Bh_{{\hat A}} e_\Bh^{\one{\hat A}}  + \redOofo r^{-n})
	\nonumber \\
	&=&
	  e^\Bh_{\one} \big(
  b_{{\hat A}{\hat B}}e_\Bh^{{\hat A}{\hat B}} \big) + \sqrt{\frac{\beta}{r^2}+\frac{1}{\ell^2}} \Big( b_{{\hat A}{\hat B}}e_\Bh^{{\hat A}{\hat B}} - (n-1) e_\Bh^{\one\one} \Big)
\nonumber
\\
 &&
     - \frac1 r{\mathring\mathcal{ D}}^\Bh_{{\hat A}} e_\Bh^{\one{\hat A}}   + \redOofo r^{-n})
	\,,
	\phantom{xxx}
	\label{8II15_2}
\end{eqnarray}
Here  ${\mathring\mathcal{ D}}^\Bh_{{\hat A}}$
 denotes the covariant derivative on $(\sectionofScri,\mathring h)$, with $ e_\Bh^{\one{\hat A}}$ being understood as a vector field on $\sectionofScri$, with $\hat A$, $\hat B$ running from $2$ to $n$.

We also have
\bean
	\frac1{8\pi} \Delta_\Bh^{\hat \alpha[\one} \Big(X_\Bh^{\zero]}\Big)_{;\hat \alpha} & = & \frac1{16\pi} \left(\Delta_\Bh^{\one\one}-\Delta_\Bh^{\zero\zero}\right) \big(X_\Bh^{\zero}\big)_{;\one}
	=
	\frac r {16\pi\ell^2} \left( \Delta_\Bh^{\one\one} - \Delta_\Bh^{\zero\zero} \right) 
	\nonumber \\
	& = &
	-\frac r {16\pi\ell^2} b_{{\hat A}{\hat B}}e_\Bh^{{\hat A}{\hat B}} + \redOofo r^{-n})
	\label{uo1oX_new}
	\,.
\end{eqnarray}
Inserting all this into \eq{toto1_new} one is finally led to the following simple expression for the \emph{Hamiltonian mass} of asymptotically Birmingham metrics:
\begin{eqnarray}
	\label{MHamBh}
	\mham^\Bh
	&=&
	\lim_{R\to\infty} \frac {r^n} {16\pi}
	\int_{\hyp \cap\{r=R\}}
		\left[
		\left(\frac{1}{\ell^2}+  \frac{\beta}{r^2}  \right)
    \left({\frac {r \partial \big(b_{{\hat A}{\hat B}}e_\Bh^{{\hat A}{\hat B}} \big)}{\partial r}} - (n-1)
   e_\Bh^{\one\one}  \right)
   \right.
		\nonumber \\
		&&
	\phantom{xxxxxxxxxxxxxxxxxxxxx}
\left.
		+
		 \frac{\beta}{r^2}  b_{{\hat A}{\hat B}}e_\Bh^{{\hat A}{\hat B}}
  \right] d^{n-1}\mu_{{\mathring h}}
	\,,
\end{eqnarray}
In space-time dimension $n+1=4$ this simplifies to the expression given in~\cite{ChruscielSimon}:
\begin{eqnarray}
	\label{MHamBh3}
	\mham^\Bh
	&=&
	\lim_{R\to\infty} \frac {R^3} {16\pi\ell^2}
	\int_{\hyp \cap\{r=R\}}
		\left[
		r \frac{ \partial e_\Bh^{{\two}{\two}} }{\partial r}
		+
		r \frac{ \partial e_\Bh^{{\three}{\three}} }{\partial r}
		-
		2 e_\Bh^{\one\one}
  \right] d^{n-1}\mu_{{\mathring h}}
	\,.
	\phantom{xxx}
\end{eqnarray}
If in addition to \eq{hamfalof_new1} we assume that
\begin{equation}
  \label{hamfalof_new1strong}
  e^{\hmu\hnu}= \redOof r^{-n} )\,, \quad e_{\hrho}(e^{\hmu\hnu})= \redOof r^{-n})
\end{equation}
(this is actually the fall-off rate for Birmingham metrics), \eq{MHamBh} can be rewritten in a form similar to \eq{MHamBh3} in higher dimensions as well:
\begin{eqnarray}
	\label{MHamBhstrong }
	\mham^\Bh
	&=&
	\lim_{R\to\infty} \frac {r^n} {16\pi\ell^2}
	\int_{\hyp \cap\{r=R\}}
		\left[
    \left({\frac {r \partial \big(b_{{\hat A}{\hat B}}e_\Bh^{{\hat A}{\hat B}} \big)}{\partial r}} - (n-1)
   e_\Bh^{\one\one}  \right)
  \right] d^{n-1}\mu_{{\mathring h}}
	\,.
   \phantom{xxxxxx}
\end{eqnarray}

As an example, if $\fullgg$ is the $3+1$-dimensional \gK\ metric \eq{6XI12.4}, we find
\begin{equation}
  \label{hammassHM}
  \mham^\HM =   \frac{\mu_{\mathring h}({\sectionofScri}) m}{4\pi}\,,
\end{equation}
where
\begin{equation}
\label{areanorm}
	\mu_{\mathring h}({\sectionofScri}) :=   \int_{\sectionofScri} d^{n-1}\mu_{{\mathring h}}
	\,.
\end{equation}
We conclude that the Hamiltonian mass is proportional to $m$, with   the same proportionality factor as the characteristic mass of null hypersurfaces asymptotic to level sets of $t$, see \eq{21XII14.5b}:
\bel{19II16.3}
 \mTB=\mham
  \,.
\ee

When
$\Mtwo=T^2$ (equivalently, $\beta=0$) with area normalized to $4\pi$ we obtain $\mham=m$. For
$\beta=\pm 1$ it follows from the Gauss--Bonnet theorem that
$\mu_{\check h}({\sectionofScri})=4\pi|1-g_\infty|$, where $g_\infty$ is the genus of $\Mtwo$,
hence
\begin{equation}
  \label{masskot2_new}
  \mham= |1-g_\infty| m\,.
\end{equation}
One recovers $\mham=m$ for $\Mtwo=S^2$,
but this will be true only up to a positive proportionality factor for $\Mtwo$'s of higher genus.

\subsection{Asymptotically HM-type metrics}
 \label{ss18II15.4}

The aim of this section is to derive a formula analogous to \eq{MHamBhstrong } for metrics with Horowitz--Myers-type asymptotics.
%
%
For this consider, as before, the  background metric
$$
 \backg^\Bh:= - e^\Bh_{\zero}\otimes e^\Bh_{\zero} + e^\Bh_{\one}\otimes e^\Bh_{\one} + \sum_{A=2}^{n}e^\Bh_{\hat A}\otimes e^\Bh_{\hat A}
  \,,
$$
where now instead of \eq{frame_new} we set
%
%
\begin{equation}
  \label{29XI14.6_new}
  e^\HM_{\zero}=  \frac{\ell }{r } \partial_0\,, \quad
  e^\HM_{\one}=  {\sqrt{\beta + \frac{r^2}{\ell^2}}} \partial_r\,, \quad
  e^\HM_{\hat A} = \frac 1r \chitwo ^\HM_{\hat A}\,,  \quad
  e^\HM_{\hat n} = \frac 1{\ell\lambda \sqrt{\beta + \frac{r^2}{\ell^2}}} \partial_\chicoordinate
	\,.
\end{equation}
Here $\chitwo ^\HM_{\hat A}$ is an ON frame for the metric ${\localch}$ as in the first line of \eq{21XII14.8}, and in this section we let
\bel{11III16.1}
 A,B \in \{2,3,...,n-1 \}
 \,,
\ee
similarly for hatted indices.

A metric $g$ will be said to be \emph{asymptotically HM along $\hyp$}, or simply \emph{asymptotically HM}, if there exists a coordinate system $(t,r,x^ A)$ and $\epsilon>0$ such that at $\hyp:=\{t=0\}$ we have
\bel{24II15.2}
 \frac{\det g_{\mu\nu}} {\det b_{\mu\nu}} = 1 + \redOof r^{-n-\epsilon})
 \,,
\ee
and if the frame components of $g$ with respect to the frame \eq{29XI14.6_new} satisfy
\bea
 &
 e^{\hat \alpha \hat \beta}:= g^{\hat \alpha \hat \beta} - b^{\hat \alpha \hat \beta} = \redOof r^{-n/2-\epsilon})
 \,,
 \qquad
  e_{\hat \mu}(e^{\hat \alpha \hat \beta})= \redOof r^{-n/2-\epsilon})
 \,.
  &
\eeal{24II15.1}
This is formally the same as \eq{hamfalof_new1}, but both the frame and the background metric are different. (As before, the volume element condition is added to guarantee convergence of mass integrals.)

The identity \eq{26XII15.1}
shows as before that conditions \eq{24II15.2}-\eq{24II15.1} guarantee a finite Hamiltonian mass in vacuum. We expect that the arguments of~\cite{ChNagy} can be adapted to this case to show that the mass is independent of the freedom of choice of coordinates and frames satisfying our conditions above, but we have not attempted to check this.
 \ptctodo{consider doing?}

Similarly to the Birmingham case, our calculations of the boundary integral will be done with \eq{24II15.1} replaced by  the slightly weaker conditions
\bel{24II15.3}
 e^{\hat \alpha \hat \beta} = \redOofo r^{-n/2})
 \,,
 \quad
  e_{\hat \mu}(e^{\hat \alpha \hat \beta})= \redOofo r^{-n/2})
 \,,
 \quad
  e =1+ \redOofo r^{-n/2})
 \,.
\ee

The connection coefficients $\omega^\HM_{\hat \mu \hat \nu \hat \rho}$ of the background metric $b$ read
%
\begin{eqnarray}
  & \omega^\HM_{\zero \one \zero} = - \frac{ \sqrt{\beta + \frac{r^2}{\ell^2}} } r = - \sqrt{\frac{\beta}{r^2}+\frac{1}{\ell^2}} = -\frac 1 \ell + \redOof r^{-2})
	\,, &
	\nonumber \\
 	& \omega^\HM_{\one \hat A \hat B} = - \frac{ \sqrt{\beta + \frac{r^2}{\ell^2}} } r
  b_{\hat A \hat B}
 		\,, &
 		\nonumber \\
	& \omega^\HM_{\one \hat n \hat n} =	- \frac r {\ell^2 \sqrt{\beta + \frac{r^2}{\ell^2}}} = - \frac{1}{\sqrt{\beta\frac{\ell^4}{r^2}+\ell^2}} = -\frac 1 \ell + \redOof r^{-2})
	\,. &
	\label{5II15_1}
\end{eqnarray}
The remaining  possibly non-vanishing connection coefficients,
 which are not obtained from the above  by permutations of indices, are the  $\omega^\Bh_{\hat A \hat B \hat C}$'s, with $A\ne B$.
As in the previous section, the exact values  of the $\omega^\HM_{\hat A \hat B \hat C}$'s are not needed in what follows. We further have,
%
%
\begin{eqnarray}
	\label{5II15_2}
	& X_\HM^{\zero} = \frac r {\ell}
	\,, &
	\\
	\label{5II15_3}
	&\hspace{-2mm} e^\HM_{\one}( X_\HM^{\zero} )= \big(X_\HM^{\zero}\big)_{;\one} = - X^\HM_{\zero;\one} = X^\HM_{\one;\zero} = \frac 1 {\ell} \sqrt{\beta + \frac{r^2}{\ell^2}} = \frac r {\ell^2} + \redOof r^{-1})
	\,,	\phantom{xxxx}	&
\end{eqnarray}
where all the remaining $X_\HM^{\hmu}$'s and $X^\HM_{\hmu;\hnu}$'s are zero.

Writing $f^\HM_0$ for $f^\HM\big|_{m=0}$, from \eq{frame_new} we see that the HM-type metrics can be written as
\bel{29XI14.8_new+}
 g_\HM^\sharp = - e^\HM_{\zero}\otimes  e^\HM_{\zero}
  +\frac{f^\HM}{f^\HM_0} e^\HM_{\hat 1 }\otimes  e^\HM_{\hat 1}
   + \sum_{\hat A=2}^{n-1}  e^\HM_{\hat A }\otimes  e^\HM_{\hat A}
    + \frac{f^\HM_0}{f^\HM} e^\HM_{\hat n }\otimes  e^\HM_{\hat n}
  \,.
\ee
This leads to
\bean
 &
  \displaystyle
 e_\HM^{\hat 1 \hat 1} = g_\HM^{\hat 1 \hat 1} - b_\HM^{\hat 1 \hat 1} =
 \frac{f^\HM}{f^\HM_0}-1 =  \frac{f^\HM - f^\HM_0}{f^\HM_0} = -\frac {2m \ell^2}{r^n} \left(1 + O (r^{-2})\right)
 \,,
 &
\\
 &
  \displaystyle
  e_\HM^{\hat 0 \hat 0} =  0 = e_\HM^{\hat A \hat B}
  \,,\qquad
  e_\HM^{\hat \mu \hat \nu} =0  \ \mbox{for $\mu\ne \nu$}
  \,,
  &
\nonumber \\
 &
   \displaystyle
    e_\HM^{\hat n \hat n} =  \frac{f^\HM_0-f^\HM }{f^\HM } =  \frac {2m \ell^2}{r^n} \left(1 + O (r^{-2})\right)
    \,,
\eeal{29XI14.7_new}
which satisfies the decay conditions set forth above.

Quite generally, for metrics satisfying \eq{24II15.3} we find as before
\be
	e = 1 + \frac 12 \Big(
	\underbrace{
	 e_\HM^{\zero\zero} - e_\HM^{\one\one} -
 b_{{\hat A}{\hat B}}e_\HM^{{\hat A}{\hat B}} - e_\HM^{{\hat n}{\hat n}}
 }\Big) + \redOofo r^{-n})
	\,,
\ee
and note that  \eqref{24II15.2} implies that the underbraced term is also $\redOofo r^{-n})$.
\Eq{23II15.10} still applies and, taking into account \eq{11III16.1}, gives
%
\begin{eqnarray}
	\frac{{16\pi}{\ourU^\HM{}^{\one\zero}}_{\zero}}{\sqrt{|\det b_{\alpha\beta}|}}
	&=&
	-  2e^\HM{}_{;\one}
 - e^{\one\hat 1}{}_{;\hat 1}
 - e^{\one\hat A}{}_{;\hat A}
 - e^{\one\hat n}{}_{;\hat n}
 + e^{\zero\zero}{}_{;\one}  +\redOofo r^{-n})
	\nonumber \\
	&=&
	e^\HM_{\one} \Big( b_{{\hat A}{\hat B}}e_\HM^{{\hat A}{\hat B}}
	+	e_\HM^{{\hat n}{\hat n}} \Big)
	-
	e^\HM_{\hat n} ( e_\HM^{{\one}{\hat n}})
	+
	\omega^\HM_{\hat A \one \hat B}   e_\HM^{\hat A \hat B}
	-
	\omega^\HM{}^{\hat A}{}_{ \one \hat A}   e_\HM^{\one \one}
	\nonumber \\
	&&
	\qquad \,
	+ \,
	\omega^\HM_{\hat n \one \hat n} \Big( e_\HM^{\hat n \hat n} - e_\HM^{\one \one} \Big)
	-
	\frac1 r{\check\mathcal{ D}}^\Bh_{{\hat A}} e_\Bh^{\one{\hat A}}
	+ \redOofo r^{-n})
	\nonumber \\
	&=&
	e^\HM_{\one} \Big( b_{{\hat A}{\hat B}}e_\HM^{{\hat A}{\hat B}}	+	e_\HM^{{\hat n}{\hat n}} \Big)
	-
	e^\HM_{\hat n} (e_\HM^{{\one}{\hat n}})
	+
	\sqrt{\frac{\beta}{r^2}+\frac{1}{\ell^2}} \Big( b_{{\hat A}{\hat B}}e_\HM^{{\hat A}{\hat B}}
    - (n-2)     e_\HM^{\one \one}
        \Big)
	\nonumber
\\
	&&
	+
	\bigg( \beta\frac{\ell^4}{r^2}+\ell^2 \bigg)^{-1/2} \Big( e_\HM^{\hat n \hat n} -  e_\HM^{\one \one} \Big)
	-
	\frac1 r{\check\mathcal{ D}}^\HM_{{\hat A}} e_\HM^{\one{\hat A}}
	+ \redOofo r^{-n})
	\,.
	\phantom{xxxxx}
	\label{8II15_4}
\end{eqnarray}
Here ${\check\mathcal{ D}}^\HM_{{\hat A}} e_\HM^{\one{\hat A}}$ is understood as the covariant divergence of the  vector field  $e_\HM^{\one{\hat A}}e_{\hat A} $ with respect to the metric ${\localch}$.

Furthermore,
\begin{eqnarray}
	\frac1{8\pi} \Delta_\HM^{\hat \alpha[\one} \Big(X_\HM^{\zero]}\Big)_{;\hat \alpha} & = & \frac1{16\pi} \left(\Delta_\HM^{\one\one}-\Delta_\HM^{\zero\zero}\right) \big(X_\HM^{\zero}\big)_{;\one}
	\nonumber \\
	& = &
	\frac 1 {16\pi\ell} \left( \Delta_\HM^{\one\one} - \Delta_\HM^{\zero\zero} \right)  \sqrt{\beta + \frac{r^2}{\ell^2}} + \redOofo r^{-n})
	\nonumber \\
	& = &
	- \frac 1 {16\pi\ell} \left( b_{{\hat A}{\hat B}}e_\HM^{{\hat A}{\hat B}} + e_\HM^{\hat n \hat n} \right)  \sqrt{\beta + \frac{r^2}{\ell^2}} + \redOofo r^{-n})
	\label{8II15_5}
	\,.
\end{eqnarray}

Inserting all the results into \eq{toto1_new} we finally find the following expression for the \emph{Hamiltonian mass} for asymptotically HM metrics, where we have used the fact that some terms integrate out to zero:
\begin{eqnarray}
	\label{MHamHM}
\lefteqn{
	\mham^\HM
	=
	\lim_{R\to\infty} \frac {r^n} {16\pi\ell^2}
	\int_{\hyp \cap\{r=R\}}
		\left[
    \frac{  e_\HM^{{\hat n}{\hat n}} - e_\HM^{\one\one} }{\sqrt{ \frac{\beta\ell^2}{r^2} + 1 }}
 \right.
 }
 &&
		\nonumber
\\
		&&
		\left.
		+
		\sqrt{1+\frac{\beta\ell^2} {r^2}} \left(
 \frac { r\partial \big(b_{{\hat A}{\hat B}}e_\HM^{{\hat A}{\hat B}}+ e_\HM^{{\hat n}{\hat n}}\big) }{\partial r}
		  - (n-2)e_\HM^{\one\one} -e_\HM^{{\hat n}{\hat n}} \right)
 \right] d^{n-1}\mu_{{\hat h}}
	\,,
	\phantom{xxxx}
\end{eqnarray}
where $d^{n-1}\mu_{{\hat h}}$ is the measure element associated with the metric \eq{24II15.11}.
In space-time dimension $n+1=4$ this coincides formally with \eq{MHamBh3}
\begin{equation}
	\label{MHamHM3}
	\mham^\HM  =
	\lim_{R\to\infty} \frac {r^3} {16\pi\ell^2}
  \times
	\int_{\hyp \cap\{r=R\}}
		\left[
		\frac{ r \partial ( e_\HM^{{\two}{\two}}  + e_\HM^{{\three}{\three}} ) }{\partial r}
		-
		2 e_\HM^{\one\one}
   \right] d^{n-1}\mu_{{\hat h}}
	\,.
\end{equation}

As an example, if $\fullgg$ is the $3+1$-dimensional Horowitz-Myers metric,  we find
\begin{equation}
  \label{hammassHM+}
  \mham^\HM = - \frac{\mu_{\hat h}({\sectionofScri}) m}{8\pi}\,,
\end{equation}
where
\begin{equation}
\label{areanorm+}
	\mu_{\hat h}({\sectionofScri}) := \lim_{R\to\infty} \int_{\hyp \cap\{r=R\}} d^{n-1}\mu_{{\hat h}}
	\,.
\end{equation}
This coincides with what we found for the coordinate mass of Horowitz--Myers metrics, where however we had to restrict ourselves to the case $\beta=0$. We see that no such restriction arises for the Hamiltonian mass.

\subsection{Fefferman-Graham asymptotics with an  ultrastatic conformal infinity}
 \label{ss23II16.1}

In this section we
assume that $n=3$ and $\Lambda<0$, unless explicitly indicated otherwise. We consider a vacuum space-time with a smooth conformal completion, thus both the background metric $b$ and $g$ have  a Fefferman-Graham  expansion as in \eq{10I15.3}  in a suitable coordinate system such that $x=0$ at $\scri$:
\bean
 b
 &=& x^{-2}\ell^2\big( dx^2+ \bbh_{ab}(x,x^c)dx^ a dx^ b\big)
 \,,
\\
 g
 &=& x^{-2}\ell^2\big(  dx^2+ \hhere_{ab}(x,x^c)dx^ a dx^ b\big)
 \,,
\eeal{10I15.3+}
where 
$(x^a)=(t,x^A)$, and
with the coordinate components $\bbh_{ab}  $ asymptotic to $\hhere_{ab} $ as $\redOof x^3)$. Here we have used the same compactifying factor
$$
 \Omega = x/\ell
$$
to pass from $g$ to $\tilde g=\Omega^2 g$ as from $b$ to $\tilde b= \Omega^2 b$.

For simplicity we will assume an \emph{ultrastatic} form of the conformal-boundary metric
\bel{14III16.3}
 \mathring \bbh\equiv \mathring \bbh_{ab} \, dx^a dx^b :=  \bbh_{ab}|_{x=0}\, dx^a dx^b = \hhere_{ab}|_{x=0}\, dx^ a dx^ b
 =: \mathring \hhere_{ab} \, dx^a dx^b
 \equiv \mathring \hhere
  \,,
\ee
namely
\bel{24II16.11}
\bbh_{0A}(0,x^c)=0
 \,,
 \qquad
 \partial_a \bbh_{00}(0,x^c)=0\,,
 \qquad
 \partial_0 \bbh_{AB}(0,x^c)=0
\ee
(compare the discussion after \eq{31VII15.3}).
Note that this is compatible both with asymptotically Birmingham and asymptotically Horowitz-Myers metrics.
More general metrics and sections of $\scri$ will be considered in future work.
 \ptctodo{to be done, in all dimensions}

Let $(\bbh_{ab})_n$ denote the coefficient of $x^n$ in a Taylor expansion of $\bbh_{ab}$ at $x=0$, similarly for $(\hhere_{ab})_n$. (The reader is warned that these coefficients do not translate \emph{as such} to expansion coefficients in e.g.\ Bondi coordinates, as $r_\Bo \ne 1/x$ in general even if $\ell=1$, see \eq{18III16.1} below.) It follows from Section~\ref{s14III16.1} that
\bel{14III16.4}
 (\bbh_{ab})_n= (\hhere_{ab})_n\,,\ n\in\{0,1,2\}
 \
 \mbox{and}\
 (\bbh_{ab})_1=0
 \,.
\ee
In the calculations below we will assume that
\bel{14III16.5}
 (\bbh_{ab})_3=0
 \,.
\ee
If this is not the case, in all the formulae below it suffices to replace $(\hhere_{ab})_3$ by $(\hhere_{ab})_3-(\bbh_{ab})_3$.

We wish to determine the \charmass\ of a null hypersurface asymptotic to a section of $\scri$ with constant $x^0$,
 \ptctodo{should be generalised}
and compare it with the Hamiltonian mass $\mham$.
Without loss of generality, after choosing a conformal gauge appropriately, we can assume that $\bbh_{AB}|_{x=0} dx^A dx^B$ has constant scalar curvature $\beta$. It then follows from \eq{11I15.4} that
\bel{14III16.7}
  (\bbh_{0 A})_2=0
  \quad
   \Longleftrightarrow
   \quad
  \bbh_{0 A} = \redOof x^3)
  \,.
\ee

We pass now to the calculation of the Hamiltonian mass of $g$. Using $e^{11}=0=e^{\hat 1 \hat 1}$ and the first line of \eq{8II15_2} (which applies here), we find quite generally in dimension $n$, without assuming \eq{24II16.11},
\be
	\frac{{ 16\pi \ourU^\Bh{}^{\one\zero}}_{\zero}}
{\sqrt{|\det b_{\alpha\beta}|}}
 =
 e_{\hat 1} (b_{\hat A \hat B} e^{\hA\hB})
  + (2 \omega^{\hat 0}{}_{\hat A \hat 1}  - \omega^{\hat 1}{}_{\hat 0 \hat A}) e^{\hat 0 \hat A}
   - \omega^{\hat 1}{}_{\hat A \hat B}  e^{\hat A\hat B}
   +\redOofo x^n)
	\,.
	\label{12III16.2}
\ee
 \ptctodo{general frame commented out}
%
Returning to the three-dimensional ultrastatic case, we choose the $b$-orthonormal frame $e_{\hat \alpha}$ as
\bel{12III16.5}
 e_{\hat 1} = \frac x \ell \partial_x
 \,,
 \quad
 e_{\hat 0} = \frac x {\ell \sqrt{|\bbh_{00}|}} \partial_0
 \,,
	\quad
  e_{\hA} = \frac  x \ell \big(\betatwo _{\hA} + \redOof x^2) \partial_B
  +  \redOof x^3)\partial_0
  	\big)
  \,,
\ee
where $\{\betatwo_{\hA}\}_{
\hA =2,\ldots, n}$ is an ON frame for $\mathring \bbh_{AB}dx^A dx^B$.
Let us denote by $\theta^\hmu$ the coframe dual to $e_\hnu$, then
 \ptcheck{13 and 14 III 16}
\bel{12III16.6}
 \theta^{\hat 1} = \frac  \ell x dx
 \,,
 \quad
\theta^{\hat 0} = \frac   {\ell \sqrt{|\bbh_{00}|}} x dx^0  + \redOof x^2) dx^ A
 \,,
	\quad
  \theta^{\hA} = \frac   \ell x
  	\big(\chitwo^{\hA} + \redOof x^2) dx^ a\big)
  \,,
\ee
where $\chitwo^{\hA}$ is a coframe dual to $\betatwo _{\hA}$. The components $g_{\hmu\hnu}$ of the metric $g$ with respect to this frame read
 \ptcheck{13 and 14 III 16} 
\beal{13III16.1}
 &
 \displaystyle
  g_{\hat 1 \hat 1}  = 1
   \,, \quad
  g_{\hat 1 \hat a}  = 0
   \,, \quad
    g_{\hat 0 \hat 0} 
     = - 1+ \frac{(\hhere_{00})_3}{|\bbh_{00}|} x^3 + o(x^3)
     \,,
     &
\\
 &
 \displaystyle
  g_{\hat 0 \hat A} =  \frac{(\hhere_{0\hat A})_3}{\sqrt{|\bbh_{00}|}} x^3 + o(x^3)
   \,, \quad
    g_{\hat A \hat B} = b_{\hat A \hat B} + {(\hhere_{\hat A \hat B})_3} x^3 + o(x^3)
     \,,
     &
\eea
where $(\hhere_{0\hat A})_3$ denotes the $\chitwo$-component of $(\hhere_{0  A})_3dx^A$, as defined through the formula
$$
 \mbox{ $(\hhere_{0  A})_3dx^A = (\hhere_{0 \hat A})_3\chitwo^\hA$; similarly  
  $ (\hhere_{  A B})_3 dx^A dx^B = (\hhere_{\hat A \hat B})_3 \chitwo^\hA \chitwo^\hB $.}
$$
This leads to
 \ptcheck{13 and 14 III 16}
\beal{13III16.1+}
 &
 e^{\hat A \hat B}
 = - b^{\hat A \hat C}b^{\hat B \hat D}(\hhere_{\hat C \hat D})_3 x^3 + o(x^3)
 = - (\hhere_{\hat A \hat B})_3 x^3 + o(x^3)
 \,,
  &
\\
 &
  \displaystyle
 e^{\hat 1 \hat \mu} =0
 \,, \quad
 e^{\hat 0 \hat 0} =
  - \frac{ (\hhere_{00})_3 }{|\bbh_{00}|}x^3 + o(x^3)
 \,, \quad
 e^{\hat 0 \hat A} =
   \frac{ (\hhere_{0\hat A})_3 }{\sqrt{|\bbh_{00}|}}x^3 + o(x^3)
 \,,
\eea
where, of course, $b_{\hA\hB} = b^{\hA\hB}=\delta^\hA_\hB$.
Note that the condition $e=1+\redOofo x^3)$, which is equivalent to the $\mathring \bbh_{ab}$-tracelessness of $(\hhere _{ab})_3$, reads
\bel{13III16.5}
 - \frac{ (\hhere_{00})_3 }{|\bbh_{00}|} + b^{\hA\hB}  (\hhere_{\hA\hB})_3 \equiv
 - \frac{ (\hhere_{00})_3 }{|\bbh_{00}|} + \mathring\bbh^{A B }  (\hhere_{A B})_3 =0
  \,.
\ee
Setting $\mathring\bbh_{AB}:= \bbh_{AB}|_{x=0}$, and using $\omega^{\hat \alpha}{}_{\hmu\hnu} = \theta^{\hat \alpha}(e_{\hmu;\hnu})$, we find
\ptcheck{13 III 16}
\beal{12III16.3}
 &
 \Gamma^1_{AB} = \frac {\mathring\bbh_{AB}}x + \redOof x^2)
  \,,
 \
 \omega^{\hat 1} {}_{\hat A \hat B} =
   \frac 1  \ell{b_{\hA \hB}}  + \redOof x^3 )
   \,,
 \
 \omega^{\hat 0} {}_{\hat A \hat 1} =    \redOof x^2 )
   \,,
  &
\\
 &
 \omega^{\hat A} {}_{\hat 0 \hat A} =    \redOof x^4 )
   \,,
   \quad
 \omega^{\hat 1} {}_{\hat 0 \hat A} =    \redOof x^4 )
   \,.
    &
\eeal{12III16.9}
This leads to the following rewriting of \eqref{12III16.2}:
 \ptcheck{13 III 16}
\begin{eqnarray}
	\frac{{ 16\pi \ourU^\Bh{}^{\one\zero}}_{\zero}}
	{\sqrt{|\det b_{\alpha\beta}|}}
 &=&
 \nn
   \frac x \ell \partial_x ( b_{\hat A \hat B}  e^{\hat A \hat B})
   -  \frac 1 \ell  b_{\hat A \hat B}  e^{\hat A\hat B} + \redOofo x^3)
\\
 &=&
 - \frac 2  \ell b^{\hat A \hat B}  (\hhere_{\hat A \hat B}) _3 x^3 + \redOofo x^3)
	\,.
   	\label{12III16.7}
\end{eqnarray}
Next, we choose $X$ to be $\partial_0$, so that
\ptcheck{16 III 16} 
\bea
 &
  \displaystyle
 X= \partial_0 = \frac{\ell \sqrt{|\bbh_{00}|}}x  e_{\hat 0} \equiv  X^{ \hat 0} e_{\hat 0}
 \,,\quad
 X^{\hat 0}{}_{;\hat 1} = - \frac { \sqrt{|\bbh_{00}|}}{\ell x} + \redOof x ) = X^{\hat 1}{}_{;\hat0}
 \,,\
  &
\\
 &
  \displaystyle
 X^{\hat 1}{}_{;\hat a} = \redOof x^3)
 \,,\quad
 X^{\hat 0}{}_{;\hat a} = \redOof x^2)
 \,,
  \quad
 \Delta^{\hat 1 \hat 1} =  \redOofo x^3)
  \,,
  \quad
 \Delta^{\hat 1 \hat a} =   0
 \,,
  &
\\
 &
  \displaystyle
 \Delta^{\hat 0 \hat 0} =  e^{\hat 0 \hat 0} + \redOofo x^3)
  \,,
  \quad
\Delta^{\hat 0 \hat A} =   e^{\hat 0 \hat A} +\redOofo x^3)
 \,,
  &
\\
 &
  \displaystyle
	\frac1{8\pi} \Delta_\Bh^{\hat \alpha[\one}  X_\Bh^{\zero]}{}_{;\hat \alpha}
	 = \frac { \sqrt{|\bbh_{00}|}} {16\pi \ell x}  e^{\hat 0 \hat 0}    + \redOofo x^{2})
	 = - \frac { \sqrt{|\bbh_{00}|}} {16\pi } \times \frac{ (\hhere_{0 0} )_3} {|\bbh_{00}|\ell } x^2  + \redOofo x^{2})
	\label{12III16.3+}
	\,.
	&
\end{eqnarray}
Hence, using \eq{toto_new}-\eq{Freud2.0_new}, for any hypersurface $\hyp$ intersecting $\scri$
in a section $\{x^0=\const\}$, after taking into account an overall minus sign because of the change of orientation when replacing $r$ by $x=\ell/r + ...$,
\bean
 \mham
   &= & - \lim_{x\to 0} \frac{\ell^2}{x^2}\int U^{\hat 1 \hat 0} \sqrt{\det \mathring\bbh_{AB} } d^2x
\\
 & = &
 \frac{\ell{\sqrt{|\bbh_{00}|}}}{16 \pi }\int
  \bigg(  2 b^{\hat A \hat B}  (\hhere_{\hat A \hat B}) _3  +  \frac{ (\hhere_{0 0} )_3} {|\bbh_{00}|} \bigg)  \sqrt{\det \mathring\bbh_{AB} } d^2x
   \nn
\\
 & = &
 \frac{3\ell {\sqrt{|\bbh_{00}|}}}{16\pi }\int
     b^{\hat A \hat B}  (\hhere_{\hat A \hat B}) _3     \sqrt{\det \mathring\bbh_{AB} } d^2x
 \,.
\eeal{12III16.11}
It is clearly convenient to normalise the asymptotic time coordinate $x^0$ so that
 (compare \eq{18II16.1})
\bel{25II16.1}
 \ringlambda_{00}  \equiv  \bbh_{00}|_{x=0}=-\ell^{-2}
 \quad
  \Longleftrightarrow
  \quad
    \ringlambda^{00}  =-\ell ^2
 \,,
\ee
leading  finally to
\bea
 \mham
 & = &
 \frac{3 }{16\pi }\int
     b^{\hat A \hat B}  (\hhere_{\hat A \hat B}) _3     \sqrt{\det \mathring\bbh_{AB} } d^2x
 \,.
\eeal{14III16.1}
Note that this coincides formally with both \eqref{MHamBhstrong } and \eqref{MHamHM}, but it was not a priori clear to us that it should.

We wish to compare \eq{12III16.11} with the characteristic mass as defined by \eq{eq:mTB}-\eq{eq:massaspectdef}.  For this, we need to determine the mass aspect function $M$ of \eqref{eq:massaspectdef}.
If the zero-level set  of $u_\Bo$ is asymptotic to the zero-level set of $t$, an asymptotic expansion of the solutions of the equations which determine the Bondi coordinates shows that
$$
  t= u_\Bo  -\ell x  - \frac1{6\ell^{3}}(\hhere_{00})_2 x^3 + \redOof x^4)
   \,,
   \quad
   x^A = x^A_\Bo + \frac 13  \ell (\bbh_{0A})_2 x^3 + \redOof x^4)
   \,.
$$
The above solution is obtained after imposing the condition that $u$ is a retarded null coordinate, hence $t$ is an increasing function of $r$ at fixed $u$, hence decreasing in $x$ at fixed $u$.

Changing coordinates, one finds
\bel{25II16.5}
 g_{u_\Bo u_\Bo} =   x^{-2}\big(-1 + \ell^2 (\hhere_{00})_2 x^2  + \ell^2 (\hhere_{00})_3 x^3 + \redOof x^4)
 \big)
  \,.
\ee
It remains to replace $x$ by a Bondi coordinate, defined through the formula
$$
 r_{\Bobo}= \left(\frac{\det g_{AB}}{\det \mathring \bbh_{AB}}\right)^{1/4}
  \,.
$$
A {\sc Mathematica} calculation gives
%
\bel{18III16.1}
 \frac 1x=\frac{r_{\Bobo}}{\ell}
  -\frac{\ell     \mathring \bbh^{AB}(\hhere_{AB})_2}{   4    r_{\Bobo}}
 -\frac{\ell^3
     \mathring \bbh^{AB}(\hhere_{AB})_3 }{4
   r_{\Bobo}^2}
    + \redOof r_{\Bobo}^{-3})
 \,.
\ee
Inserting into \eq{25II16.5}, one finds
that the mass aspect is
\bel{25II16.4}
   M\equiv \frac{( g_{u_\Bo u_\Bo})_1} 2 = \frac 12 \Big( \ell^3 (\hhere_{00})_3
   +\frac 1 2  b^{\hA\hB} (\hhere_{\hA\hB})_3 \Big)
    = \frac {3 \ell} 4   b^{\hA\hB} (\hhere_{\hA\hB})_3
   \,,
\ee
where we have used \eqref{13III16.5}. Comparing \eq {14III16.1} and \eq{eq:mTB}, we conclude that
$$
 \mham = \mTB
 \,,
$$
as desired.

\ptctodo{file FGforBandHM.tex contains a calculation of the FG expansions for B and HM metrics}

 \ptctodo{Iin the section explicit there is a calculation that a FG expansion implies that $R/x$ is in $L^1$ on slices of constant time, can be used to something later but is irrelevant here}

\section{Conclusions}
 \label{s16II16.1}

We have introduced a natural notion of total mass for characteristic hypersurfaces in space-times with non-vanishing cosmological constant. The mass is a natural generalisation of the Trautman-Bondi mass, as defined for $\Lambda=0$. We have proved a generalisation of the positivity identity of~\cite{ChPaetzBondi}. The identity introduces the renormalised volume as a new global quantity associated to characteristic initial data sets. In the simplest case of light-cones in vacuum this is the identity~\eqref{eq:massfinalConeVac}, which we rewrite as
\bean
	\lefteqn{
 \mTB
 - \frac \Lambda{8\pi}\left(  \frac{1}{12}\int_{\sectionofscri}{
    \tau_2\bigg(\frac{ \tau_2^2}{2 }  - \frac{|\sigma|_4^2}{3}\bigg) } \hringmeasure
    + 	{ V_\renm}
 \right)
 }
 &&
	\nonumber   \\
	&&
 =
	\frac{1}{16\pi}	\int_0^{\infty} \int_{\sectionofscri} \bigg( \frac{1}{2}
    |\xi|^2
	+
	 |\sigma|^2  \text{e}^{ \int_r^{\infty} \frac{\tilde{r}\tau-2}{2\tilde{r}} d\tilde{r} } \bigg) d\mu_{\coneg} dr
	\,.
\eeal{16II16.1}
The left-hand side involves the renormalised volume together with objects which can be determined by looking at the asymptotic behaviour of the fields.
This provides a new global positivity statment, proving indeed that the left-hand side of \eq{16II16.1} is positive. It follows from \cite{CCG} that the left-hand side vanishes if and only if the space-time is de Sitter or anti-de Sitter to the future of the light-cone.

The balance formula \eq{16II16.1} raises the question of the right definition of mass when $\Lambda \ne 0$. Recall that we used \eqs{eq:mTB}{eq:massaspectdef} to  define $\mTB$:
\bel{16II16.2}
	 \mTB = \frac{1}{8\pi} \int_{\sectionofScri }    \left(\ol g_{00}^\Bo \right)_1\, \hringmeasure
	\,.
\ee
A first naive idea would be to define instead the left-hand side of \eq{16II16.1} as the mass, obtaining positivity as a corollary of \eq{16II16.1}. But the calculations in
Appendix~\ref{s15II16.1} strongly suggest that a splitting of the left-hand side of  \eq{16II16.1} in a renormalised-volume contribution and a mass contribution is meaningful.

The next idea would be to define the characteristic mass as
\bel{18III16.4}
 m_c =
 \mTB
     - \frac \Lambda{8\pi}\left(  \frac{1}{12}\int_{\sectionofscri}{
    \tau_2\bigg(\frac{ \tau_2^2}{2 }  - \frac{|\sigma|_4^2}{3}\bigg) } \hringmeasure
 \right)   \,,
\ee
leading to the more elegant identity:
\bea
 m_c
 - \frac \Lambda{8\pi} 	{ {V}_\renm}
 =
	\frac{1}{16\pi}	\int_0^{\infty} \int_{\sectionofscri} \bigg( \frac{1}{2}
    |\xi|^2
	+
	 |\sigma|^2  \text{e}^{ \int_r^{\infty} \frac{\tilde{r}\tau-2}{2\tilde{r}} d\tilde{r} } \bigg) d\mu_{\coneg} dr
	\,.
\eeal{18III16.5}

Alternatively, one could add an integral expression involving $\tau_2$ and $|\sigma|_4$ to the definition of $\mTB$, adjusting  \eq{16II16.1} accordingly. Recall that \eq{16II16.2} is equivalent to \eq{eq:solamssbondi}, which for a smooth conformal completion reads
\begin{equation}
\label{16II16.3}
	\mTB = \frac{1}{16\pi}\int_{\sectionofScri} (\zeta^\Bo)_2 \, \hringmeasure
	\,.
\end{equation}
In the asymptotically flat case and with  spherical cross-sections of $\scri$, the gauge-invariant version of this formula is~\cite{ChPaetzBondi}
\begin{equation}
\label{16II16.4}
	 \frac{1}{16\pi}\int_{\sectionofScri} \big(\zeta_2 + \tau_2 \big)\, \hringmeasure
	\,,
\end{equation}
and one could use this formula as a definition of characteristic mass. (Whether or not, and in which sense, this is gauge-invariant when $\Lambda \ne 0$ remains to be seen).
\ptctodo{should be done...}
Recall that we have seen (cf.\ \eqref{eq:solmassbondicharecteristic} in vacuum and with a smooth conformal completion) that \eq{16II16.3} translates instead into
\begin{equation}
\label{16II16.7}
	\mTB = \frac{1}{16\pi}\int_{\sectionofScri} \big( \zeta_2 + \tau_2 \big) \, \hringmeasure
	+ \frac{\Lambda}{16\pi} \int_{\sectionofScri}   \tau_2  |\sigma|^2_4   \, \hringmeasure
 \phantom{xxxx}
\end{equation}
(note that the multiplicative factor
${\zR}/{2}$ in front of $\tau_2$ in our formula equals $1$  for a sphere),
when  $\Lambda\ne 0$ and an affine parameter $r$ is used.

As seen in Appendix~\ref{s15II16.1}, we have $\tau_2\ne 0$ for asymptotically Horowitz-Myers metrics, which suggests strongly
that using $\tau_2^3$-terms to redefine the mass is not a good idea. Whether or not adding some $\tau_2|\sigma|^2_4$-terms is meaningful requires
further analysis. We plan to return to this question in the future.
\ptctodo{In all the examples we have analysed it holds that $\tau_2=0$, so none can be used to decide which one out of $m_c$, $
\mTB$ or \eq{16II16.4} is more relevant. We plan to return to this issue in future work. (this is wrong, HM has tau2 nonvanishing,
so clearly the tau2 power three term should rather be included in the volume and not in the mass)}

Yet another alternative is to define the renormalized volume as the whole expression in brackets at the left-hand side of \eq{16II16.1},
\bel{16OO16/5}
  V_\renm\mapsto
    	\widetilde{ V}{_\renm}
    :=	{ V_\renm}
    +
    \frac{1}{12}\int_{\sectionofscri}{
    \tau_2\bigg(\frac{ \tau_2^2}{2 }  - \frac{|\sigma|_4^2}{3}\bigg) } \hringmeasure
    \,,
\ee
leading similarly to a nicer identity:
\bea
 \mTB
 - \frac \Lambda{8\pi} 	{ \widetilde{V}_\renm}
 =
	\frac{1}{16\pi}	\int_0^{\infty} \int_{\sectionofscri} \bigg( \frac{1}{2}
    |\xi|^2
	+
	 |\sigma|^2  \text{e}^{ \int_r^{\infty} \frac{\tilde{r}\tau-2}{2\tilde{r}} d\tilde{r} } \bigg) d\mu_{\coneg} dr
	\,.
\eeal{16II16.6}

Possibly, a mixture of the above will provide the most meaningful definitions.

Incidentally, $\widetilde{ V}{_\renm}$ can be obtained by replacing $r$ in the original definition of $V(r)$, where $r$ is the affine coordinate normalised as before, by
\bel{23II16.1}
 r =\tilde r + \frac {\tau_2} 2 + \frac a {\tilde r^2} - \frac {2 \tau_2 |\sigma|^2_4}{9\tilde r^2}
 \,,
\ee
where $a$ is any function of the angular coordinates.
In other words, set
$$
 \widetilde V (\tilde r):= V\big(\tilde r + \frac {\tau_2} 2 + \frac a {\tilde r^2} - \frac {2 \tau_2 |\sigma|^2_4}{9\tilde r^2}\big)
  \,.
$$
Then  $\widetilde{ V}{_\renm}$ is the limit, as $\tilde r$ goes to infinity, of $\widetilde{ V} (\tilde r)$ minus the sum of the terms with positive powers of $\tilde r$ and the $\ln \tilde r$ term. In fact, a change of variables of the form
%
\bel{29III16.1}
 r =\tilde r + \frac {\tau_2} 2 + \frac {|\sigma|^2_4} {2\tilde r}
  - \frac {2 \tau_2 |\sigma|^2_4}{9\tilde r^2}
  + \frac { \tau_2 |\sigma|^2_4}{3\tilde r^2}\ln \tilde r + o(\tilde r^{-2})
\ee
gives
\bel{29III16.2}
 V(r) = \frac {4 \pi}{3}\tilde r^3 + V_\renm + o (\tilde r^{-1})
 \,.
\ee

One wonders about the nature of \eqref{29III16.1}. The naive guess would be that $\tilde r$ is  the Bondi coordinate. However, in our case we have (in vacuum, but allowing $|\sigma|_5\ne 0$)
\bel{23II16.I}
	r_{\Bobo} = r - \frac{\tau_2}{2} + \frac{|\sigma|^2_4}{4r} + \frac{1}{6} \frac{|\sigma|^2_5 - \frac{1}{4}|\sigma|^2_4 \tau_2}{r^2} + O({r^{-3}})
	\,,
\ee
with inverse transformation
\bel{23II16.IV}
	r = r_{\Bobo} + \frac{\tau_2}{2} - \frac{|\sigma|^2_4}{4
   r_{\Bobo}} + \frac{|\sigma|^2_4 \tau_2 - |\sigma|^2_5}{6r_{\Bobo}^2} + O({r_{\Bobo}^{-3}})
	\,.
\ee
We see that $\tilde r $ coincides with the Bondi coordinate  at  order zero, but differs at the next order.

Which definition is most relevant, or indeed whether there exists a most relevant definition at all, requires further studies.

In any case, we have shown in some well understood general cases with $\Lambda \le 0$, as well as on specific examples, that the characteristic mass $\mTB$ defined by \eq{16II16.7} coincides with previously accepted definitions of mass.

We emphasise that both the definition of mass and the balance formula \eq{16II16.1} have a clear, geometric and gauge-independent, meaning; compare Remark~\ref{R12II16.1}.

\appendix
\section{Null geometry of Horowitz-Myers metrics and the balance equation}
 \label{s15II16.1}

The mass identity \eqref{eq:massfinal} can be viewed as a balance formula. It is instructive to work-out the contribution of each of the terms appearing there to the total mass for the Horowitz-Myers metrics. For this we need to derive the asymptotics both for small and large $r$ of the fields appearing there. We consider the metric \eq{6XI12.4x} with $f$ given by \eq{6XI12.5x} and $\beta=0$ in space-dimension equal to three.

Choosing $s(r_0)=0$, where $r_0$ is the largest zero of $f$, from \eq{10VII14.2} we obtain a small-$r$ expansion, for $r\ge r_0$:
\bel{10VII14.3}
 s = \frac{2 r_0}{\ell \sqrt{f'(r_0)}}\sqrt{r-r_0} +O\big((r-r_0)^{\frac 32}\big)
 \ \mbox{ for $r-r_0>0$ small}
 \,,
\ee
where we have assumed that $f'(r_0)\ne 0$.
This implies, for small $s$,
\beal{10VII14.4}
 r-r_0 & = & \frac{\ell^2 f'(r_0)}{4 r_0^2} s^2 + \redOof s^4)
  \,,
\\
 f(r)
  & = &
     \frac{\big(\ell f'(r_0)\big)^2}{4 r_0^2} s^2 + \redOof s^4)
 \,.
\eeal{10VII14.5}

Recall that the large-$s$ behaviour of $r$ has been derived in \eq{10VII14.6c}-\eq{29I15.12}.

\newcommand{\localchg}{\check g}%
As in Section~\ref{s24XII15.1}, we denote by ${\localchg}$ the metric induced by $g$ on the level sets of $u$ and $r$:
\bea
{\localchg} &= &  f(r) \ell^2  \lambda^2 d\chicoordinate^2 + r^2 d\varphi^2
 \,.
\eea
Let $x^A$ denote the coordinates $\chicoordinate$ and $\varphi$.
In the affine parameterisation and in the region where $f$ is non-negative it holds that%
\footnote{We use the notation of \cite{CCM2},  except that we denote here by $\localchg$ the tensor field denoted by $\tilde g$ there.}
 \ptclevoca{all these formulae are perfectly unneeded in a set of lecture notes about black holes, relevant for the energy notes perhaps, though}
\bea
 \label{13XII14.12}
 \sqrt {\det \localchg_{AB}}
  & = &
  r \ell \lambda \sqrt {f(r)}
 = \lambda \frac {\ell^2 f'(r_0)}2 (s+\redOof s^3))
  \quad \mbox{for small $s$}
  \,,
\\
 \nu_A  & = & 0
  \,,
\\
 \xi_A  & = & 0
  \,,
\\
 \nonumber
 \chitensor  &:= &
 \frac 12 \partial_s{\localchg}
  =  \frac {\ell \sqrt {f(r)}}{2r}  \partial_r \big(f(r) \lambda ^2 \ell^2  d\chicoordinate^2 + r^2 d\varphi^2\big)
\\
 & = &    {\ell \sqrt {f(r)}}  \left(\frac{\partial_r f(r)}{2r} \lambda ^2 \ell^2  d\chicoordinate^2 +  d\varphi^2\right)
\\
 & \approx &    \frac{\ell^2 {f'(r_0)}s +\redOof s^3)}{2r_0}  \left(\frac{ f'(r_0)}{2r_0} \lambda ^2 \ell^2  d\chicoordinate^2 +  d\varphi^2\right)
  \ \mbox{for small $s$}
  \,,
   \phantom{xxx}
\\
 \tau
  & := &
   g^{AB}\chitensor _{AB}
    = {\ell}  \left(\frac{\partial_r f(r)}{2 r \sqrt {f(r)}}  +  r^{-2} \sqrt {f(r)}\right)
\\
 & = & \left\{
         \begin{array}{ll}
 \displaystyle
  s^{-1} +\redOof s)
      , & \hbox{\mbox{for small $s$} ;} \\
       \displaystyle    2 s^{-1} + 2(s_*-r_0) s^{-2} + \redOof s^{-3}), & \hbox{ \mbox{for large $s$}, }
 \end{array}
       \right.
\label{22IV15.1}
\eea
with $s_* = (2m\ell)^{1/3} x_*$, where $x_*$ is given by \eqref{29I15.14}. Further
\bean
 \nonumber
\sigma
 & : = &
 \chitensor  - \frac 12 \tau \, g_{AB}dx^A dx^B= {\ell \sqrt {f(r)}}  \left(\frac{\partial_r f(r)}{2r} \lambda ^2 \ell^2  d\chicoordinate^2 +  d\varphi^2\right)
 \\
   & &
- \frac 12
 {\ell}  \left(\frac{\partial_r f(r)}{2 r \sqrt {f(r)}}  +  r^{-2} \sqrt {f(r)}\right)
 (f \lambda ^2 \ell^2  d\chicoordinate^2 + r^2 d\varphi^2)
 \\
   &= &
 \frac {\ell} 2
  \left(\frac{\partial_r f(r)}{2 r \sqrt {f(r)}}  -  r^{-2} \sqrt {f(r)}\right)
 \left( f \lambda ^2 \ell^2  d\chicoordinate^2 -r^2
   d\varphi^2
 \right)
 \,,
 \\
 \nonumber
 \sigma^\sharp  &:= & g^{AC}\sigma_{CB} \partial_A \otimes dx^ B
\\
 & = &
 \frac {\ell} 2
  \left(\frac{\partial_r f(r)}{2 r \sqrt {f(r)}}  -  r^{-2} \sqrt {f(r)}\right)
        (\partial_\chicoordinate\otimes d\chicoordinate- \partial_\varphi \otimes d\varphi)
\\
 & = &\left\{
              \begin{array}{ll}
 \frac {(1+\redOof s^2))} {2s}
        (\partial_\chicoordinate\otimes d\chicoordinate- \partial_\varphi \otimes d\varphi), & \hbox{for small $s$;} \\
 \frac {(3 \ell^2 m +\redOof s^{-1}))} {2s^4}
        (\partial_\chicoordinate\otimes d\chicoordinate- \partial_\varphi \otimes d\varphi), & \hbox{for large $s$,}
              \end{array}
            \right.
\\
 |\sigma^2|
& = &
 2 \times
 \frac {\ell^2} 4
  \left(\frac{\partial_r f(r)}{2 r \sqrt {f(r)}}  -  r^{-2} \sqrt {f(r)}\right) ^2
\\
 & = &\left\{
              \begin{array}{ll}
 \frac {(1+\redOof s^2))} {2s^2} , & \hbox{for small $s$;} \\
 \frac {(9 \ell^4 m^2 +\redOof s^{-1}))} {2s^8}, & \hbox{for large $s$,}
              \end{array}
            \right.
  \label{16XII14.1}
\\
 g^{ss} & = & g^{\mu\nu}\frac{\partial s}{\partial x^\mu}
 \frac{\partial s}{\partial x^\nu}
=  g^{rr}\left( \frac{ds}{dr}\right)^2 =f \left( \frac{r}{\ell \sqrt f }\right)^2 =  \frac{r^2}{\ell^2}
 \,,
\eea
%
%
\newcommand{\localchecknabla}{\check\nabla}
We will also need the following objects from \cite{CCM2}, denoting by
$\localchecknabla$ the derivative operator associated with   $\localchg $, when $\localchg$ is viewed as a metric on the level sets of $u$ and $r$:
\begin{eqnarray}
%
 \nonumber
\overline{\Gamma}^{1}
&:=&
  \nu^{0}  \overline{g}^{AB}\localchecknabla _ B \nu_{A}
- \frac{\partial_{1} (\nu_{0}\overline{g}^{11}\sqrt{\det\localchg })}{\nu_{0}\sqrt{\det\localchg }}
- \frac{1}{2} \nu^{0} \overline{g}^{AB} \overline{\partial_{0} g_{AB}}
+ \frac{1}{2} \nu^{0} \overline{g}^{11} \overline{\partial_{0} g_{11}}
 \,,
\\
 \nu_0
 & := & g_{us} =g_{ur} \frac{dr}{ds}=-   \frac {r}  {\ell \sqrt{f(r)}}  \frac{dr}{ds}
  =   -   \frac {r}  {\ell \sqrt{f(r)}} \frac {\ell \sqrt{f(r)}} {r} = -1
  \,,
\\
 \nu^0
 & := &   \frac {1}  {\nu_0}= -1
  \,,
\\
\overline{\Gamma}^{s}
 & =
 &
- \frac{\partial_{s} (\nu_{0}\overline{g}^{ss}\sqrt{\det\localchg })}{\nu_{0}\sqrt{\det\localchg }}
 =
 -\tau \ol g^{ss} - {\partial_{s} \overline{g}^{ss} }
\; .
\eea
From the definition of $\zeta$  (compare~\cite[Equations~(10.33) and (10.36)]{CCM2},%
\footnote{Note a missing term $\tau \ol g^{11}/2$ in the rightermost term of~\cite[Equation~(10.36)]{CCM2}, which however does not affect the formula we use.}
and note that $\kappa\equiv\ol \Gamma^s_{ss}\equiv \ol \Gamma^1_{11}$ vanishes in affine parameterisation) we have
%
\bea
 \zeta & := & (2 \partial_s +2\ol \Gamma^s_{ss}+ \tau) {\overline g}{}^{ss} + 2 \overline \Gamma^s
 =  - \tau  {\overline g}{}^{ss}
 \nonumber
\\
 \label{15II16.1}
 &
 =
 &  - \ell^{-1} \left(\frac{r \partial_r f(r)}{2  \sqrt {f(r)}}  +   \sqrt {f(r)}\right)
\\
 & = &    \left\{
         \begin{array}{ll}
 - {r_0^2}{\ell^{ -2}s^{-1}}  +\redOof s)
 \quad
      , &
\hbox{\mbox{for small $s$};} \\
 - {2
   r}{\ell ^{-2}}- {m}{r^{-2}}+O
   \left( {r}^{-5}\right)
&
\\
\phantom{xx}
    = -{ 2 (s+ r_0-s_*)}{\ell^{-2}} - {2m}{s^{-2}}+O
    \left( {s}^{-5}\right),
 & \hbox{ \mbox{for large $s$}. }
 \end{array}
       \right.
 \phantom{xxx}
\label{28I15.1}
\end{eqnarray}
%
%

Recall that the vacuum Raychaudhuri equation with affine parameter $s$,
\bel{10VII14.7}
    \partial_s\tau 
 + \frac{\tau^2}{2} + |\sigma|^2 
 =0
 \,,
\ee
can be solved as
\bea
 \tau (s) & = & \frac 2 s -s^{-2}\Psi^{-1}   \int_{0}^{s}  |\sigma|^2
\Psi s^2 ds
 \,,
\eeal{5I14.1}
%
where
%
\bel{9XII13.9}
 \Psi(s,\psi,\varphi) =  \exp\bigg(  -\int_s^\infty \frac {\tilde s\tau(\tilde s, \psi,\varphi)-2} {2 \tilde s} \,d\tilde s \bigg)
 \,.
\ee
%
%
%
As $s$ tends to zero, the integral in \eq{9XII13.9} approaches  infinity as $\frac 12 \ln s$,
hence the  weight-factor $\Psi$ behaves as a constant times $s^{-1/2}$. This, together with the $1/(2s^2)$-behaviour of $|\sigma|^2$ for small $s$,  leads in \eq{5I14.1} to the  required $1/s$-behaviour of $\tau$ for $s$ approaching zero.

An alternative derivation of \eq{15II16.1} proceeds by solving directly  \eq{eq:wavegaugeconstraint5}:
\bea
	\left( \partial_s + \tau   \right) \zeta
 &= &
  \frac{1}{2} |\xi|^2 - \conenabla_A \xi^A - \coneR + 8\pi \left(\overline g^{AB} \ol{T}_{AB} - \ol{T} \right)  + 2 \Lambda
 	\, .
\label{14XII14.3a}
\eea
In the current case \eq{14XII14.3a} reads
\bea
	\left( \frac{dr}{ds}\partial_r + \tau   \right) \zeta%
 & = &
 2 \Lambda
	\, .
\label{14XII14.3b}
\eea
It follows from \eq{14XII14.3b} that
\be
 \frac{d( \zeta \sqrt {\det \localchg})}{ds} = 2 \Lambda  \sqrt {\det \localchg}
\,.
\ee
Integrating in $s$, we find
\bean
 \left( \zeta \sqrt {\det \localchg}\right)(s)
 &= &
  \underbrace{
    \lim_{s\to 0} \left( \zeta \sqrt {\det \localchg} \right)(s)
 }_{-3 m \lambda}+
 2 \Lambda \int_{0}^s  \sqrt {\det \localchg}
 \; ds
\\
 & = &
  \lambda
 \big(-3 m  +
 \frac{2 \Lambda}3 (r^3-r_0^3)
 \big)
 \nonumber
\\
 & = &
  \lambda
 \big(-3 m  +
 \frac{2 \Lambda}3 (r^3-2 m \ell^2)
 \big)
 \,,
\eea
which coincides indeed with \eq{15II16.1}:
$$
 \zeta \sqrt {\det \localchg}= \lambda  \left(m-\frac{2 r^3}{\ell ^2}\right)=
\lambda  \left(m+\frac{2 \Lambda r^3}{3}\right)
 \,.
$$

We are ready now  to check the contribution of various terms to the mass identity \eqref{eq:massfinal} for
the Horowitz-Myers metrics  \eq{6XI12.4xy}.  For these metrics we have $T_{\mu\nu}\equiv 0 \equiv \zR \equiv \xi \equiv |\sigma|^2_4 \equiv |\sigma|^2_5$ (compare Equation \eqref{16XII14.1}),  and from \eq{22XII14.1} and \eq{eq:massfinal} we find
 \ptctodo{what is the geometric significance of the integrand over the core geodesic after the limit has been taken?}
\bel{14XII14.1}
	-2 m \mu_{\hat h}(\sectionofScri)
	=
	16 \pi \mTB
	=
	\lim_{s \rightarrow 0}
  \int_\sectionofScri  \zeta d\mu_{\coneg}
	+
	\frac{\Lambda}{12} \int_{\sectionofscri}{ \tau_2^3 } d\mu_{\hat h}
  +
	2{\Lambda V_\renm}
	\,.
\ee
Recall that $\coneg$ is defined as the angular part of the metric on the light-cone,
$$
\coneg =\ol g_{AB}dx^A dx^B =  f \lambda ^2 \ell^2  d\chicoordinate^2 + r^2 d\varphi^2
\,.
$$
and that  the limiting metric $\hat h$ defined in \eq{24II15.11} is
$$
 \hat h = \lim_{r\to\infty}r^{-2}\coneg =  \lambda^2   d\chicoordinate^2  +
 d\varphi^2
 \,.
$$
Keeping in mind that the measure associated with $\hat h$ is
$$
 d\mu_{\hat h}= \lambda \,
 d\chicoordinate
 \,d\varphi
 \,,
$$
\eq{13XII14.12} and \eq{28I15.1} lead to
\bean
 \lim_{r\to r_0} \int_{\sectionofScri } \zeta d\mu_{\coneg}
 & = &
  -  \frac{r_0^2}{\ell^{ 2}} \times \frac {\ell^2 f'(r_0)}2 \times \mu_{\hat h}(\sectionofScri)
\\
 & = &
  -  \frac{r_0^2}{\ell^{ 2}} \times \frac {\ell^2 f'(r_0)}2 \times
 \left( \frac 1 {2 \ell f'(r_0)}\times (2\pi)^2\right)
 \,.
\eeal{12VII14.1}
\Eq{28I15.2} gives
\bel{28I15.3}
 f'(r_0) = \frac 1{r_0}\left(2 \frac{r_0^2}{\ell^2} + \frac {2 m}{r_0 }\right)
 =
 \frac {6 m}{r_0 ^2}
 =
 \frac {3 (2 m \ell^2)^{\frac 13} }{\ell ^2}
 \,.
\ee
We can thus rewrite \eq{12VII14.1} as
\bean
 \lim_{r\to r_0} \int_{\sectionofScri } \zeta d\mu_{\coneg}
 & = &
  -  {3m} \times \mu_{\hat h}(\sectionofScri)
\\
 & = &
  -  {3m}
   \times
    \frac {2\pi^2  \ell } {3 (2 m\ell^2)^{\frac 13}}
    =
  -  {3m}
   \times
    \frac {2 \pi^2 } {3}
    \left(\frac {   \ell } { 2 m }\right)^{\frac 13}
 \,.
\eeal{12VII14.3}
The relation $\Lambda = - 3/\ell^2$
and \eq{14XII14.1} give the \emph{balance formula}
\bel{14XII14.1ag}
	2 m \ell^2 \mu_{\hat h}(\sectionofScri)
 =
    {3m\ell^2}  \mu_{\hat h}(\sectionofScri)	
  +
	6 V_\renm
	+
	\frac{1}{4} \int_{\sectionofscri}{ \tau_2^3 } d\mu_{\hat h}
	\, .
\ee
%
%

We note that \eqref{22IV15.1} gives $\tau_2 = 2(s_* - r_0)$, and that \eqref{29I15.13} can be rewritten as
\bel{22IV15.2}
	V_\renm = - \frac{1}{6} \left(   { m \ell^2} \mu_{\hat h}(\sectionofScri) + 2 \int_{\sectionofscri}{ (s_* - r_0)^3 } d\mu_{\hat h} \right)
	\,,
\ee
in agreement with \eq{14XII14.1ag}.

\def\polhk#1{\setbox0=\hbox{#1}{\ooalign{\hidewidth
  \lower1.5ex\hbox{`}\hidewidth\crcr\unhbox0}}} \def\cprime{$'$}
  \def\cprime{$'$}
\providecommand{\bysame}{\leavevmode\hbox to3em{\hrulefill}\thinspace}
\providecommand{\MR}{\relax\ifhmode\unskip\space\fi MR }
\providecommand{\MRhref}[2]{%
  \href{http://www.ams.org/mathscinet-getitem?mr=#1}{#2}
}
\providecommand{\href}[2]{#2}


\begin{thebibliography}{10}

\bibitem{AbbottDeser}
L.F. Abbott and S.~Deser, \emph{Stability of gravity with a cosmological
  constant}, Nucl.\ Phys. \textbf{B195} (1982), 76--96.

\bibitem{AshtekarBongaKesavanI}
A.~Ashtekar, B.~Bonga, and A.~Kesavan, \emph{Asymptotics with a positive
  cosmological constant: {I}. {B}asic framework}, Class.\ Quantum Grav.
  \textbf{32} (2015), 025004, 41 pp., arXiv:1409.3816 [gr-qc]. \MR{3291776}

\bibitem{AshtekarBK}
\bysame, \emph{{Asymptotics with a positive cosmological constant: II. Linear
  fields on de Sitter space-time}}, Phys. Rev. \textbf{D92} (2015), 044011,
  arXiv:1506.06152 [gr-qc].

\bibitem{Birmingham}
D.~Birmingham, \emph{Topological black holes in anti-de {Sitter} space},
  Class.\ Quantum Grav. \textbf{16} (1999), 1197--1205, arXiv:hep-th/9808032.
  \MR{MR1696149 (2000c:83062)}

\bibitem{BBM}
H.~Bondi, M.G.J. van~der Burg, and A.W.K. Metzner, \emph{Gravitational waves in
  general relativity {VII}: Waves from axi--symmetric isolated systems}, Proc.\
  Roy.\ Soc.\ London A \textbf{269} (1962), 21--52. \MR{MR0147276 (26 \#4793)}

\bibitem{CCG}
Y.~Choquet-Bruhat, P.T. Chru\'{s}ciel, and J.M. Mart\'in-Garc\'ia, \emph{{The
  light-cone theorem}}, Class.\ Quantum Grav. \textbf{26} (2009), 135011 (22
  pp), arXiv:0905.2133 [gr-qc]. \MR{2515694 (2010g:53131)}

\bibitem{CCM2}
\bysame, \emph{{The Cauchy problem on a characteristic cone for the Einstein
  equations in arbitrary dimensions}}, Ann.\ H.\ Poincar\'e \textbf{12} (2011),
  419--482, arXiv:1006.4467 [gr-qc]. \MR{2785136}

\bibitem{ChEnergy}
P.T. Chru\'{s}ciel, \emph{Lectures on energy in general relativity},
  \url{http://homepage.univie.ac.at/piotr.chrusciel/teaching/Energy/Energy.pdf}.

\bibitem{ChAIHP}
P.T. Chru\'{s}ciel, \emph{On the relation between the {Einstein} and the
  {Komar} expressions for the energy of the gravitational field}, Ann.\ Inst.\
  Henri Poincar\'e \textbf{42} (1985), 267--282. \MR{797276 (86k:83018)}

\bibitem{ChBamberg}
\bysame, \emph{The {H}amiltonian mass and asymptotically anti-de {S}itter
  space-times}, Proceedings of the Symposium ``100 Years Werner
  Heisenberg---Works and Impact'' (Bamberg, 2001), vol.~50, 2002, pp.~624--629.
  \MR{MR1909102 (2003h:83034)}

\bibitem{ChHerzlich}
P.T. Chru\'{s}ciel and M.~Herzlich, \emph{The mass of asymptotically hyperbolic
  {R}iemannian manifolds}, Pacific J. Math. \textbf{212} (2003), 231--264,
  arXiv:dg-ga/0110035. \MR{MR2038048 (2005d:53052)}

\bibitem{CJK2}
P.T. Chru\'{s}ciel, J.~Jezierski, and J.~Kijowski, \emph{{The Hamiltonian mass
  of asymptotically Schwarzschild-de Sitter space-times}}, Phys.\ Rev.
  \textbf{D87} (2013), 124015 (11 pp.), arXiv:1305.1014 [gr-qc].

\bibitem{CJL}
P.T. Chru\'{s}ciel, J.~Jezierski, and S.~\L\c{e}ski, \emph{The{ Trautman-Bondi}
  mass of hyperboloidal initial data sets}, Adv.\ Theor.\ Math.\ Phys.
  \textbf{8} (2004), 83--139, arXiv:gr-qc/0307109. \MR{MR2086675 (2005j:83027)}

\bibitem{ChMS}
P.T. Chru{\'s}ciel, M.A.H. MacCallum, and D.~Singleton, \emph{Gravitational
  waves in general relativity. {XIV}: {B}ondi expansions and the
  ``polyhomogeneity'' of {S}cri}, Philos.\ Trans.\ Roy.\ Soc.\ London Ser. A
  \textbf{350} (1995), 113--141, arXiv:gr-qc/9305021. \MR{MR1325206
  (97f:83025)}

\bibitem{ChNagy}
P.T. Chru\'{s}ciel and G.~Nagy, \emph{The {H}amiltonian mass of asymptotically
  {anti-de Sitter} space-times}, Class.\ Quantum Grav. \textbf{18} (2001),
  L61--L68, hep-th/0011270.

\bibitem{ChNagyATMP}
\bysame, \emph{The mass of spacelike hypersurfaces in asymptotically {anti-de
  Sitter} space-times}, Adv.\ Theor.\ Math.\ Phys. \textbf{5} (2002), 697--754,
  arXiv:gr-qc/0110014.

\bibitem{ChPaetzBondi}
P.T. Chru\'{s}ciel and T.-T. Paetz, \emph{The mass of light-cones}, Class.\
  Quantum Grav.\ \textbf{31} (2014), 102001, arXiv1401.3789 [gr-qc].

\bibitem{ChPaetz3}
\bysame, \emph{{Light-cone initial data and smoothness of Scri. I. Formalism
  and results}}, Ann.\ H.\ Poincar\'e \textbf{16} (2015), 2131--2162,
  arXiv:1403.3558 [gr-qc]. \MR{3383324}

\bibitem{ChruscielSimon}
P.T. Chru\'{s}ciel and W.~Simon, \emph{Towards the classification of static
  vacuum space-times with negative cosmological constant}, Jour.\ Math.\ Phys.
  \textbf{42} (2001), 1779--1817, arXiv:gr-qc/0004032. \MR{1820431
  (2002j:83013)}

\bibitem{FG}
C.~Fefferman and C.R. Graham, \emph{{Conformal invariants}}, {\'Elie Cartan et
  les math\'ematiques d'aujourd'hui, The mathematical heritage of \'Elie
  Cartan, S\'emin. Lyon 1984, Ast\'erisque, No.Hors S\'er. 1985, 95-116}.

\bibitem{FriedrichCMP}
H.~Friedrich, \emph{On the hyperbolicity of {E}instein's and other gauge field
  equations}, Commun.\ Math.\ Phys. \textbf{100} (1985), 525--543. \MR{MR806251
  (86m:83009)}

\bibitem{Friedrich}
\bysame, \emph{On the existence of n--geodesically complete or future complete
  solutions of {E}instein's field equations with smooth asymptotic structure},
  Commun.\ Math.\ Phys. \textbf{107} (1986), 587--609.

\bibitem{friedrich:JDG}
\bysame, \emph{On the global existence and the asymptotic behavior of solutions
  to the {E}instein-{M}axwell-{Y}ang-{M}ills equations}, Jour.\ Diff.\ Geom.
  \textbf{34} (1991), 275--345. \MR{MR1131434 (92i:58191)}

\bibitem{FriedrichMassive}
\bysame, \emph{{Smooth non-zero rest-mass evolution across time-like
  infinity}}, Ann.\ Henri Poincar\'e \textbf{16} (2015), 2215--2238,
  arXiv:1311.0700 [gr-qc].

\bibitem{FriedrichDust}
\bysame, \emph{{Sharp asymptotics for Einstein-$\lambda$-dust flows}}, Commun.\
  Math.\ Phys. (2016), in press, arXiv:1601.04506 [gr-qc].

\bibitem{GrahamVolume}
C.R. Graham, \emph{Volume and area renormalizations for conformally compact
  {E}instein metrics}, The {P}roceedings of the 19th {W}inter {S}chool
  ``{G}eometry and {P}hysics'' ({S}rn\'\i, 1999), no.~63, 2000,
  arXiv:math/9909042 [math.DG], pp.~31--42. \MR{1758076 (2002c:53073)}

\bibitem{HorowitzMyers}
G.T. Horowitz and R.C. Myers, \emph{The {AdS/CFT} correspondence and a new
  positive energy conjecture for general relativity}, Phys. Rev. \textbf{D59}
  (1999), 026005 (12 pp.), arXiv:hep-th/9808079.

\bibitem{TimAsymptotics}
T.-T. Paetz, \emph{Characteristic initial data and smoothness of {Scri}. {II.
  Asymptotic} expansions and construction of conformally smooth data sets},
  Jour.\ Math.\ Phys. \textbf{55} (2014), 102503, arXiv:1403.3560 [gr-qc].

\bibitem{TimDiss}
\bysame, \emph{{On characteristic Cauchy problems in general relativity}},
  Ph.D. thesis, University of Vienna, 2014,
  \url{http://homepage.univie.ac.at/piotr.chrusciel/papers/Tim.pdf}.

\bibitem{TimConformal}
\bysame, \emph{{Conformally covariant systems of wave equations and their
  equivalence to Einstein's field equations}}, Ann. H.~Poincar\'e \textbf{16}
  (2015), 2059--2129, arXiv:1306.6204 [gr-qc]. \MR{3383323}

\bibitem{Sachs}
R.K. Sachs, \emph{Gravitational waves in general relativity {VIII.} {Waves} in
  asymptotically flat space-time}, Proc.\ Roy.\ Soc.\ London A \textbf{270}
  (1962), 103--126. \MR{MR0149908 (26 \#7393)}

\bibitem{TodSzabados}
L.B. Szabados and P.~Tod, \emph{{A positive Bondi--type mass in asymptotically
  de Sitter spacetimes}}, Class. Quant. Grav. \textbf{32} (2015), 205011,
  arXiv:1505.06637 [gr-qc].

\bibitem{TafelBondi2}
J.~Tafel, \emph{{On the energy of a null cone}}, Class.\ Quantum Grav.
  \textbf{31} (2014), 235011, [Class. Quant. Grav.31,235011(2014)].

\bibitem{TamburinoWinicour}
L.A. Tamburino and J.H. Winicour, \emph{Gravitational fields in finite and
  conformal {B}ondi frames}, Phys.\ Rev. \textbf{150} (1966), 1039--1053.

\bibitem{T}
A.~Trautman, \emph{Radiation and boundary conditions in the theory of
  gravitation}, Bull. Acad. Pol. Sci., S\'erie sci. math., astr. et phys.
  \textbf{VI} (1958), 407--412.

\bibitem{Wang}
X.~Wang, \emph{Mass for asymptotically hyperbolic manifolds}, Jour.\ Diff.\
  Geom. \textbf{57} (2001), 273--299. \MR{MR1879228 (2003c:53044)}

\bibitem{WoolgarRigidityHM}
E.~Woolgar, \emph{{The rigid Horowitz-Myers conjecture}},  (2016),
  arXiv:1602.06197.

\end{thebibliography}
\end{document}